\documentclass[PAPER, atlasdraft=false, UKenglish, texmf, cernpreprint, orcidlogo, texmf, orcidlogo]{atlasdoc} %
\usepackage[subfigure]{atlaspackage}
\usepackage{atlasbiblatex}

\usepackage{atlasphysics}
\graphicspath{{logos/}{figures/}}

\usepackage{ANA-TOPQ-2020-23-PAPER-defs}

\usepackage{multirow}
\usepackage{cancel}
\usepackage{makecell}
\usepackage{float}
\usepackage{array}

\AtlasTitle{
Search for flavour-changing neutral-current couplings between the top quark
and the Higgs boson in multi-lepton final states in \SI{13}{\text{T\text{e\kern-0.1em V}}} $pp$ collisions
with the ATLAS detector
}

\AtlasAbstract{
A search is presented for flavour-changing neutral-current interactions involving the top quark,
the Higgs boson and an up-type quark (\(q=u,c\)) with the ATLAS detector at
the Large Hadron Collider.
The analysis considers leptonic decays of the top quark along
with Higgs boson decays into two \(W\) bosons, two \(Z\) bosons or a
\(\tau^{+}\tau^{-}\) pair.
It focuses on final states containing either two leptons (electrons or muons)
of the same charge or three leptons.
The considered processes are $t\bar{t}$ and $Ht$ production.
For the $t\bar{t}$ production, one top quark decays via \(t\to Hq\).
The proton--proton collision data set analysed amounts to \(\SI{140}{\text{fb}^{-1}}\)
at \(\sqrt{s}=\SI{13}{\text{T\text{e\kern-0.1em V}}}\).
No significant excess beyond Standard Model expectations is observed and upper
limits are set on the \(t\to Hq\) branching ratios at \SI{95}{\percent}
confidence level, amounting to observed (expected) limits of
\(\mathcal{B}(t\to Hu)<2.8\,(3.0) \times 10^{-4}\) and
\(\mathcal{B}(t\to Hc)<3.3\,(3.8) \times 10^{-4}\).
Combining this search with other searches for \(tHq\) flavour-changing
neutral-current interactions previously conducted by ATLAS, considering \(H\to b\bar{b}\) and \(H\to\gamma\gamma\) decays, as
well as \(H\to\tau^{+}\tau^{-}\) decays with one or two hadronically decaying $\tau$-leptons,
yields observed (expected) upper limits on the branching ratios of
\(\mathcal{B}(t\to Hu)<2.6\,(1.8) \times 10^{-4}\) and
\(\mathcal{B}(t\to Hc)<3.4\,(2.3) \times 10^{-4}\).
}

\AtlasRefCode{TOPQ-2020-23}

\PreprintIdNumber{CERN-EP-2024-070}

\AtlasJournalRef{Eur.\ Phys.\ J. C 84 (2024) 757}
\AtlasDOI{10.1140/epjc/s10052-024-12994-1}

\hypersetup{pdftitle={ATLAS document},pdfauthor={The ATLAS Collaboration}}

\begin{document}

\maketitle

\tableofcontents

\section{Introduction}
%

% The next lines are included from the .//tex/01_introduction.tex input file

Following the discovery of the Higgs boson at the Large Hadron Collider (LHC)
by the ATLAS and CMS experiments in 2012~\cite{HIGG-2012-27,CMS-HIG-12-028}, several
measurements have probed properties of the particle.
In addition to Higgs boson couplings to fermions and gauge bosons, which agree
well with the Standard Model (SM) predictions~\cite{HIGG-2021-23,CMS-HIG-22-001}, further
interactions can be investigated to search for evidence of possible new physics
in the Higgs sector.
A possibility is that the Higgs boson has an interaction involving two up-type
quarks of different generations, denoted by \tHq where $q=(u,c)$, leading to
\(gq\to Ht\) production and \(t\to Hq\) decay.
These are flavour-changing neutral-current (FCNC) interactions, forbidden at
tree-level in the SM and suppressed at higher orders due to the
Glashow-Iliopoulos-Maiani (GIM) mechanism~\cite{GIM_mechanism}.
The SM predictions for the $t \to Hq$ branching ratios \(\BR(\ttoHq)\) are very
small, around $10^{-15}$~\cite{Eilam:1990zc,Mele:1998ag,AguilarSaavedra:2004wm,Zhang:2013xya,PhysRevD.100.015003},
and beyond the sensitivity of the LHC\@.
However, in the framework of new physics models such as two-Higgs-doublet models
(2HDMs)~\cite{Branco:2hdm2012}, the minimal supersymmetric SM
(MSSM)~\cite{Bejar:2000ub,Guasch:1999jp,Cao:2007dk,Cao:2014udj}, supersymmetric
models with R-parity violation~\cite{Eilam:2001dh}, quark-singlet
models~\cite{AguilarSaavedra:2002kr}, and warped extra dimensions~\cite{Azatov:2009na},
the branching ratios of these processes are modified making them large enough to be
measured at the LHC\@.
For example, branching ratios of up to $10^{-3}$ are possible in 2HDMs
without explicit flavour symmetry, since there are no symmetries explicitly
forbidding tree-level FCNC interactions~\cite{Cheng:1987rs,Baum:2008qm,Chen:2013qta,Chiang:2015cba,Crivellin:2015hha,Botella:2015hoa,Gori:2017tvg,Chiang:2017fjr} in this model.
The FCNC interaction can also mediate the production of a Higgs boson
with a single top quark ($pp \to tH$)~\cite{Greljo:2014dka}.
The SM prediction of the $pp \to tH$ cross-section is $74.3^{+0.4}_{-0.3}$\,fb at
$\sqrt{s} = \SI{13}{\TeV}$~\cite{deFlorian:2016spz}.

The FCNC interactions are introduced using an effective field theory (EFT) framework,
which is used for indirect searches for new physics~\cite{FCNC_EFT}.
Here the SM is regarded as a low-energy approximation of an ultraviolet complete theory
containing new particles, whose masses are characterised by an energy scale \(\Lambda = \SI{1}{\TeV}\).
The new physics contributions are parameterised in terms of operators with mass
dimension greater than four containing only the SM fields, scaled by dimensionless
Wilson coefficients and inverse powers of \(\Lambda\).
In the case where only the \tHu and \tHc interactions are considered, the relevant operators are

\begin{align}
\opLh & =\left(\phi^\dagger\phi-\frac{v^2}{2}\right)(\bar{q}_L\,t_R)\tilde{\phi} & \opRh=\left(\phi^\dagger\phi-\frac{v^2}{2}\right)(\bar{t}_L\,q_R)\tilde{\phi},
\end{align}

where $q$ corresponds to an up or charm quark, depending on the FCNC coupling.
The index $u$ is the coupling to any up-type quark, $t$ is the top-quark, $\phi$
denotes the Higgs boson field with $v$ corresponding to the absolute value of
its vacuum expectation value.
The two left-handed quark doublet fields are $\bar{q}_L$ and $\bar{t}_L$,
with $q_R$ and $t_R$ being the corresponding right-handed singlets.
The operators are scaled with Wilson coefficients $\Cuphi^{qt}$ and $\Cuphi^{tq}$,
and $1/\Lambda^2$ to give the relevant Lagrangian:

\begin{align}
\mathcal{L}_{\text{EFT}} & =\mathcal{L}_{\text{SM}}+\sum\limits_{q=u,c}\left[\frac{\CuphiLh}{\Lambda^2}\opLh+\frac{\CuphiRh}{\Lambda^2}\opRh\right].
\end{align}

In the case of \ttbar production with a FCNC \(t\to Hq\) decay, no kinematic
differences between \opLh and \opRh are expected because the top quarks are
produced unpolarised and the Higgs boson is a scalar particle.
For \(t\to Hq\) production, comparisons of simulations using either \opLh or
\opRh have shown that differences between production rates and kinematic distributions
are negligible in the phase space considered.
Thus, the mean of both the couplings \(\Cuphiavg=\frac{\CuphiLh+\CuphiRh}{2}\) is
considered.

Both the ATLAS and CMS Collaborations have undertaken various searches for
\tHq FCNC processes, split by Higgs boson final state~\cite{TOPQ-2017-07,HIGG-2016-26,TOPQ-2017-07,TOPQ-2017-15,TOPQ-2019-17,HDBS-2020-16,TOPQ-2019-04_alt,CMS-TOP-17-003,CMS-TOP-19-002,CMS-TOP-20-007_fixed}.
In this paper, a search is conducted using final states containing either two leptons
(electrons or muons) of the same charge (\dil, $\ell=e,\mu$) or three leptons, exactly two of
which have the same charge (\tri, $\ell=e,\mu$).
The ATLAS Collaboration searched for \tHq FCNC couplings in \dil and \tri
final states  using a partial Run 2 data sample of \SI{36.1}{\ifb}~\cite{TOPQ-2017-15},
resulting in observed (expected) 95\% CL upper limits of \(\BR(t \to Hu)<19(15)\times10^{-4}\) and
\(\BR(t \to Hc)<16(15)\times10^{-4}\).
The strongest limits set by the ATLAS Collaboration come from a combination of searches
in the \(H \to b\bar{b}\), \(H \to \gamma\gamma\) and \(H\to\tau^{+}\tau^{-}\)
channels using the full Run 2 data sample~\cite{TOPQ-2019-04_alt}.
The obtained limits are \(\BR(t \to Hu)<4.0(2.4)\times10^{-4}\) and
\(\BR(t \to Hc)<5.8(3.0)\times10^{-4}\).
For the \tHc coupling, these are  the most stringent limits published to date.
However, for the \tHu coupling the strongest limits are set by a search conducted
with the CMS detector in the \(H\to\gamma\gamma\) channel~\cite{CMS-TOP-20-007_fixed},
amounting to \(\BR(t \to Hu)<1.9(3.1)\times10^{-4}\).

\begin{figure}
\centering
\subfigure[]{\includegraphics[width=0.3\linewidth]{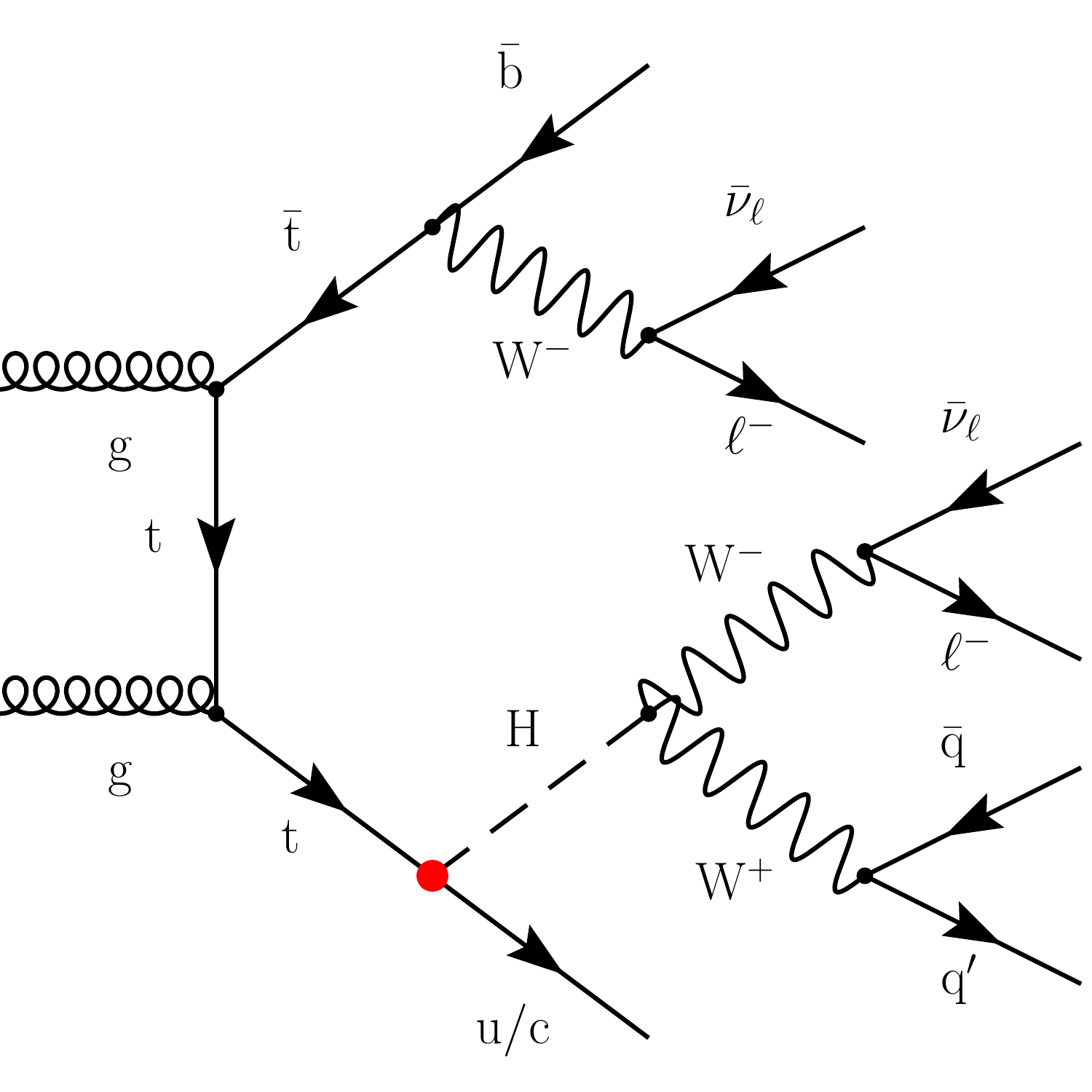}}
\hspace{0.15\linewidth}
\subfigure[]{\includegraphics[width=0.3\linewidth]{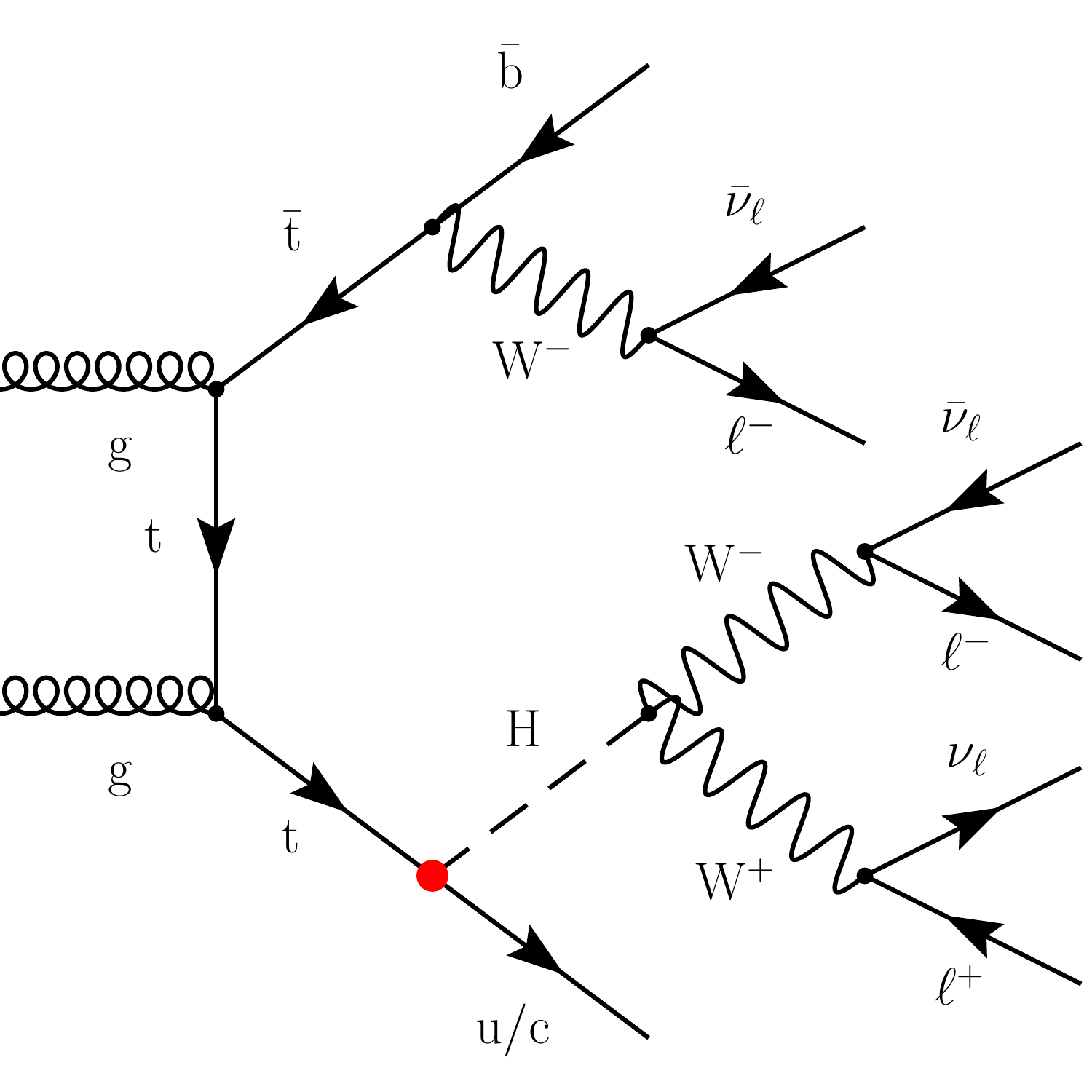}}
\caption{
Example Feynman diagrams of the \(\ttbar(t\to Hq)\) decay signal process resulting in (a) the
\dil final state and (b) the \tri final state.
The FCNC vertex is highlighted in red.
}\label{fig:feynman_decay_signal}
\end{figure}

\begin{figure}
\centering
\subfigure[]{\includegraphics[width=0.3\linewidth]{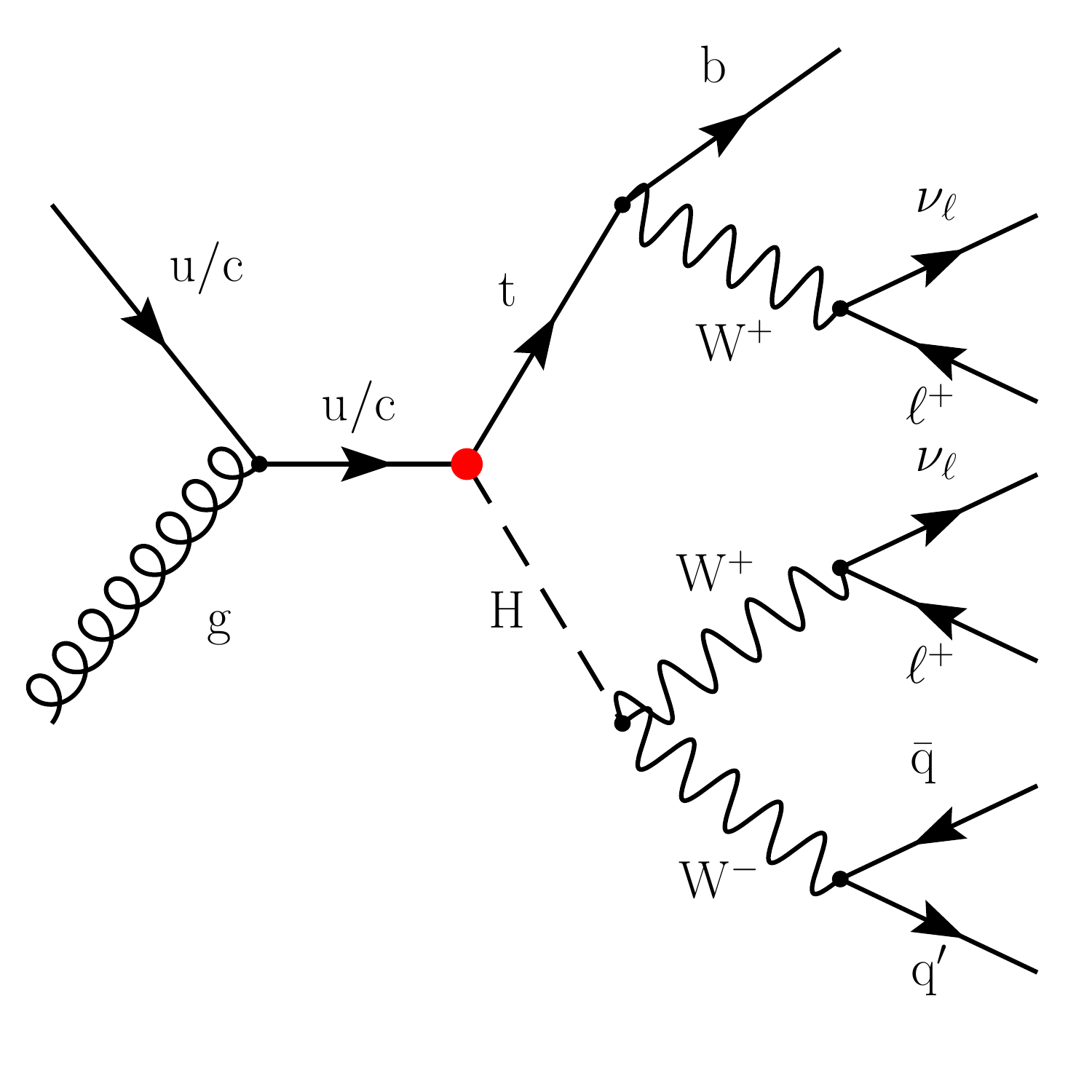}}
\hspace{0.15\linewidth}
\subfigure[]{\includegraphics[width=0.3\linewidth]{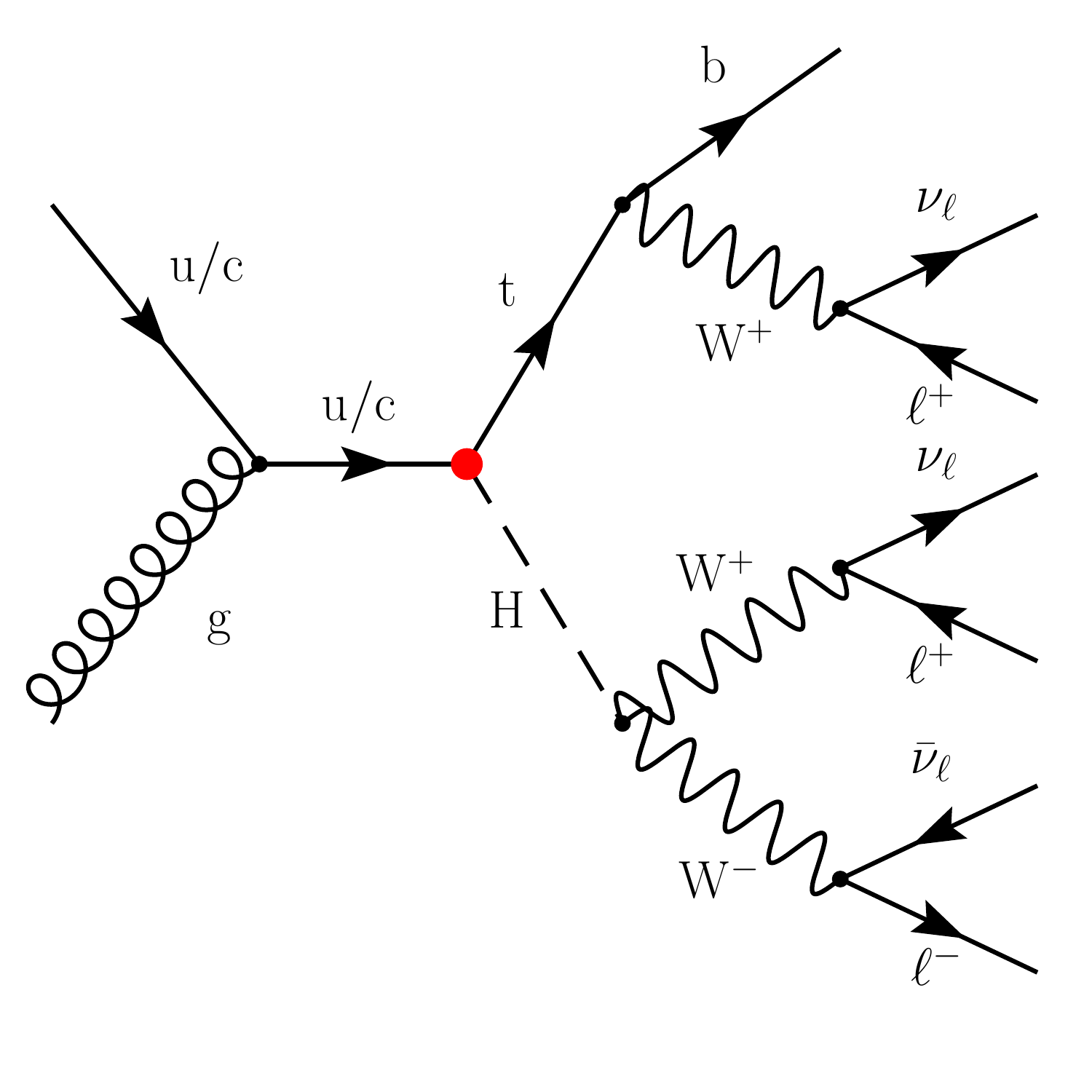}}
\caption{
Example Feynman diagrams of the \(gq\to Ht\) production signal process resulting in (a) the
\dil final state and (b) the \tri final state.
The FCNC vertex is highlighted in red.
}\label{fig:feynman_production_signal}
\end{figure}

The dominant Higgs-boson decay mode resulting in \dil and \tri final states across
production and decay processes is $H \to WW^*$, with $WW^* \to \ell \nu jj$
or $WW^* \to \ell \nu \ell \nu$ for the \dil and \tri final state respectively.
The corresponding Feynman diagrams are shown in \Cref{fig:feynman_decay_signal,fig:feynman_production_signal}.
Other Higgs-boson decay modes such as $H \to ZZ^*$ and $H \to \tau^{+}\tau^{-}$ can
also meet the selection criteria.
Based on the SM branching ratios of the Higgs boson and the top quark, approximately
\SI{73}{\percent} of all \dil signal events are expected from the $H\to WW^*$
decay mode, while the rest originate from leptonic $H\to \tautau$ decays.
Less than \SI{1}{\percent} of events are attributable to the $H\to ZZ^*$ decay mode.
In the \tri final state, the latter contributes more significantly, being responsible
for \SI{14}{\percent} of all events.
The majority, accounting for \SI{54}{\percent} of the \tri signal events,
originate from $H\to WW^*$ decays, while the remaining \SI{32}{\percent} arise
from $H\to\tautau$ events.
These numbers do not take into account detector acceptance effects or selection
requirements.
Although the total number of events meeting the \dil and \tri criteria is modest,
the low statistical precision is balanced by minimal background contributions.

The strategy of the analysis is to first identify kinematic regions enriched in
the signal process.
In these regions, various reconstruction algorithms are implemented to create
variables that can distinguish between the signal and several background processes.
These reconstructed variables are combined into a single discriminant using a
feed-forward neural network~\cite{Feed_forward_NN}.
The same kinematic regions are used for the \tHu and the \tHc channel of the
analysis, while individual neural networks are trained for each of the channels.
Aside from this, the analysis strategy is identical for the \tHu and the \tHc
channel.
The distribution of the neural network output is then used as input to a
maximum-likelihood fit, which considers statistical and systematic uncertainties.
In addition to the signal-enriched regions, several regions enriched in specific
background processes are included in the fit, to constrain the normalisation of
these processes.
If the fit shows evidence of a signal, the corresponding significance is
determined.
Otherwise, upper limits are set on the FCNC branching ratios and the Wilson
coefficients of the EFT dimension-6 operators.
Finally, the results of this analysis are statistically combined with those from
other ATLAS searches for \tHq FCNC interactions in different final states.

% End of text imported from the .//tex/01_introduction.tex input file

%
\section{The ATLAS detector}
%

% The next lines are included from the .//tex/02_atlas-detector.tex input file
%
\newcommand{\AtlasCoordFootnote}{%
ATLAS uses a right-handed coordinate system with its origin at the nominal interaction point (IP)
in the centre of the detector and the \(z\)-axis along the beam pipe.
The \(x\)-axis points from the IP to the centre of the LHC ring,
and the \(y\)-axis points upwards.
Cylindrical coordinates \((r,\phi)\) are used in the transverse plane,
\(\phi\) being the azimuthal angle around the \(z\)-axis.
The pseudorapidity is defined in terms of the polar angle \(\theta\) as \(\eta = -\ln \tan(\theta/2)\).
Angular distance is measured in units of \(\Delta R \equiv \sqrt{(\Delta\eta)^{2} + (\Delta\phi)^{2}}\).}

The ATLAS detector~\cite{PERF-2007-01} at the LHC covers nearly the entire solid angle around the collision point.\footnote{\AtlasCoordFootnote}
It consists of an inner tracking detector surrounded by a thin superconducting solenoid, electromagnetic and hadron calorimeters,
and a muon spectrometer incorporating three large superconducting air-core toroidal magnets.

The inner-detector system (ID) is immersed in a \qty{2}{\tesla} axial magnetic field
and provides charged-particle tracking in the range of \(|\eta| < 2.5\).
The high-granularity silicon pixel detector covers the vertex region and typically provides four measurements per track,
the first hit normally being in the insertable B-layer (IBL) installed before Run~2~\cite{ATLAS-TDR-19,PIX-2018-001}.
It is followed by the silicon microstrip tracker (SCT), which usually provides eight measurements per track.
These silicon detectors are complemented by the transition radiation tracker (TRT),
which enables radially extended track reconstruction up to \(|\eta| = 2.0\).
The TRT also provides electron identification information
based on the fraction of hits (typically 30 in total) above a higher energy-deposit threshold corresponding to transition radiation.

The calorimeter system covers the pseudorapidity range \(|\eta| < 4.9\).
Within the region \(|\eta|< 3.2\), electromagnetic calorimetry is provided by barrel and
endcap high-granularity lead/liquid-argon (LAr) calorimeters,
with an additional thin LAr presampler covering \(|\eta| < 1.8\)
to correct for energy loss in material upstream of the calorimeters.
Hadron calorimetry is provided by the steel/scintillator-tile calorimeter,
segmented into three barrel structures within \(|\eta| = 1.7\), and two copper/LAr hadron endcap calorimeters.
The solid angle coverage is completed with forward copper/LAr and tungsten/LAr calorimeter modules
optimised for electromagnetic and hadronic energy measurements respectively.

The muon spectrometer (MS) comprises separate trigger and
high-precision tracking chambers measuring the deflection of muons in a magnetic field generated by the superconducting air-core toroidal magnets.
The field integral of the toroids ranges between \num{2.0} and \qty{6.0}{\tesla\metre}
across most of the detector.
Three layers of precision chambers, each consisting of layers of monitored drift tubes, cover the region \(|\eta| < 2.7\)
and are complemented by cathode-strip chambers in the forward region, where the detector occupancy is highest.
The muon trigger system covers the range of \(|\eta| < 2.4\) with resistive-plate chambers in the barrel, and thin-gap chambers in the endcap regions.

Recorded events are selected by the first-level trigger system implemented in custom hardware,
followed by selections made by algorithms implemented in software in the high-level trigger~\cite{TRIG-2016-01}.
The first-level trigger accepts events from the \qty{40}{\MHz} bunch crossings at a rate below \qty{100}{\kHz},
which the high-level trigger further reduces  to record events to disk at about \qty{1}{\kHz}.

An extensive software suite~\cite{ATL-SOFT-PUB-2021-001} is used in data simulation, in the reconstruction
and analysis of real and simulated data, in detector operations, and in the trigger and data acquisition
systems of the experiment.

% End of text imported from the .//tex/02_atlas-detector.tex input file

%
\section{Samples of data and simulated events}
%

% The next lines are included from the .//tex/03_samples.tex input file
%
Proton--proton ($pp$) collision data are used recorded with the ATLAS detector in the years 2015 to 2018
at a centre-of-mass energy of 13$~\TeV$.
After applying data-quality requirements~\cite{DAPR-2018-01}, the data sample corresponds to an integrated luminosity
of \SI{140.1(12)}{\ifb}~\cite{DAPR-2021-01}. The LUCID-2 detector~\cite{LUCID-2}
was used for the primary luminosity measurements, complemented by measurements using the inner detector and calorimeters.

Events were selected online during data taking by single-electron or single-muon triggers~\cite{TRIG-2018-05,TRIG-2018-01}.
Multiple unprescaled triggers were combined in a logical OR to increase the selection efficiency.
The lowest-threshold triggers utilised isolation requirements to reduce the trigger rate.
The higher-level lepton triggers had transverse momentum (\pT) thresholds of $20\,\GeV$ for muons and $24\,\GeV$ for electrons in 2015 data,
and $26\,\GeV$ for both the  lepton types in 2016, 2017 and 2018 data.
They were complemented by other triggers with higher \pT thresholds but no isolation requirements
to increase the trigger efficiency.

Large sets of simulated events from signal and background processes were produced with
Monte Carlo (MC) event generator programs to model the selected data.
After event simulation, the response of the ATLAS detector was simulated using the
\GEANT toolkit~\cite{Agostinelli:2002hh} with a full detector model~\cite{SOFT-2010-01}
or a fast simulation~\cite{ATL-PHYS-PUB-2010-013,SOFT-2010-01} which employed a parameterisation of the calorimeter response.
Samples relying on fast simulation were used to evaluate systematic uncertainties in the
event generators and for the modelling of \ttbar production with two bosons.

To account for additional inelastic \(pp\) collisions in the same and neighbouring bunch crossings (pile-up),
minimum-bias interactions were overlaid on the hard-scattering events at the level of energy depositions
simulated using \GEANT.
The minimum-bias events were simulated using \PYTHIA[8.186]~\cite{Sjostrand:2014zea} with the
A3~\cite{ATL-PHYS-PUB-2016-017} set of tuned parameters and the \NNPDF[2.3lo] set of parton distribution functions (PDF)~\cite{Ball:2012cx}.
The resulting events were weighted to reproduce the observed pile-up distribution.
The average number of interactions per bunch crossing during the entire data-taking period from 2015 to 2018
is 33.7.

Finally, the simulated events were reconstructed using the same software as applied to the collision data.
The same event selection requirements were applied and the selected events were passed through the same analysis chain.
Small corrections were applied to simulated events such that the efficiencies of the utilised
reconstruction methods are in better agreement with the response observed in data.
More details of the simulated event samples are provided in the following subsections.
Except for the events simulated with the \SHERPA generator~\cite{Gleisberg:2008ta,Bothmann:2019yzt},
the \EVTGEN~\cite{Lange:2001uf} program was used to simulate bottom and charm hadron decays.
If not mentioned otherwise, the top-quark mass was set to \(m_t=\SI{172.5}{\GeV}\)
and the Higgs-boson mass to $m_H=\SI{125}{\GeV}$.
All processes for which the parton shower is simulated using \Pythia[8] use the A14 set of tuned
parameters~\cite{ATL-PHYS-PUB-2014-021} and the \NNPDF[2.3lo] PDF set.

\subsection[Simulation of \tHq FCNC signal samples]{
Simulation of \(\boldsymbol{\tHq}\) FCNC signal samples
}
Two different signal processes are studied: the \(\ttbar(t\to Hq)\)
decay and the \(gq\to Ht\) production process.
The EFT operators parametrising the FCNC coupling are implemented in the \TopFCNC~\cite{fcnc_production_theory}
model using the \FEYNRULES[2.0] framework~\cite{ALLOUL20142250}, allowing for
next-to-leading-order (\NLO) calculations of the postulated processes.
The decay process events were simulated using the \NLO matrix-element generator
\POWHEGBOX[v2]~\cite{Nason:2004rx,Frixione:2007nw,Frixione:2007vw,Alioli:2009je,Alioli:2010xd,Re:2010bp,Frederix:2012dh},
with the \NNPDF[3.0nlo] PDF set~\cite{Ball:2014uwa}.
The production of a top quark--antiquark pair is performed according to SM
calculations in the five-flavour scheme, setting the masses of
all quarks except for the top quark to zero.
One top quark is required to decay via the SM decay mode \(t\to Wb\), while the other performs the FCNC decay \(t\to Hq\).
Both top-quark decays are modelled using \MADSPIN~\cite{Frixione:2007zp,Artoisenet:2012st}.
Only leptonic final states for the SM decay of the top quark are considered.
The Higgs-boson decay, and parton showers, hadronisation, and the
underlying event, were modelled using \PYTHIA[8.308]~\cite{10.21468/SciPostPhysCodeb.8}.
The matrix-element-to-parton-shower matching is steered by the $\hdamp$ parameter,
which controls the \pt of the first additional gluon emission beyond the leading-order (\LO)
Feynman diagram in the parton shower and therefore regulates the high-\pt
emission against which the \ttbar system recoils.
The event generation was performed with $\hdamp=1.5m_t$~\cite{ATL-PHYS-PUB-2016-020}.
Four separate samples are produced, modelling the \tHu/ \tHc coupling with either
the top quark or the top antiquark decaying via the FCNC interaction.
For consistency across searches, all \tHq FCNC samples are normalised to the
cross-section corresponding to a branching ratio of \(\BR(t\to Hq)=\SI{0.1}{\percent}\).
As explained previously, kinematic differences between the left- and right-handed
couplings are negligible in the case of  \(t\to Hq\) decay process.
Therefore, the same samples are used to model the left-handed and the
right-handed FCNC couplings.
The matrix element generator \MGNLO[2.9.9]~\cite{Alwall:2014hca} was used to
model the \(gq\to Ht\) production signal using the \TopFCNC model with
the \NNPDF[3.0nlo] PDF set and the five-flavour scheme.
Separate samples are produced, each with exactly one of the Wilson coefficients
set to 1.0 and the others set to zero, all of them at \NLO in QCD\@.

The  SM top-quark and Higgs-boson decays are
simulated using \MADSPIN\@.
The three relevant Higgs-boson decays ($H\to WW^*$, $H\to ZZ^*$,
$H\to \tau^{+}\tau^{-}$) are modelled in separate samples.
All Higgs-boson decays, parton shower, hadronisation, and the
underlying event, were modelled using \PYTHIA[8.307].
The samples are normalised to the cross-section calculated with \MGNLO\@.

\subsection{Simulation of background processes}
\label{sec:mc_samples_bg}
Samples of simulated events from SM \ttbar and single-top-quark production were
simulated using the \NLO matrix-element generator \POWHEGBOX[v2].
For \ttbar and \(tW\) production and $s$-channel single-top-quark
production ($t\bar{b}$ production) the \NNPDF[3.0nlo] PDF set was used with the
five-flavour scheme.
Following a recommendation given in Ref.~\cite{Frederix:2012dh}, single top-quark
production in the \(t\)-channel ($tq$ production) was simulated with the
\NNPDF[3.0nlo\_nf4] PDF set, which implements the four-flavour scheme.
Parton showers, hadronisation, and the underlying event were modelled using
\PYTHIA[8.230].
The event simulation used a value of $\hdamp=1.5m_t$.

The \ttbar production cross-section was scaled to \(\sigma(\ttbar)=832\,\)pb,
the  value obtained from
next-to-next-to-leading-order  predictions from the \TOPpp{2.0}
program (see Ref.~\cite{Czakon:2011xx} and references therein), and it includes
the resummation of next-to-next-to-leading logarithmic soft-gluon terms.
The predicted cross-sections of \(tq\) and \(\tbar q\) production used to scale
the corresponding samples of simulated events are \(\sigma(tq)=136\,\)pb
and \(\sigma(\bar{t}q)=81\,\)pb and were calculated at \NLO in QCD with
the \HATHOR[2.1] program~\cite{Aliev:2010zk,Kant:2014oha}.
The total cross-section for $t\bar{b}$ production was also computed at \NLO in QCD with
the \HATHOR[2.1] program and the corresponding sample
of simulated events was scaled to \(\sigma(t\bbar+\tbar b)=10.32\,\)pb.
The cross-section used for normalising the \(tW\) sample is
\(\sigma(tW+\tbar W)=71.7\,\)pb~\cite{Kidonakis:2015nna}.

The production of a vector boson in association with jets ($V$+jets, $V=W,Z$), including
$b$- and $c$-jets, was simulated with the \SHERPA[2.2.11] generator.
NLO-accurate matrix elements for up to two partons and LO-accurate matrix
elements for three to five partons are calculated in the five-flavour scheme
using the \textsc{Comix}~\cite{Gleisberg:2008fv} and
\OPENLOOPS~\cite{Buccioni:2019sur,Cascioli:2011va,Denner:2016kdg} libraries.
The default \SHERPA parton shower~\cite{Schumann:2007mg} based on Catani--Seymour
dipole factorisation and the cluster hadronisation model~\cite{Winter:2003tt} are used.
The samples are simulated using a dedicated set of tuned parameters developed by
the \SHERPA authors and use the \NNPDF[3.0nnlo] PDF set.

The \ttW process was simulated using the \SHERPA[2.2.10] generator.
The matrix elements were calculated for up to one additional parton at \NLO and
up to two partons at \LO using \textsc{Comix} and \OPENLOOPS, and merged with
the \SHERPA parton shower using the \MEPSatNLO prescription~\cite{Hoeche:2012yf}.
In addition to the nominal prediction at \NLO in QCD, higher-order corrections
related to electroweak (EWK) contributions are also included.
Event-by-event correction factors are applied that provide virtual \NLO EWK
corrections to $\mathcal{O}(\alpha^{2}\alphas^{2})$ and \LO corrections to
$\mathcal{O}(\alpha^3)$~\cite{Kallweit:2015dum,Bothmann:2019yzt,Gutschow:2018tuk}.
An independent \SHERPA[2.2.10] sample was produced at \LO to account for the
sub-leading EWK corrections to $\mathcal{O}(\alpha^{3}\alphas)$~\cite{Frederix:2017wme}.
The \NLO QCD and \NLO EWK contributions from \SHERPA are combined following the
method of  Ref.~\cite{Frederix:2021agh}.
The \ttW samples are normalised using a cross-section of \SI{722}{\fb} computed
at \NLO including the hard non-logarithmically enhanced radiation at \NLO in
QCD~\cite{Frederix:2021agh}.

The modelling  of the  production of a top quark--antiquark pair with a
$Z$ boson decaying leptonically is done using the \MGNLO[2.8.1]
generator.
It provides matrix elements at \NLO in QCD, employing the \NNPDF[2.3lo] PDF set.
The generated events are interfaced with \PYTHIA[8.244]
for the parton shower and hadronisation stages.
The $\ttZ/\gamma^*(\to \ell^+\ell^-)$ prediction was normalised to the calculation at
\NLO QCD and NLO EWK accuracy based on Ref.~\cite{deFlorian:2016spz} with the
additional inclusion of off-shell effects, corresponding to \SI{162}{\fb}.

The production of \ttH events was modelled using the \POWHEGBOX[v2]
generator at \NLO in QCD with the \NNPDF[3.0nlo] PDF set.
The events were interfaced to \PYTHIA[8.230].
The cross-section was calculated at \NLO QCD and \NLO EWK  using
\MGNLO and amounts to \(507\,\si{\fb}\).

The production of two or three weak vector bosons (\VV, \VVV) in $2\ell$, $3\ell$
and $4\ell$ final states was simulated using the \SHERPA[2.2.12] generator including the
off-shell contributions.
Matrix elements including additional partons were calculated at \NLO in QCD for
up to one parton and at \LO accuracy for two or three additional partons
using \textsc{Comix} and \OPENLOOPS.
For the parton shower simulation, the same method as in the \wjets and
\zjets samples is used.
The \VV and \VVV samples are normalised to the total cross-sections provided by
\SHERPA.

The \tZq process was modelled at \NLO with the \MGNLO[2.9.5] generator and the
\NNPDF[3.0nlo] PDF\@.
The generated events were interfaced with \PYTHIA[8.230].
Matrix elements of the \tWZ process were also calculated with \MGNLO[2.2.2], using the
same PDF set.
The events were interfaced with \PYTHIA[8.212] using the same tune and PDF
set as used for \tZq production.
The samples were normalised to the theoretical cross-section at \NLO QCD\@.

Some rare processes are considered: \ttt, \tttt, \tHW, \tHq, \ttWW,\ttHH, \ttWH, \ttZZ, \ttWZ and \VH production.
The processes involving at least one top quark were simulated with \MGNLO, while \VH was produced with \POWHEGBOX[v2].
Simulated events of all processes were interfaced to \PYTHIA[8] to simulate
parton showers and hadronisation.
The processes are normalised to their predicted \NLO cross-sections.
Their combined contribution is nearly negligible, ranging from
\SIrange{0.1}{1}{\percent}, depending on the kinematic region considered.

% End of text imported from the .//tex/03_samples.tex input file

%
\section{Object reconstruction and event preselection}

% The next lines are included from the .//tex/04_object-reconstruction_preselection.tex input file
%
%
%
%
Events are required to have at least one vertex reconstructed from at least two ID tracks with transverse momenta
of $\pT>0.5\,\GeV$. The primary vertex of an event is defined as the vertex with the highest  $\pT^2$ summed over
the ID tracks~\cite{ATL-PHYS-PUB-2015-026} matched to it.

Electron candidates are reconstructed by matching a track in the ID to clusters of energy deposits in
the electromagnetic calorimeter~\cite{EGAM-2018-01}.
The pseudorapidity of clusters, $\eta_{\mathrm{cluster}}$, is required to be in the range of
$|\eta_{\mathrm{cluster}}|<2.47$.
However, clusters are excluded if they are in the transition region $1.37<|\eta_{\mathrm{cluster}}|<1.52$
between the barrel and endcap electromagnetic calorimeters.
Electron candidates must have $\pt>10\,\GeV$.
A likelihood-based discriminant is constructed to simultaneously evaluate several properties of
electron candidates, including shower shapes in the electromagnetic calorimeter, track quality, and
the detection of transition radiation produced in the TRT~\cite{EGAM-2021-01}.
Applying a discriminant requirement enhances the selection of prompt electrons
produced in \(W\)-, \(Z\)- or Higgs-boson decays and leptonic $\tau$-lepton decays, while
effectively rejecting photon conversions and hadrons misidentified as electrons.
All selected electrons must meet the \textit{Tight} selection criteria as defined in
Ref.~\cite{EGAM-2021-01}.
ig
Muon candidates are reconstructed by combining tracks in the MS with tracks in the ID~\cite{MUON-2018-03}.
The tracks must be in the range of $|\eta|<2.5$ and have $\pt>10\,\GeV$.
Similarly to electrons, likelihood-based identification criteria are applied~\cite{MUON-2018-03}.
Muons are required to satisfy the \textit{Medium} selection defined in Ref.~\cite{MUON-2018-03}.

The tracks matched to electron and muon candidates must point to the primary vertex, which is ensured
by requirements imposed on the transverse impact-parameter significance,
$|d_{0}/\sigma(d_{0})|<5.0$ for electrons and $|d_{0}/\sigma(d_{0})|<3.0$ for muons,
and on the longitudinal impact parameter, $z_0$, for which
$|z_0\sin(\theta)|<0.5$~mm has to be satisfied for both lepton flavours.
Non-prompt electrons and muons are leptons produced by mechanisms other than
\(W\)-, \(Z\)- or Higgs-boson decays and leptonic $\tau$-lepton decays.
They are effectively rejected by using
multivariate discriminants computed with boosted decision trees (BDT), which
integrate electromagnetic shower shapes and track information from the ID~\cite{HIGG-2017-02}.
Separate BDTs are trained for electrons in the barrel ($|\eta|<1.37$) and endcap
($|\eta|>1.37$) regions, while a single BDT is employed for muons.
The efficiency for correctly identifying prompt muons (electrons) ranges from
approximately \SI{80}{\percent} to \SI{95}{\percent} (\SI{65}{\percent} to
\SI{90}{\percent}) for \pT values between \SI{20}{\GeV} and \SI{45}{\GeV}, with
the efficiency plateauing beyond \SI{45}{\GeV}.
Overall, the BDTs achieve a \SI{71}{\percent} (\SI{90}{\percent}) rejection rate
for muons (electrons) from \(B\)-hadron decay.
Additionally, the contribution of electrons reconstructed with an incorrectly reconstructed charge
is significantly reduced by using an additional BDT~\cite{EGAM-2018-01}, which
consolidates information about an electron candidate's charge, impact parameter,
energy, and ID track into a single discriminant.
This achieves a \SI{95}{\percent} efficiency for electrons with a correct charge
assignment while rejecting \SI{94}{\percent} of charge-misidentified electrons.

Scale factors are used to correct the efficiencies in simulation to match the efficiencies
measured for the electron~\cite{TRIG-2018-05} and muon~\cite{TRIG-2018-01} triggers,
and the reconstruction, identification and isolation criteria~\cite{EGAM-2018-01,MUON-2018-03}.

Jets are reconstructed from particle-flow objects~\cite{PERF-2015-09} with the \antikt\ clustering
algorithm~\cite{Cacciari:2008gp,Fastjet} using a radius parameter of 0.4.
This algorithm matches topological energy clusters~\cite{PERF-2014-07} in the calorimeters to selected
tracks in the ID\@. The energy of tracks is subtracted from the matched topological clusters and
both the tracks and the energy-subtracted topological clusters are used as input to the jet reconstruction.
The jet energy is calibrated by applying several simulation-based corrections and techniques
correcting for differences between simulation and data~\cite{JETM-2018-05}.
The jets must satisfy $\pt > \SI{20}{\GeV}$ and $|\eta|<2.5$.

To suppress jets originating from pile-up collisions, several track-based variables are combined
with a multivariate technique to form the jet-vertex-tagger~(JVT) discriminant~\cite{PERF-2014-03}.
Jets with $\pt < \SI{60}{\GeV}$ and $|\eta|<2.4$ are required to have a JVT discriminant above $0.5$,
which corresponds to an efficiency of $92\%$ for jets from the primary vertex,
while \SI{98}{\%} of jets from pile-up events are rejected.

Identification of jets containing \(B\) hadrons (\(b\)-tagging) is performed  with the
\texttt{DL1r} algorithm,
which uses a deep feed-forward neural network with several \(b\)-tagging algorithms
as inputs~\cite{FTAG-2019-07}.
These input algorithms exploit the impact parameters of charged-particle tracks, the properties of reconstructed
secondary vertices and the topology of \(b\)- and \(c\)-hadron decays inside the jets.
The requirement on the \texttt{DL1r} discriminant is chosen such that the average tagging efficiency of \(b\)-jets
from simulated dileptonic \(t \bar t\) events is \(\SI{70}{\percent}\).
The corresponding \pT-dependent $c$-jet rejection factors range from 10 to 14,
while those for light-flavour jets range from 100 to 900.
Differences between the \(b\)-tagging efficiency between collision data and simulation are corrected using
simulation-to-data scale factors derived from \(t\bar{t}\) events. The scale factors depend on the \(\pt\) of the
jets and are consistent with unity within the uncertainties.
The obtained scale factors depend on the parton-shower generator used to produce the \(\ttbar\) samples.
When using samples produced with a different parton-shower generator,
for example \SHERPA, to model \wjets events, or when evaluating systematic uncertainties with a setup
based on \HERWIG, additional correction factors called MC-to-MC scale factors are applied.
Since the \texttt{DL1r} algorithm uses measurements from the ID,
the identification of \bjets is limited to the region with \(\abseta < 2.5\).

With the above definitions of physical objects it is possible for ambiguities
to arise, since all objects are reconstructed independently of each other.
To avoid the double-counting of candidates which satisfy more than one selection
criterion, a procedure called \textit{overlap removal} is applied.
Reconstructed physical objects are removed in the following order:
electrons sharing an ID track with a muon;
jets within \(\Delta R = 0.2\) of an electron, thereby avoiding double-counting electron energy deposits as jets;
electrons within \(\Delta R = 0.4\) of a remaining jet, for reducing the impact of non-prompt electrons;
jets within \(\Delta R = 0.2\) of a muon if they have two or fewer matched tracks;
muons within \(\Delta R = 0.4\) of a remaining jet, reducing the rate of non-prompt muons.

The missing transverse momentum \vecpmiss is reconstructed as the negative vector sum of the transverse momentum of
the reconstructed leptons and jets, and ID tracks that point to the primary vertex
but are not matched to a reconstructed object~\cite{ATLAS:2024cmj}.
The magnitude of \vecpmiss is denoted by \met.

Events selected by this analysis are required to have at least one charged lepton with
\(\pT>\SI{28}{\GeV}\).
Additional leptons must satisfy a \pT threshold of \SI{10}{\GeV}.
In the \dil final state, exactly two leptons of the same charge are required, while in
the \tri final state the requirement is three leptons with a total
charge of \(\pm1e\).
The lepton with the highest \pT is referred to as $\ell_0$.
The lepton with the second highest \pT is denoted by $\ell_1$ and, for events in
the \tri final state, the lepton with the lowest \pT is called $\ell_2$.
In the \tri final state an alternative labelling of leptons is introduced based
on their charge.
In any given \tri event two leptons have the same charge while the remaining
lepton has an opposite charge.
The latter is labelled \(\ell_{\text{OS}}\), the former \(\ell_{\text{SS,0}}\)
and \(\ell_{\text{SS,1}}\) with the number indexing the lepton in order of
decreasing \pT.
In addition to the requirements on leptons, considered events need to contain at
least one jet, at least one of which is $b$-tagged.
A summary of the preselection applied to events  is given in
\Cref{tab:preselection_overview}.

\begin{table}
\centering
\caption{
Overview of the preselections applied in the analysis. These selections ensure jet and $b$-tag multiplicities, lepton momenta, and define the \dil and \tri final states.}\label{tab:preselection_overview}
\begin{tabular}{ccc}
\toprule
\multicolumn{3}{c}{Preselection}                                                                        \\
\midrule
\(\njets\)                                     & \multicolumn{2}{c}{\(\geq1\)}                          \\
\(\nbtags\) & \multicolumn{2}{c}{\(\geq1\)}                          \\
\(\pT(\text{jet}) \)                           & \multicolumn{2}{c}{\(\geq\SI{20}{\GeV}\)}              \\
\(\pT(\ell)\)                                  & \multicolumn{2}{c}{\(\geq\SI{10}{\GeV}\)}              \\
\(\pT(\ell_0)\)                                & \multicolumn{2}{c}{\(\geq\SI{28}{\GeV}\)}              \\
\midrule
& \dil                                      & \tri       \\
\midrule
\(N_{\ell}\)                                   & \(=2\)                                    & \(=3\)     \\
\(\sum q(\ell_i)\)                             & \(=\pm2e\)                                & \(=\pm1e\) \\
\bottomrule
\end{tabular}
\end{table}

% End of text imported from the .//tex/04_object-reconstruction_preselection.tex input file

%
\section{Background estimate}

% The next lines are included from the .//tex/05_background-estimation.tex input file
Several SM background processes are present in the signal regions.
Many of these processes can be simulated by using MC,
others require dedicated treatment.
Particularly the modelling of background from non-prompt leptons must be
validated and potentially corrected.
This background is separated into various categories,
encompassing electrons from prompt muon decay, electrons from photon conversion
and leptons produced in a hadronic jet, where further differentiation
depending on the jet primary particle are made.
Jets initiated by \(b\)-quarks, by other quarks or gluons, and by $\tau$-leptons are
considered, where leptons from \(b\)-jets are further categorised into
electrons and muons.
The production of these non-prompt
leptons is fully accounted for by the simulated SM samples.
By navigating through the generator-level parent particles of a given lepton,
this lepton is sorted into one of the previously discussed lepton-origin categories.
Based on this, several MC templates are defined containing the events
corresponding to exactly one non-prompt lepton-origin category.
The remaining events include only prompt leptons and remain in their original
SM background template.
All background estimate methods discussed in the following are based on these
prompt and non-prompt MC templates.
Several of the processes discussed in this section are assigned unconstrained (\textit{free-floating}) normalisation factors
in the final maximum-likelihood fit.
These normalisation factors are constrained in the fit through dedicated control regions, enriched in the respective processes.
It should be noted that the QCD multijet background is negligible in this analysis,
owing to the rarity of the considered \dil and \tri final states.

\subsection{Non-prompt lepton background}
The primary source of non-prompt lepton background is the decay of \(B\)-hadrons.
These particular objects are also referred to as \textit{heavy-flavour (HF)
decay leptons}.
They mainly consist of events from \ttbar production (\SI{75}{\percent}), while
the remaining events are split evenly between \vjets and single-top-quark production.
The \textit{Template fit method}~\cite{TOPQ-2021-08} is used to determine the
rates of HF-decay electrons and HF-decay muons.
It assumes the kinematic distributions of the
corresponding processes are well described, while their normalisation may
require corrections.
The normalisation factors are determined in the final maximum-likelihood fit,
which is performed simultaneously in the \dil and the \tri final states.

As explained above, various other non-prompt lepton processes contribute to this
analysis.
Their contribution is minor compared with the \HFe and \HFmu processes.
In the fit, they are assigned normalisation uncertainties of \SI{50}{\percent}.

\subsection{Charge misidentification}
In addition to the processes discussed above, the phase space of the \dil final
state is contaminated by events containing prompt electrons with a
misidentified charge.
The modelling of such objects is strongly dependent on their \pT.
Furthermore, electrons can radiate hard bremsstrahlung through interaction with the
detector material, which then undergoes asymmetric conversion into an electron-positron-pair.
The probability for electrons to interact with the detector material increases
with the absolute value of the pseudorapidity \(\abseta\).
Both, charge misidentified electrons and electrons from bremsstrahlung, are
considered as a single background labelled \qmisid, which is
modelled using a data-driven method based on the one described in
Ref.~\cite{TOPQ-2021-08}.
To avoid double-counting \qmisid events, MC events featuring an electron from
any of the considered processes are identified using generator level information and
are excluded from the analysis.

The data-driven \qmisid background estimate is performed as follows.
Events containing two isolated electrons with an invariant mass around the mass
of the \(Z\) boson are considered, where the electrons can have either an
opposite-sign charge (OS) or the same-sign charge (SS).
To maintain orthogonality to the kinematic regions of the main analysis, there
is an \(\nbtags = 0\) requirement.
The \(m_{ee}\) distribution in the vicinity of \(m_Z=\SI{91.19}{\GeV}\)~\cite{Zboson_mass}
is fitted separately for OS and SS events using Breit--Wigner functions.
The fitted centres \(m_{\text{peak}}\) and widths \(\sigma\) are used to
define the \(Z\)-windows up to \(4\sigma\) from the centres, and sideband regions
\([m_{\text{peak}} -8\sigma,\,m_{\text{peak}} - 4\sigma]\,\cup\,[m_{\text{peak}} + 4\sigma,\,m_{\text{peak}} + 8\sigma]\).
The assumption is made, that events in the \(Z\)-window consist of prompt
\(Z\to\ee\) events and a uniform background.
The uniform background is estimated from the average yield in the sideband regions,
separately for the OS and SS events, and are subtracted in the \(Z\)-window.

\qmisid efficiencies are extracted in bins of \(\abseta\) and \pT, corresponding to the
probability for an electron in a given bin to be charge-misidentified.
These relate the number of background-corrected SS and OS events in the
\(Z\)-window and can thus be determined from the obtained event yields.
The \qmisid events in a given region of the \dil phase space are then
modelled by selecting data events with two opposite-charge leptons while keeping
all other selection requirements of the region unmodified, and weighting these
data events with the obtained efficiencies.
It should be noted that due to the high accuracy of the ATLAS muon spectrometer,
the muon charge misidentification is negligible in this analysis.

\subsection{Prompt-lepton background}
The \ttW and \ttZ processes represent the primary background sources from prompt
leptons.
While measurements of \ttW production deviate from theoretical
predictions by $1.4\sigma$~\cite{ATLAS-TOPQ-2019-30},
measurements of \ttZ production align closely with SM expectations.
However, it's worth noting that the latest measurement of \ttZ production is
conducted in regions with a jet multiplicity of at least four~\cite{ATLAS-TOPQ-2020-20}.
This analysis  extends to kinematic regions with fewer
jets, which may not be adequately modelled by MC simulations.
To address this discrepancy, the normalisation of \ttW and \ttZ production is
left unconstrained in the maximum-likelihood fit.
Similar to the non-prompt templates, a single normalisation factor is applied
per process across the entire phase space, encompassing both the \dil and
the \tri final states.

The \dil and \tri event selection requires at least one $b$-jet in the final state.
In the \VV MC samples used in the analysis these quarks must be produced by the shower generator.
To account for potential mismodeling of the jet flavour composition, the \VV background is split
according to the number of leptons ($2\ell$, $3\ell$, $4\ell$)
and the generator-level flavour of additional jets.
If the event contains at least one jet originating from the decay of either a
bottom or a charm hadron is sorted into the $b/c$ category.
If it exclusively contains jets from light quarks or $\tau$-leptons it is
included in the $l/\tau$ category.
The \(\VVsplit{3}{b/c}\) template has the largest contribution in the phase space
and its normalisation is left free-floating in the maximum-likelihood fit.
The remaining templates (\(\VVsplit{2}{l/\tau}\), \(\VVsplit{2}{b/c}\),
\(\VVsplit{3}{l/\tau}\), \(\VVsplit{4}{b/c}\) and \(\VVsplit{4}{l/\tau}\)) are
assigned individual \SI{50}{\percent} normalisation uncertainties.
These minor \VV templates, alongside the minor non-prompt templates and several
SM processes are combined into one \textit{Others} category in the following.
The rates of the majority of processes in this category are similar in the
analysis phase space.

% End of text imported from the .//tex/05_background-estimation.tex input file

%
\section{Event categorisation}

% The next lines are included from the .//tex/06_event-categorisation.tex input file
Preselected events are categorised into several kinematic regions.
These regions are either enriched in signal events, to improve the overall
sensitivity of the analysis, or in certain background events, to control the
normalisation of specific processes.

\subsection{Definition of signal regions}

After event preselection, four signal regions are defined in this analysis, two in the \dil final state and
two in the \tri final state.
One region per final state is enriched in the \(t\to Hq\) decay signal.
These two regions are referred to as \SRdildec and \SRtridec.
The other two signal regions contain a larger fraction of \(gq\to tH\) production
signal and are correspondingly named \SRdilprod and \SRtriprod.

Both the \dil signal regions require exactly one $b$-tagged jet.
Given that a higher jet multiplicity is anticipated for the decay signal compared to
the production signal, the \SRdildec is defined by requiring $\njets\geq4$.
A cut of $\njets\leq3$ in the \SRdilprod ensures the orthogonality of the two
regions.
To minimise the contribution from \HF leptons, the \pT of $\ell_1$ is
required to be larger than \SI{12}{\GeV} in the \SRdildec and \SI{16}{\GeV} in the \SRdilprod.
As \qmisid events strongly contaminate the \dil final state, a cut on the
invariant mass of the two leptons is imposed in both the regions.
If both the leptons are identified as electrons,
\(|m(e,e)-m_Z|\geq\SI{10}{\GeV}\) must be satisfied.

\begin{figure}
\centering
\subfigure[]{ \includegraphics[width=0.4\linewidth]{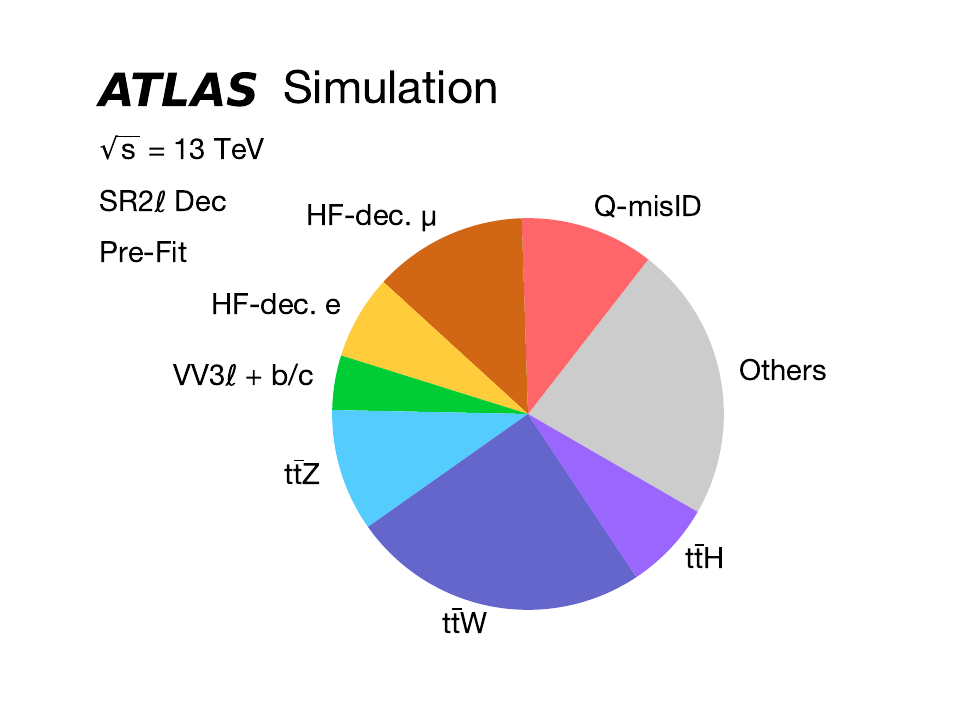} }
\subfigure[]{ \includegraphics[width=0.4\linewidth]{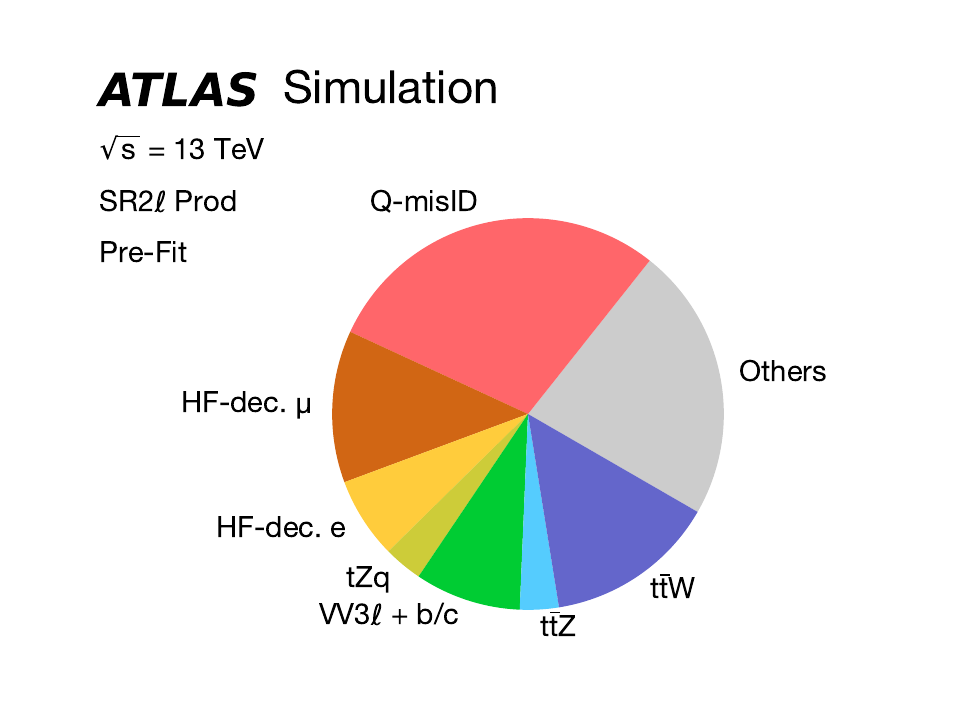} }
\subfigure[]{ \includegraphics[width=0.4\linewidth]{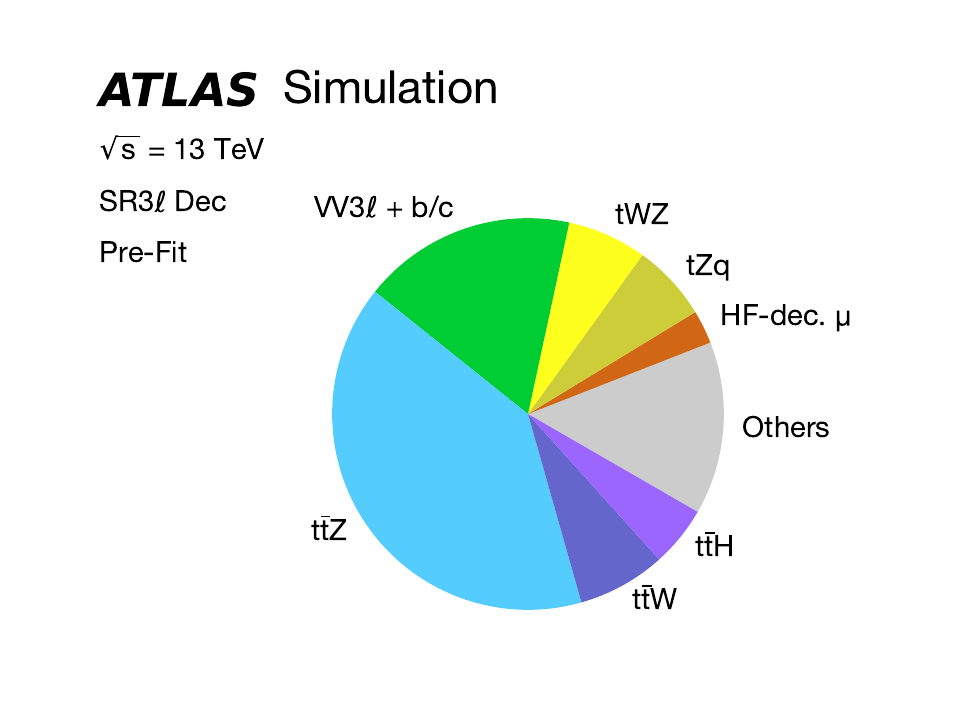} }
\subfigure[]{ \includegraphics[width=0.4\linewidth]{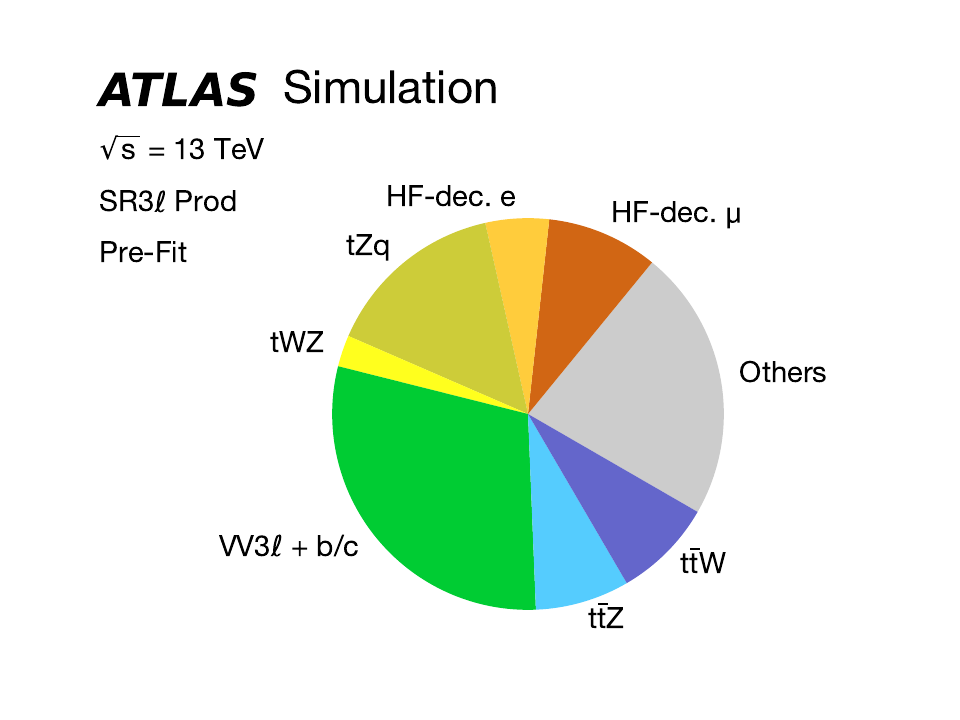} }
\caption{
The composition of background processes in the (a) \SRdildec, (b) \SRdilprod,
(c) \SRtridec and (d) \SRtriprod.
The \textit{Others} template contains various minor processes, which
individually only have small contributions.
No correction to the normalisation of any process is applied.
}\label{fig:SR_pie_charts}
\end{figure}

Similar to the \dil final state, both the signal regions in the \tri final state require
exactly one $b$-tagged jet.
The primary difference between the two regions is again the jet multiplicity.
Overall fewer jets are expected for the signal in the \tri final state.
Hence, $\njets\geq3$ is required for the \SRtridec, while $\njets\leq2$ is
required for the \SRtriprod.
To reduce the \HF contribution, the definition of both the regions require cuts on the \pT of the two
sub-leading leptons:
\(\pT(\ell_1)\geq\SI{20}{\GeV}\) and \(\pT(\ell_2)\geq\SI{16}{\GeV}\).
As the \qmisid process does not contribute to the \tri final state, no further
restrictions to the event selection are made.
The definition of all signal regions is summarised in \Cref{tab:SR_overview},
while their respective composition in terms of background processes is shown in
\Cref{fig:SR_pie_charts}.

\begin{table}
\centering
\caption{
Overview of the event selections applied to define the signal regions of
the analysis.
}\label{tab:SR_overview}
\begin{tabular}{ccccc}
\toprule
& \SRdildec             & \SRdilprod            & \SRtridec             & \SRtriprod            \\
\midrule
\(\njets\)       & \(\geq4\)             & \(\leq3\)             & \(\geq3\)             & \(\leq2\)             \\
\(\nbtags\)      & \( = 1\)              & \( = 1\)              & \( = 1\)              & \( = 1\)              \\
\(\pT(\ell_1)\)  & \(\geq\SI{12}{\GeV}\) & \(\geq\SI{16}{\GeV}\) & \(\geq\SI{20}{\GeV}\) & \(\geq\SI{20}{\GeV}\) \\
\(\pT(\ell_2)\)  & --                     & --                     & \(\geq\SI{16}{\GeV}\) & \(\geq\SI{16}{\GeV}\) \\
\(|m(e,e)-m_Z|\) & \(\geq\SI{10}{\GeV}\) & \(\geq\SI{10}{\GeV}\) & --                     & --                     \\
\bottomrule
\end{tabular}
\end{table}

\subsection{Control regions for non-prompt-lepton backgrounds}
The template-fit method assumes that corrective normalisation factors are
applicable independently of the considered final state.
To ensure this assumption is correct, two control regions are defined for each
of the \HF processes: one in the \dil final state (\CRdilHFe and \CRdilHFmu)
and one in the \tri final state (\CRtriHFe and \CRtriHFmu).

Overall, the average \pT of leptons originating from a HF decay is smaller than
that of prompt leptons.
Thus, all \HF control regions are defined by imposing upper boundaries on the
\pT of the lowest-\pT lepton.
The boundaries are chosen such that an orthogonality to the signal regions of
the respective final state is ensured.
Requiring the flavour of the lowest-\pT lepton to be either $e$ or $\mu$ ensures
that the region is exclusively enriched in \HFe or \HFmu events.
In the \CRdilHFe and the \CRdilHFmu the \pT-leading lepton is required to be a
muon, reducing the contamination of \qmisid events.
Due to high statistical uncertainty, only the event yields of the \CRdilHFe and the \CRdilHFmu
enter the maximum-likelihood fit.
For the \CRtriHFe and the \CRtriHFmu, the distributions of the \(\pT(\ell_1)\)
variable are included.
A summary of the definition of all non-prompt control regions can be found in
\Cref{tab:CR_nonprompt_overview}.

\begin{table}
\centering
\caption{
Overview of the event selections applied to define the non-prompt lepton
background control regions of the analysis.
}\label{tab:CR_nonprompt_overview}
\begin{tabular}{cccccc}
\toprule
& \CRdilHFe          & \CRdilHFmu         & \CRtriHFe             & \CRtriHFmu            \\
\midrule
\(\njets\)               & \(\leq 3\)         & \(\leq 3\)         & \(\geq 1\)            & \(\geq 1\)            \\
\(\nbtags\)              & \( \geq 1\)        & \( \geq 1\)        & \( = 1\)              & \( = 1\)              \\
\(\ell_0\) flavour & \(\mu\)           & \(\mu\)           & --                     & --                     \\
\(\ell_1\) flavour & \(e\)             & \(\mu\)           & --                     & --                     \\
\(\pT(\ell_1)\)          & \(<\SI{16}{\GeV}\) & \(<\SI{16}{\GeV}\) & \(\geq\SI{20}{\GeV}\) & \(\geq\SI{20}{\GeV}\) \\
\(\ell_2\) flavour& --                  & --                  & \(e\)                & \(\mu\)              \\
\(\pT(\ell_2)\)          & --                  & --                  & \(<\SI{16}{\GeV}\)    & \(<\SI{16}{\GeV}\)    \\
\bottomrule
\end{tabular}
\end{table}

\subsection{
Control regions for \texorpdfstring{\ttW}{ttW} and \texorpdfstring{\ttZ}{ttZ} backgrounds
}
Three additional control regions are defined to determine the normalisation of
the background from \ttW and \ttZ production.
Studies on various kinematic distributions in the signal regions of the \dil
final state show a strong similarity between the two processes.
Therefore, in this final state only one control region (denoted \CRdilttV) is
defined to constrain the sum of \ttW and \ttZ.
This region requires at least four jets, exactly two of which are
$b$-tagged, because both the top quarks in the process are expected
to produce a $b$-quark.
An additional requirement of \(\pT(\ell_1)\geq\SI{18}{\GeV}\) further increases
the purity of the region in \ttV events by reducing contamination from HF-decay
events.
The $\pT(\ell_1)$ variable is used as input for the maximum-likelihood fit.

In the \tri final state, distinct shape differences between \ttW and \ttZ
production are observed due to the absence of any requirement on the invariant mass of
lepton pairs.
Therefore, two control regions, \CRtrittW and \CRtrittZ, are defined.
Both the regions require at least two jets, two of which are $b$-tagged.
Events are then assigned to the \CRtrittZ if at least one pair of leptons with
the same flavour and opposite charge can be found satisfying
\(|m(\ell^+,\ell^-)-m_Z|<\SI{10}{\GeV}\).
In cases where this criterion is not met, events are assigned to \CRtrittW.
In both the control regions, the transverse momentum of the \pt-leading \(b\)-tagged
jet \(\pt(b\text{-jet}_0)\) is used as input for the maximum-likelihood fit.
A summary of the definition of all \ttV control regions can be found in
\Cref{tab:CR_ttV_overview}.

\begin{table}
\centering
\caption{
Overview of the event selections applied to define the \ttW and \ttZ
control regions of the analysis.
}\label{tab:CR_ttV_overview}
\begin{tabular}{cccc}
\toprule
& \CRdilttV             & \CRtrittW             & \CRtrittZ             \\
\midrule
\(\njets\)                 & \(\geq 4\)            & \(\geq 2\)            & \(\geq 2\)            \\
\(\nbtags\)                & \(=2\)                & \(=2\)                & \(=2\)                \\
\(\ell_0\) flavour  & \(\mu\)              & --                     & --                     \\
\(\pT(\ell_1)\)            & \(\geq\SI{18}{\GeV}\) & \(\geq\SI{20}{\GeV}\) & \(\geq\SI{20}{\GeV}\) \\
\(\pT(\ell_2)\)            & --                     & \(\geq\SI{16}{\GeV}\) & \(\geq\SI{16}{\GeV}\) \\
\(|m(\ell^+,\ell^-)-m_Z|\) & --                     & \(\geq\SI{10}{\GeV}\) & \(<\SI{10}{\GeV}\)    \\
\bottomrule
\end{tabular}
\end{table}

% End of text imported from the .//tex/06_event-categorisation.tex input file

%
\section{Separation of signal and background}
%

% The next lines are included from the .//tex/07_signal-background-separation.tex input file
Artificial neural networks (NNs) are used to separate signal and background in all
four signal regions combining several kinematic variables into optimised
NN discriminants.
In addition to variables derived from the reconstructed objects the NNs build on
a diverse range of reconstructed kinematic variables.
Detailed information regarding the reconstruction algorithms is provided in
\Cref{sec:reconstruction}, while a comprehensive description of the NN training
process can be found in \Cref{sec:NN_description}.

\subsection{Event reconstruction in signal regions}
\label{sec:reconstruction}
The reconstruction algorithms employed  are described below.
In \tHq FCNC interactions, multiple decay modes of the Higgs boson that contribute to
the \dil and \tri final states are considered.
However, the \(H\to WW^*\) decay mode has by far the largest contribution with
\(\geq\SI{75}{\percent}\) in all signal regions.
Thus, all reconstruction efforts presented here focus on this specific
Higgs-boson decay mode.
Most algorithms are customised to fit one of the two considered final states.
The relevant final states for each algorithm are given in parentheses.

\paragraph{Recursive Jigsaw Reconstruction (\dil and \tri)}

The Recursive Jigsaw Reconstruction (RJR) technique~\cite{RJR} is a method for
resolving combinatorial and kinematic ambiguities which arise in hadron collider events.
RJR relies on decay trees which describe the topologies of interest, and relates
the reference frames of the intermediate particles in the decay.
Lepton and jet assignment, along with the splitting of \MET, is governed by a set
of jigsaw rules which rely on known properties of the decay tree.
For instance the top-quark and Higgs-boson masses can be used to constrain the
assignment of the jets, or the $W$-boson mass can be used to facilitate \MET splitting
and matching to leptons.

\begin{figure}
\subfigure[]{\includegraphics[width=0.49\textwidth]{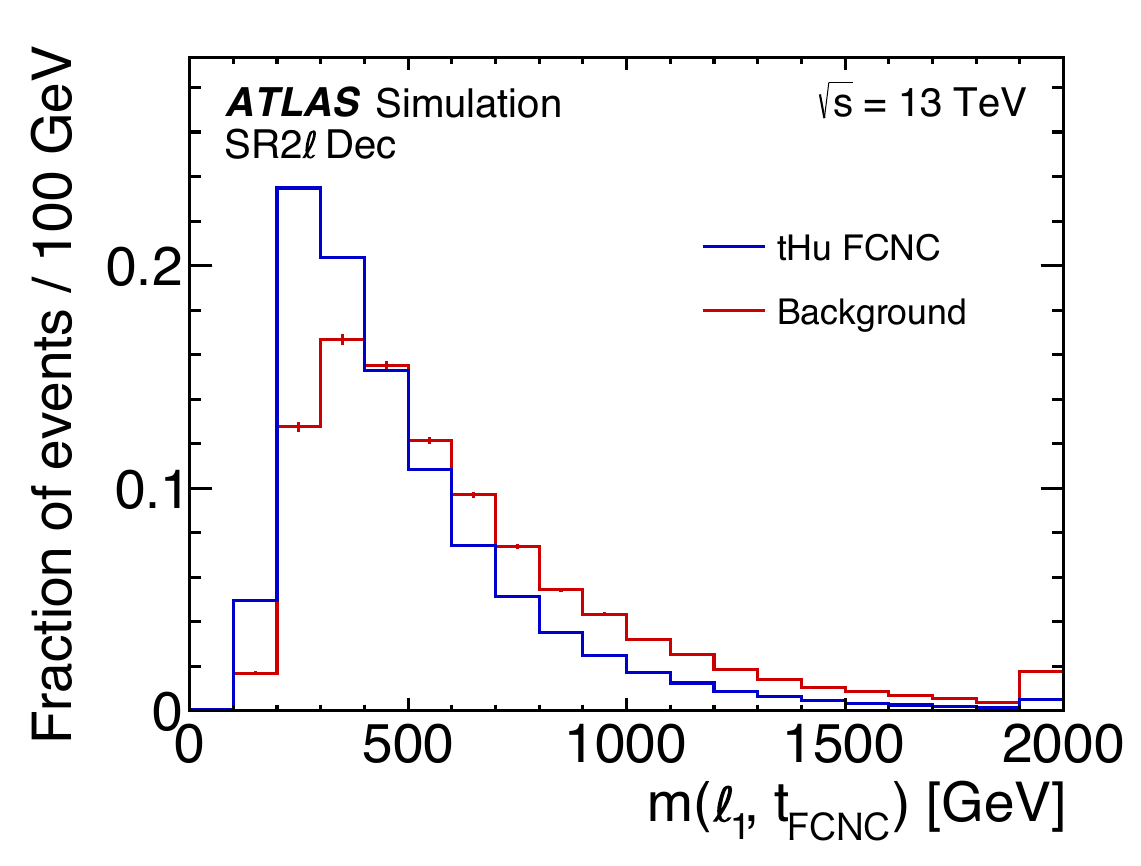}}
\subfigure[]{\includegraphics[width=0.49\textwidth]{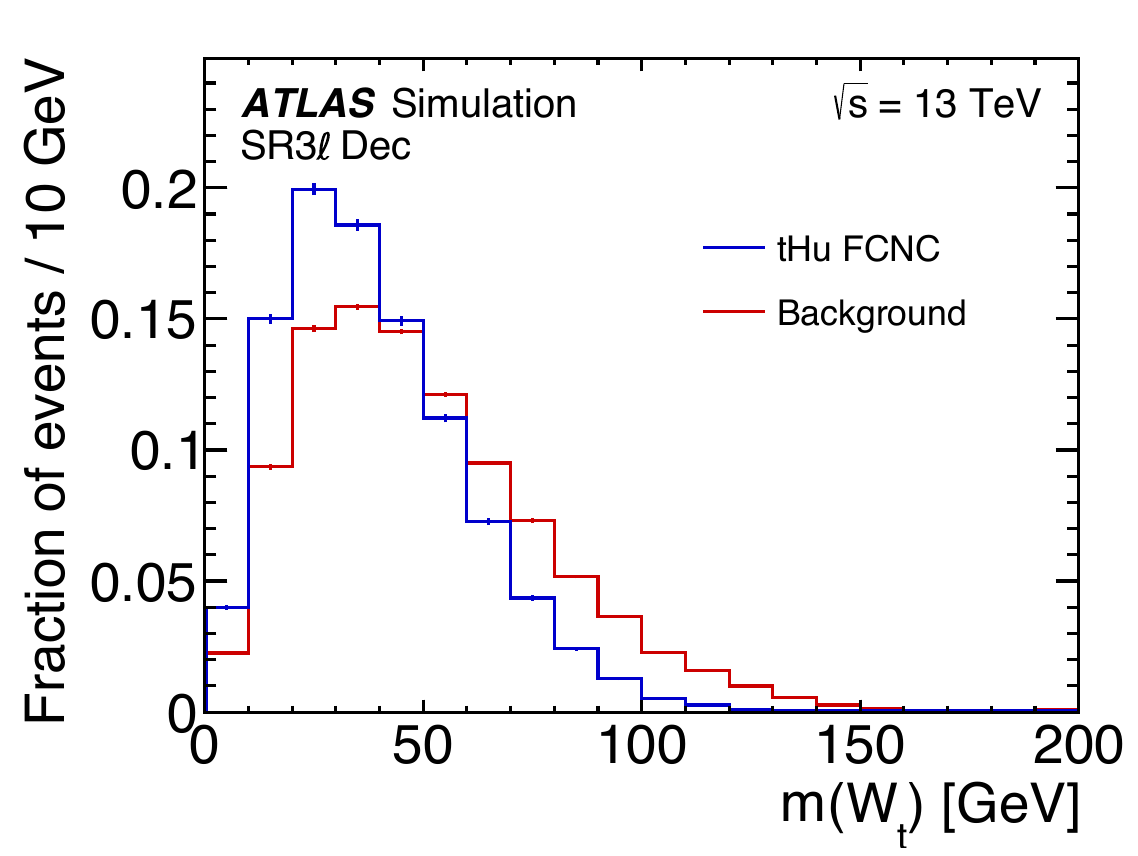}}
\caption{
Probability densities of (a) \(m(\ell_1,\tFCNC)\) in the \SRdildec and (b) \(m(W_t)\)
in the \SRtridec for the \tHu signal process and the sum of all
background processes.
The vertical lines on bins represent MC-statistical uncertainties. The last bin includes the overflow.
}\label{fig:RJR_example_vars}
\end{figure}

This process leads to the computation of four-vectors of the reconstructed
intermediate states, accessible in any of the frames defined in the decay tree.
For each of the four topologies (\(2\ell1b2j\), \(2\ell1b3j\), \(3\ell1b0j\),
\(3\ell1b1j\), where $b$ refers to \nbjets and $j$ to \njets) there are different
intermediate states, split by production versus decay and again by final state.
The particles that are reconstructed for each individual region are listed in
\Cref{tab:RJR_particles_reconstructed}. These include $W$ and Higgs bosons, and top quarks.
\Cref{fig:RJR_example_vars} depicts the probability densities of two RJR
reconstructed variables for the \tHu signal and the combination of all
background processes, showing clear shape differences between the two.

\begin{table}
\centering
\caption{
Particles from the \tHq processes which are reconstructed using the Recursive Jigsaw Reconstruction (RJR) in
the respective signal regions.
}\label{tab:RJR_particles_reconstructed}
\begin{tabular}{lll}\toprule
Regions                       & Particle name & Description                                                        \\\midrule
& \tSM          & The top quark decaying via $t\to Wb$                               \\
\SRdildec /                   & $W_t$         & The $W$ boson from the SM top-quark decay                          \\
\SRtridec                     & \tFCNC        & The top quark decaying via $t\to Hq$                               \\
& $H$           & The Higgs boson originating from th FCNC top-quark decay           \\\midrule
\SRdildec                     & \whad         & The hadronically decaying $W$ boson from the $H\to WW^*$ decay \\\midrule
\multirow{2}{*}{\SRdilprod /} & \tSM          & The top quark produced in the $gq\to Ht$ process                    \\
\multirow{2}{*}{\SRtriprod}   & $W_t$         & The $W$ boson from the top-quark decay                             \\
& $H$           & The Higgs boson produced in the $gq\to Ht$ process                  \\\bottomrule
\end{tabular}
\end{table}

\begin{figure}
\subfigure[]{\includegraphics[width=0.49\textwidth]{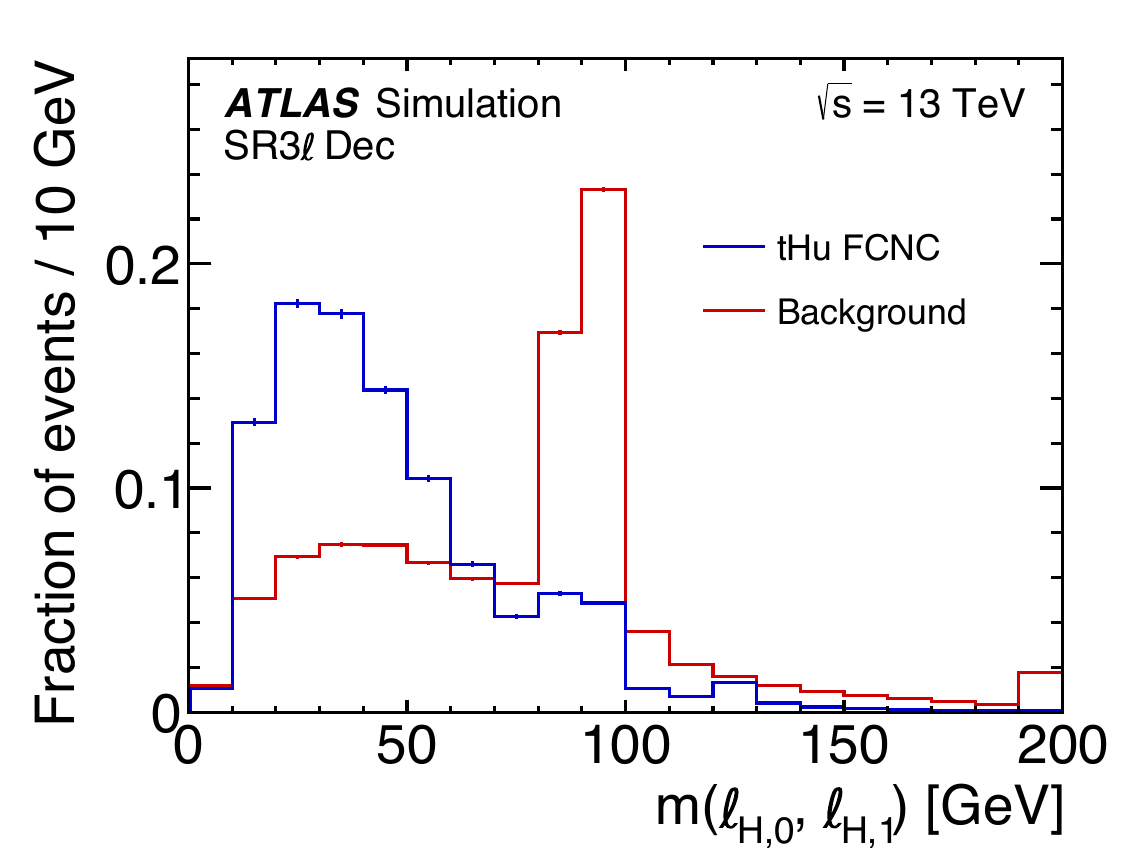}}
\subfigure[]{\includegraphics[width=0.49\textwidth]{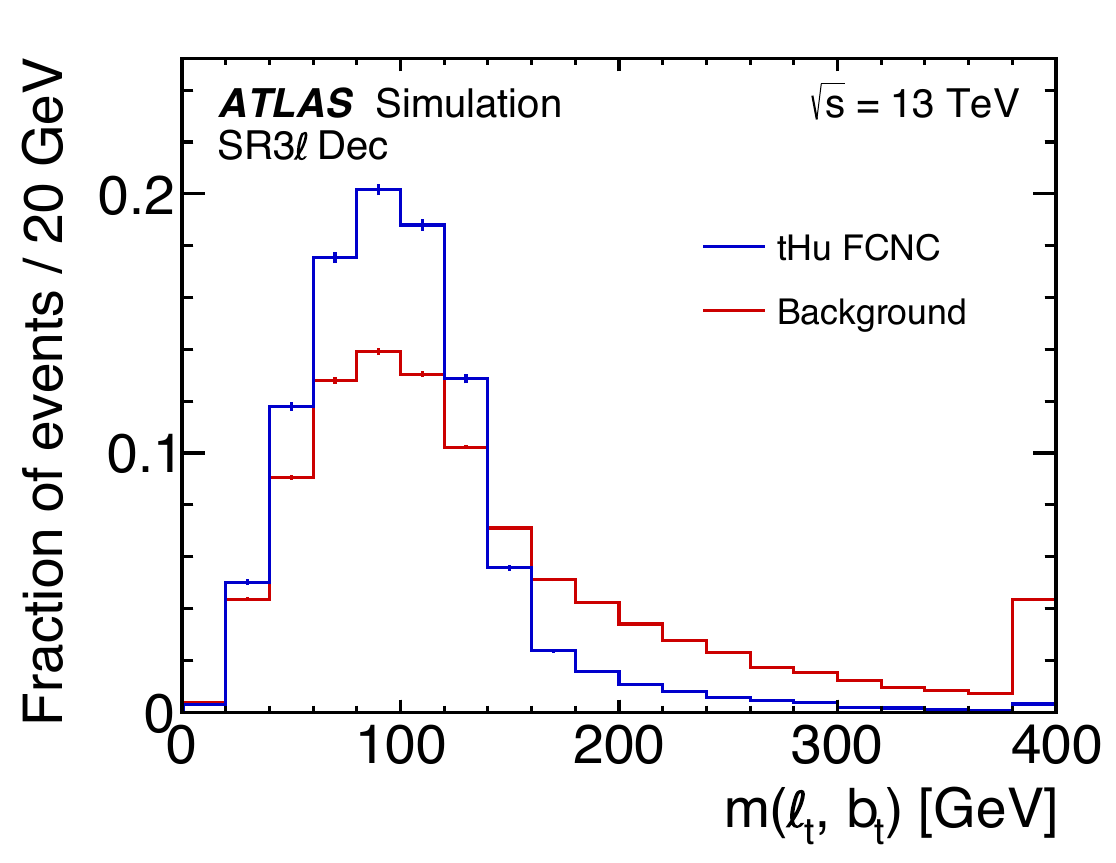}}
\caption{
Probability densities of (a) \(m(\ell_{H,0},\ell_{H,1})\) and (b) \(m(\ell_t,b_t)\)
in the \SRtridec for the \tHu signal process and the sum of all
background processes.
The vertical lines on bins represent MC-statistical uncertainties. The last bin includes the overflow.
}\label{fig:NICEReco_example_vars}
\end{figure}

\paragraph*{Neutrino Independent Combinatorics Estimator (\tri)}
The considered decay scenario of a signal event in the \tri final state
involves one top quark decaying into a $b$-quark, a lepton and a neutrino, while
the Higgs-boson decay is expected to yield two leptons and two neutrinos.
Studies of the MC signal samples show that on average the two leptons from the
Higgs-boson decay have the smallest angular separation out of any lepton pair in
the event.
Based on this information, the two leptons with the smallest angular separation
\(\Delta R\) are labelled \(\ell_{H,0}\) and \(\ell_{H,1}\), ordered by their
transverse momentum.
The remaining lepton is labelled \(\ell_{t}\), while the $b$-tagged jet is
denoted \(b_t\).
Various kinematic parameters are calculated for these objects, two of
which are shown as examples in \Cref{fig:NICEReco_example_vars} for the full \tHu
signal and the combination of all background processes, demonstrating their high
separation power.
However, these parameters are discarded, if the angular separation of \(b_t\) from
either \(\ell_{H,i}\) is smaller than the separation from \(\ell_t\).
Conversely, when the angular separation of \(b_t\) from \(\ell_t\) is found to be
smaller than from any other lepton, and further, when \(\ell_{H,0}\) and \(\ell_{H,1}\)
have opposite charges the
\textit{Neutrino independent combinatorics estimator} reconstruction (NICE Reco)
condition is defined as satisfied.
In this scenario, the calculated kinematic parameters have a particularly high
separation power between signal and background processes.
This information is provided as additional input to the NNs.

\subsection{Training of feed-forward neural networks}
\label{sec:NN_description}
The NNs are implemented using the NeuroBayes
package~\cite{feindt-2004,Feindt:2006pm}, which combines a three-layer
feed-forward NN with a complex and robust preprocessing of the input variables
before they are presented to the NN\@.
The preprocessing produces a ranking of the input variables based on an
algorithm employing the total correlation of a set of variables to the target
function which assumes the value \(1\) for signal and \(0\) for background
events.
The input variable selection is performed once per signal region.

The selected variables of all signal regions are listed in Tables \ref{tab:NNvars_SR3lprod}--\ref{tab:NNvars_SR2ldec}.
Overall, invariant masses of reconstructed objects, including RJR objects,
compose the largest fraction.
The most significant variables in both the \tri signal regions are the invariant mass
of the opposite charge and either of the two same charge leptons
\(m(\ell_{\text{OS}}, \ell_{\text{SS,0}})\) and \(m(\ell_{\text{OS}}, \ell_{\text{SS,1}})\).
In the \SRdilprod, the invariant mass of the subleading-\pT lepton and the RJR
Higgs boson \(m(\ell_1,H)\) is the most important variable, followed by the number
of jets \njets.
The most significant variable of the \SRdildec is the scalar \pT sum of all jets
\(H_T(\text{jets})\), a quantity that is also highly ranked in other trainings.
The invariant mass of the leading-\pT lepton and the event's \(b\)-jet
\(m(\ell_0,b\text{-jet})\) is the second-highest ranked in the training.

Individual NNs are created for both the signal processes and every signal region,
summing up to a total of eight NN architectures.
The training uses the data-driven \qmisid estimate together with unmodified MC
templates as input, before  any corrections to their normalisation by the
maximum-likelihood fit.
Cross-training is employed to prevent overfitting, where the signal and background MC events are split into half.
The splitting is performed in a pseudo-random way based on the parity of the
unique event identification number.
For each half, a separate NN model is trained, which is then applied to the
other half of MC events in the final analysis.

During the training process of all the NNs, the signal is trained against all
considered backgrounds with a fraction of \SI{50}{\percent} signal events and
\SI{50}{\percent} background events.
The different background processes are weighted relative to each other according
to their expected number of events.
NeuroBayes uses Bayesian regularisation techniques for the training process to
improve the generalisation performance and to avoid overfitting. The network
infrastructure consists of one input node for each input variable, a single hidden
layer, and one output node giving a continuous output in the interval \((-1,+1)\).
The number of nodes in the hidden layer is optimised individually for each NN
model.
As a non-linear activation function NeuroBayes uses the symmetric sigmoid function.
In the region close to zero, the sigmoid function has a linear response.
The final discriminant \dnn is obtained by linearly scaling the output of the NNs
to the interval \((0,1)\).

% End of text imported from the .//tex/07_signal-background-separation.tex input file

%
\section{Systematic uncertainties}
%

% The next lines are included from the .//tex/08_systematics.tex input file
Several sources of systematic uncertainty affect the expected event yield from signal and
background processes and the shape of the NN discriminants used in the maximum-likelihood fits.
The systematic uncertainties are divided into two major categories.
There are experimental uncertainties in the reconstruction of the four-momenta of
the final-state objects: electrons, muons, jets, \btagged jets, and $\MET$
as a sign of a high-\pt neutrino.
The second category of uncertainties is related to the modelling of scattering processes
with event generators and of the interaction of particles with the detector.
All uncertainties are propagated through the analysis and their effects on the expected event
yields and discriminant distributions are accounted for
by including corresponding nuisance parameters in the fit.
In the following, the estimate of experimental and modelling uncertainties is explained in more detail.

\subsection{Experimental uncertainties}
The uncertainty in the integrated luminosity of the combined 2015--2018 data set is 0.83\% and
is based on a calibration of the luminosity scale using $x$--$y$ beam-separation
scans~\cite{DAPR-2021-01}. The luminosity uncertainty is applied to the expected signal and background
event yields.

Scale factors are applied to simulated events to correct for reconstruction, identification,
isolation and trigger performance differences between data and detector simulation for electrons and muons.
These scale factors and their systematic uncertainties, and the lepton momentum scale and resolution,
were assessed using $Z\rightarrow e^+e^-$ and $Z\rightarrow \mu^+\mu^-$ events in simulation
and data~\cite{EGAM-2018-01,MUON-2018-03}.

The jet energy scale (JES) was calibrated using a combination of test-beam data, simulation and
in situ techniques~\cite{JETM-2018-05}.
The JES is parameterised in bins of jet \pt and $\eta$.
Its uncertainty is decomposed into a set of 30 uncorrelated components,
of which 29 are non-zero in a given event depending on the type of simulation used.
Sources of uncertainty contributing to the JES uncertainties include
the $\eta$ intercalibration of forward jets in the range of $0.8< |\eta_\mathrm{det}| < 4.5$ with those in the
central barrel region ($|\eta_\mathrm{det}| < 0.8$), pile-up modelling, jet flavour composition and response,
differences between jets induced by \bquarks and those from gluons or light-quarks,
single-particle response, detector modelling, non-closure, and effects of jets not fully contained
in the calorimeter.

The uncertainty of the jet energy resolution (JER) is evaluated by smearing jet
energies according to a Gaussian function~\cite{JETM-2018-05}.
Thirteen orthogonal components account
for jet-$\pt$ and $\eta$-dependent differences between simulation and data
which were determined using dijet events and noise measurements based on random cones.
The smearing is applied to simulated events if the resolution in data is larger than in MC simulation,
and to pseudo-data when the resolution is larger in simulation than in collision data.
The JER uncertainties are defined by comparing both types of smearing and thereby taking the anti-correlation
between different components into account.
The nominal data remains unchanged.
The uncertainty in the efficiency to satisfy the JVT requirement for pile-up suppression
was derived in \(Z(\rightarrow \mu^+\mu^-)\)+jets events and is also considered~\cite{PERF-2014-03}.
The uncertainty in \met due to a possible miscalibration of its soft-track component was derived
from data--simulation comparisons of the \pT balance between the hard and soft \met
components~\cite{ATLAS:2024cmj}.

The \btag requirement made in the measurement has uncertainties in the \btag efficiency of
true \bjets and in the mistagging rates of light-quark jets and \(c\)-jets.
The \btag efficiency is measured in dileptonic \(t\tbar\) events. Differences between data
and detector simulation are corrected by \pt-dependent scale factors applied to simulated events.
The uncertainty in the scale factors is decomposed into 45 orthogonal components~\cite{FTAG-2018-01}.
The uncertainties are propagated through the analysis as weights.
The set of uncertainties used covers both the \SI{70}{\percent} working point used in this analysis,
as well as other working points of the \texttt{DL1r} discriminant.
The rate of mistagging \(c\)-jets as \bjets was measured in semileptonic \(t\tbar\) events, where one
of the $W$ bosons decays into an electron or a muon and a neutrino and the other decays into a
quark--antiquark pair~\cite{FTAG-2020-08}.
This event sample allows to utilise the relatively large and known
\(W\rightarrow cs\) contribution. The mistagging rate of \(c\)-jets depends on the jet \pt and
has a total uncertainty in the range of \SIrange{3}{17}{\percent}. The uncertainties are decomposed into 20
orthogonal components.
The misidentification rate of light-quark jets was evaluated based
on the techniques described in Ref.~\cite{FTAG-2019-02}.
The resulting calibration factors are in the range of about
1.5 -- 3.0 with uncertainties up to \SI{50}{\percent}.
The uncertainties are
decomposed into 20 independent eigenvectors.

To account for differences between simulation and data in the pile-up distribution,
the pile-up profile in the simulation is corrected to match the one in data.
The uncertainty in the correction factor is applied in the measurement
as a variation of the event weight.

\subsection{Modelling uncertainties}
Normalisation uncertainties are applied to all processes, except for the ones
whose normalisation is a free parameter of the maximum-likelihood fit,
specifically for the \HF templates, for \ttW and \ttZ production, and for
the largest \VV template \VVsplit{3}{b/c}.
Thus, no normalisation uncertainty is applied to \ttbar, single-top-quark, \zjets
and \wjets production, all of which exclusively contribute to non-prompt templates.
Minor non-prompt templates are assigned \SI{50}{\percent} normalisation uncertainties.
The same is true for the minor \VV templates and all rare processes discussed in
\Cref{sec:mc_samples_bg}.
For \tZq and \tWZ production, a respective uncertainty of \SI{30}{\percent}
is assigned based on a previous ATLAS measurement of the processes~\cite{TOPQ-2016-14}.
Asymmetric uncertainties of \(^{+\SI{7}{\percent}}_{-\SI{10}{\percent}}\) are assigned to \ttH
production, estimated by varying the renormalisation and factorisation scales,
and the combined PDF+\alphas uncertainties~\cite{deFlorian:2016spz}.

Uncertainties in modelling parton showers and hadronisation are assigned to the
\tHq signal samples, and the background originating from \ttbar, \ttH,
\ttZ and \ttW production.
For the signal and \ttbar, \ttH and \ttZ production the nominal samples are
compared with alternative samples, for which the individual matrix element
generators were interfaced to \Herwig[7.1.3]~\cite{Bahr:2008pv,Bellm:2015jjp}
(for \ttbar production), \Herwig[7.0.4] (for \ttH production) and  \Herwig[7.2.1] (for \ttZ production)
instead of the respective versions of \Pythia[8].
For \ttW production, two alternative samples are generated.
The matrix element of both the samples is generated with \POWHEGBOX[v2], which is
interfaced to \Pythia[8] for one sample and to \Herwig[7] for the other.
The relative differences between the two samples is determined for all bins in
the analysis and applied to the nominal \Sherpa[2.2.10] \ttW sample.

The impact of the generator choice for \ttW is evaluated using an alternative sample and is applied
as systematic uncertainties for both the normalisation and shape. This alternative sample is generated with up to one additional
parton in the final state at \NLO accuracy using \MGNLO[2.9.3], with different parton multiplicities
being merged using the FxFx \NLO matrix-element and parton-shower merging prescription~\cite{Frederix:2021agh}
with a merging scale of \SI{30}{\GeV}.
The events are interfaced with \PYTHIA[8.245] using the A14 set of tuned parameters and the \NNPDF[2.3lo] PDF
set to simulate hadronisation and showering.
The uncertainty is split into three components: a shape uncertainty in each region,
a region migration uncertainty and a global normalisation uncertainty.

Uncertainties related to the choice of the renormalisation scale \(\muR\) and the factorisation scale \(\muF\) for the matrix-element
calculations are evaluated by varying the scales independently by factors of
\(2\) and \(0.5\), separately for \VV production and each of the top-quark
production processes, including  production with an additional
boson and the two signal processes.
The scale variations are implemented as generator weights in the nominal sample.
The uncertainties are treated as uncorrelated across individual processes.

The uncertainty in matching the NLO matrix elements to the parton shower when generating
\(t\tbar\) events is evaluated by comparing the nominal samples of simulated events
to samples with an alternative setting of the \(\pt^{\mathrm{hard}}\) parameter in the
matching code, using one instead of the default setting of zero. This parameter regulates the
definition of the vetoed region of the parton shower, which is needed to avoid overlap in the
phase space filled by \POWHEG and \PYTHIA. This estimate of the uncertainty follows the
description in Ref.~\cite{Hoche:2021mkv}.
The uncertainty in the choice of the $h_\mathrm{damp}$ parameter for the \(t\tbar\) event
generation is estimated by using an additional \ttbar sample produced with the $h_\mathrm{damp}$
parameter set to $3 m_t$,
while keeping all other generator settings the same as used for the nominal sample of events.

Three uncertainties are defined for the modelling of the \qmisid template.
The statistical uncertainty in the \qmisid efficiencies obtained from the
likelihood fit is taken into account.
An uncertainty in the selection of the $Z$-window is calculated by redefining
the $Z$-window with $3\sigma$ and $5\sigma$, where $\sigma$ corresponds to the
width of the fitted Breit--Wigner distribution.
The width of the  $Z$-window  is varied and the efficiencies are determined again, with the
symmetrised difference to the nominal values comprising the uncertainty, which
is then propagated through the analysis.
Finally, a non-closure uncertainty is implemented by applying the \qmisid estimation
method to a \(Z\to ee\) MC sample and comparing the resulting misidentification
rates to the MC prediction.
The symmetrised difference applied to the nominal \qmisid rates constitutes the
uncertainty.

Uncertainties in the amount of initial-state radiation are
defined for the top-quark production processes \ttbar, \ttZ and \ttH.
They are assessed by varying the eigentune
\texttt{Var3c} of the A14 \Pythia[8] tune consistent with the uncertainties of the
tune.
An uncertainty in the final-state radiation is introduced by varying the
renormalisation scale \(\muR\) in the parton shower, at which the strong coupling
constant $\alpha_\mathrm{s}$ is evaluated, by factors of \(0.5\) and \(2.0\).

In all uncertainty evaluations mentioned above the alternative samples or
reweighted samples are normalised to the total cross-section of the nominal samples.

Uncertainties in the PDFs are evaluated for the background processes simulated
using \POWHEGBOX, meaning \ttbar, \singlet and \ttH production,
using the PDF4LHC15 prescription with 30 eigenvectors~\cite{Butterworth:2015oua}.
Simulated events are reweighted to the central value and the eigenvectors of the combined PDF set.
Systematically varied templates are constructed by taking the differences between the samples
reweighted to the central value and those reweighted to the eigenvectors.
In the likelihood fit, the PDF uncertainties are treated as correlated across the
top-quark production processes.

The uncertainties due to the finite number of simulated events, also called the MC statistical uncertainty,
is accounted for by adding a nuisance parameter for each bin of the \dnn distributions and
the distributions in the CRs, implementing the Barlow--Beeston approach~\cite{Barlow:1993dm}.

% End of text imported from the .//tex/08_systematics.tex input file

%
\section{Statistical analysis}
%

% The next lines are included from the .//tex/09_statistical-analysis.tex input file
The normalisation \(\mu\) of the \tHu and \tHc signal couplings is determined in
binned profile maximum-likelihood fits.
The fits are performed simultaneously in all signal and control regions.
In the signal regions, the respective \dnn distributions optimised for the
considered signal coupling are used.
For the \CRdilHFe and \CRdilHFmu only event yields are employed in the fit,
whereas for the \CRtriHFe and \CRtriHFmu the transverse momentum of the
third-leading-\pT lepton is fitted.
In the \CRdilttV, the distribution of the subleading-\pT lepton's transverse
momentum is used, as it provides separation between prompt and non-prompt leptons.
The \CRtrittW and \CRtrittZ are incorporated into the fit with the distribution
of the transverse momentum of the leading-\pT \(b\)-tagged jet.
This distribution separates the \ttZ background from remaining \VV events in the
\CRtrittZ and is also adopted in the \CRtrittW for reasons of consistency.
In the fits, each of the signal processes are treated as maximally anti-correlated,
implying that the presence of one signal automatically precludes the presence of
the other signal.

The likelihood function is constructed as a product of Poisson probability
terms over all considered bins.
The fitted event yields in the bins depend on nuisance parameters \(\theta\)
which include the effects of systematic uncertainties.
Each nuisance parameter, except those representing the MC statistical
uncertainties, is constrained by a Gaussian distribution term in the likelihood function.
Some systematically varied discriminant distributions are smoothed and nuisance
parameters of systematic uncertainties with negligible impact on the parameter
of interest are entirely removed  to reduce spurious effects in
minimisation, improve convergence of the fit, and reduce the computing time.
Normalisation and shape effects from a source of systematic uncertainty are
treated separately in this removal process.
Single-sided systematic variations are turned into symmetric variations by
taking the difference  between the nominal model
and the alternative model and mirroring this difference in the opposite direction.
This is done for the event yield and for the event shape, where the distributions are subtracted from the nominal one.
For most sources with two variations, their effects are made symmetric by using
the average deviation from the nominal prediction.
Exceptions are the uncertainties in the JER, for which the asymmetric variations
are kept because the underlying effects are known to be asymmetric.
Free-floating normalisation factors are assigned to the templates modelling the
\HFe and \HFmu background, as well as the \(\VV3\ell\)+\(b\)/\(c\), \ttW, and \ttZ templates.

The test statistic \(q_\mu\) is defined as the profile likelihood ratio
\[
q_\mu=-2 \ln \frac{
\mathcal{L}(\mu, \hat{\hat{\theta}})
}{
\mathcal{L}(\hat\mu, \hat\theta)
},
\]
where \(\hat\mu\) and \(\hat\theta\) are the values of the parameters that
maximise the likelihood function and \(\hat{\hat{\theta}}\) are the values of the
nuisance parameters that maximise the likelihood function for a given value of
\(\mu\).
The test statistic is evaluated with the RooFit package~\cite{RooFit}.
If the observed signal normalisation \(\mu\) is compatible with the background
hypothesis, that is \(\mu=0\), the test statistic \(q_\mu\) is used in the \CLs
method~\cite{Read:2002hq} to obtain exclusion limits on the signal normalisation.
For a given signal scenario, values of \(\mu\) yielding \(\CLs\leq0.05\), where
\CLs is computed using the asymptotic approximation~\cite{Cowan:2010js}, are
excluded at $\geq\SI{95}{\percent}$ CL\@.
The obtained upper limits on the signal strength \(\mu\) are
then transformed into limits on the respective Wilson coefficient \Cuphi
and the corresponding branching ratio \(\BR(t\to Hq)\) through their
dimension-6 operators~\cite{fcnc_production_theory}.

% End of text imported from the .//tex/09_statistical-analysis.tex input file

%
\section{Results}
%

% The next lines are included from the .//tex/10_results.tex input file
This section presents the results of the profile-likelihood fit.
Prior to the full fit to extract the signal normalisation, a background-only fit
using only regions with little signal contribution is performed to check the
modelling of NN input variables.
Once the modelling is checked and validity of the fit model is established, the
signal normalisation is determined in a full fit to data using the entire phase
space of the analysis.

\subsection{Cross-checks on the modelling of the neural network input variables}

\begin{figure}
\centering
\subfigure[]{
\includegraphics[width=0.4\linewidth]{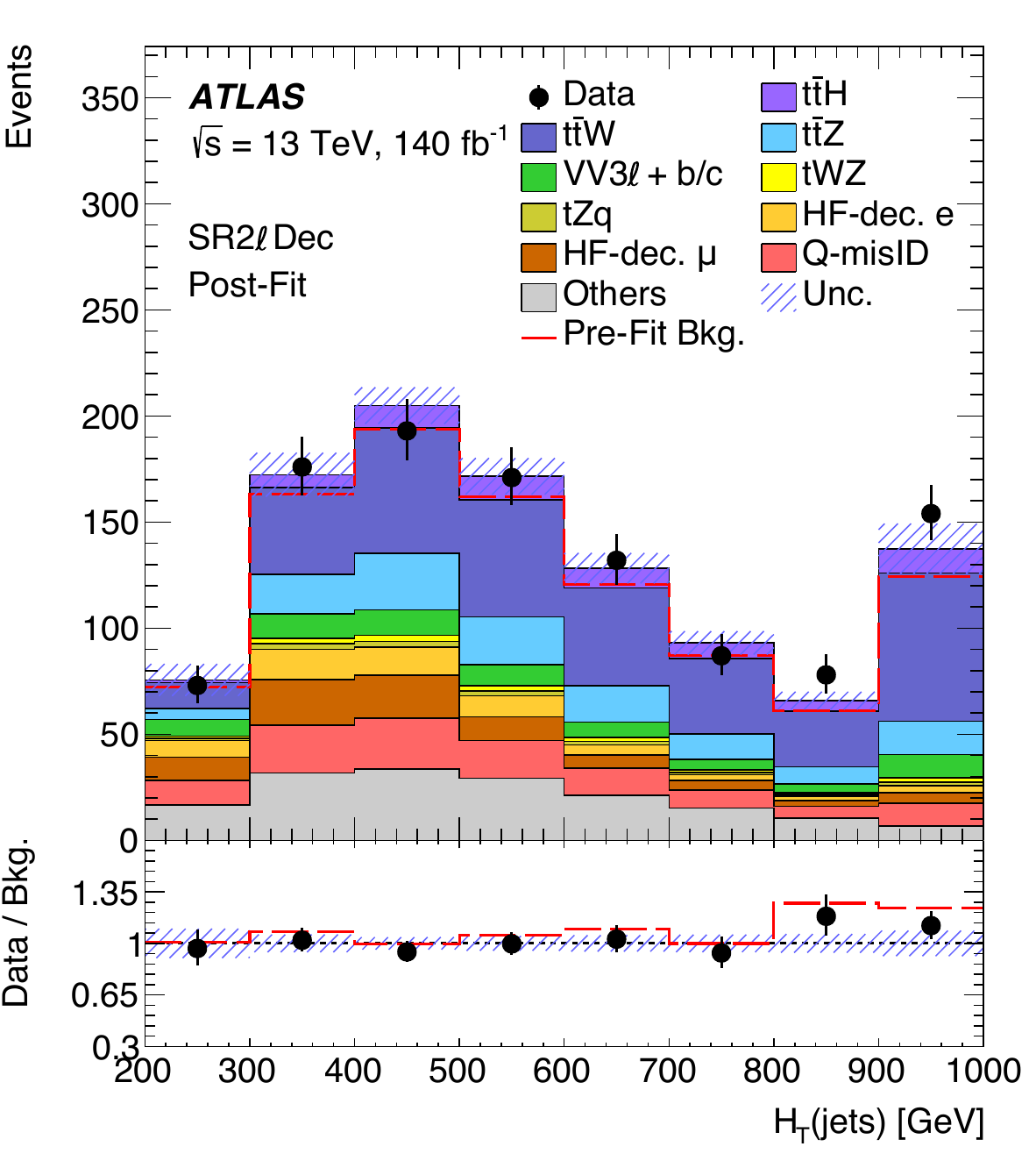}
}
\subfigure[]{
\includegraphics[width=0.4\linewidth]{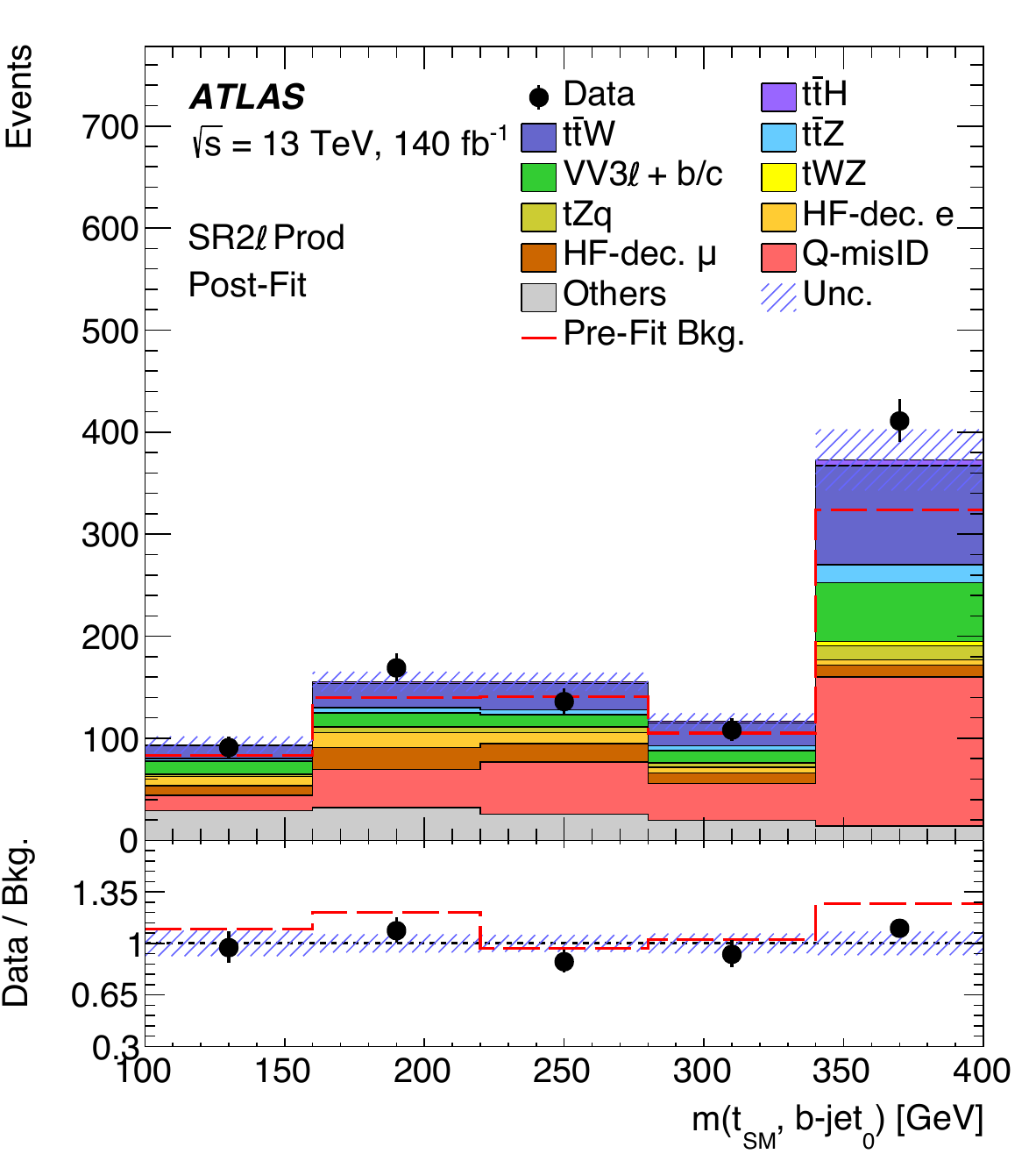}
}
\subfigure[]{
\includegraphics[width=0.4\linewidth]{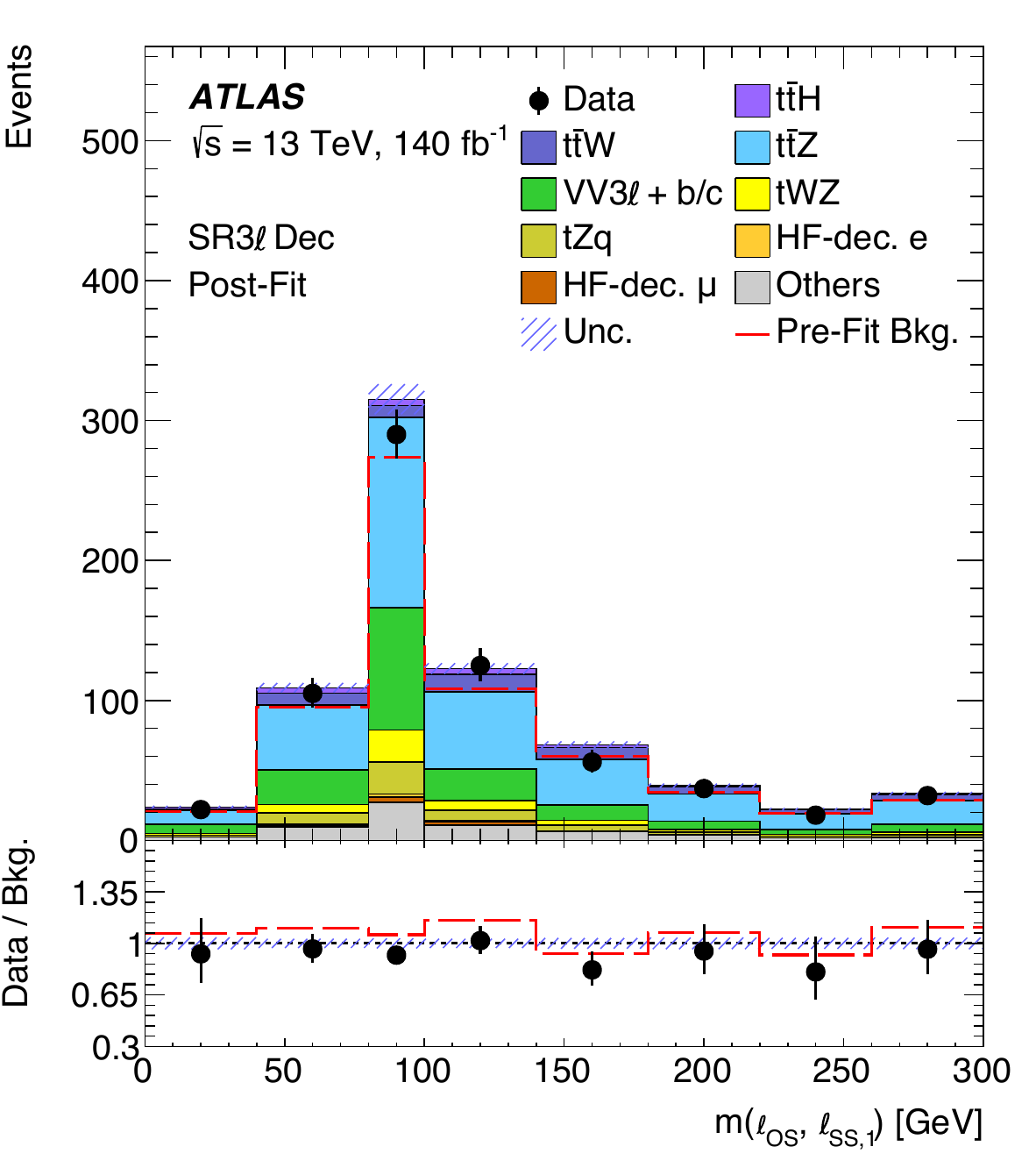}
}
\subfigure[]{
\includegraphics[width=0.4\linewidth]{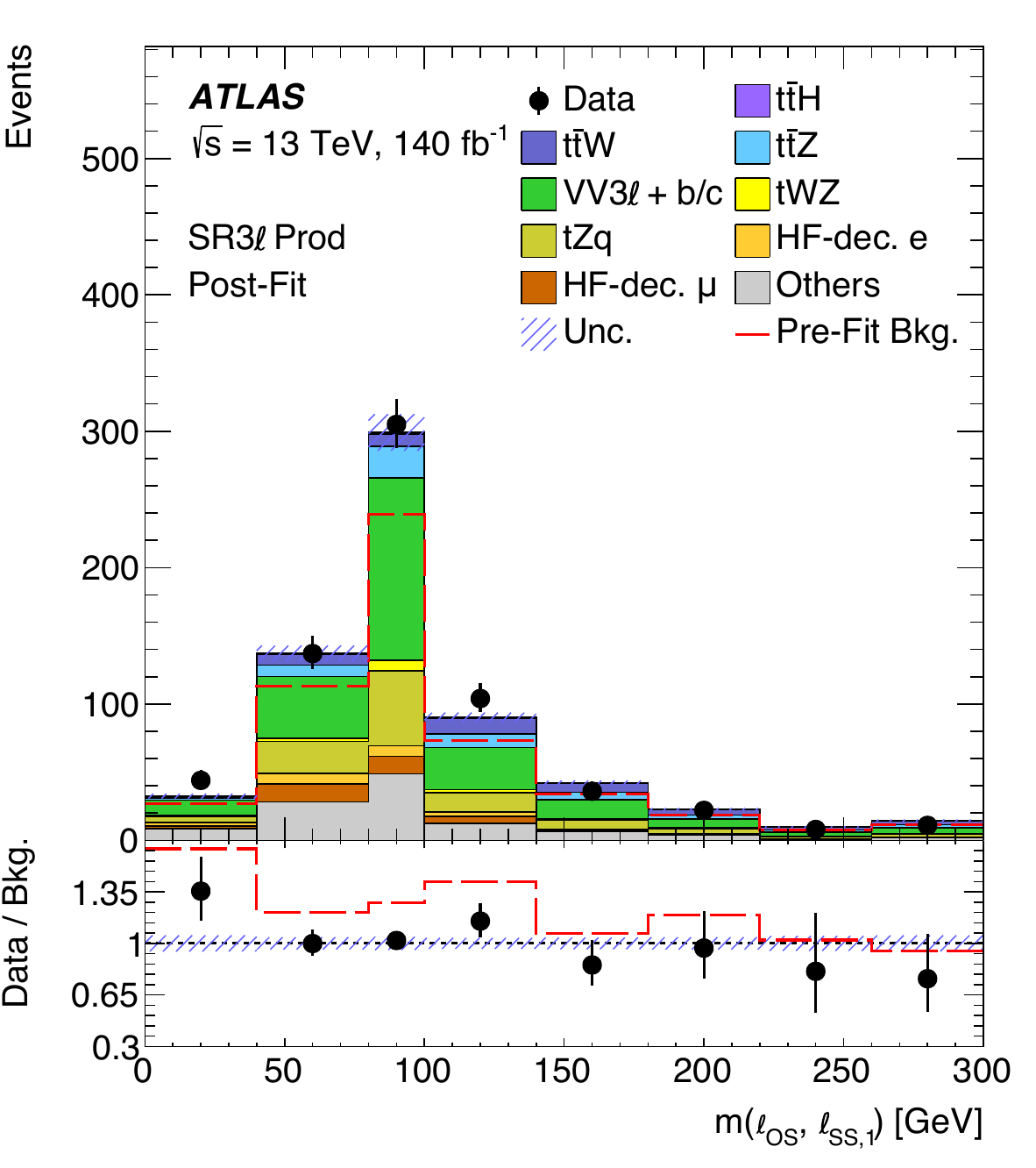}
}
\caption{
Distributions of the most important NN input variable for the signal regions (a) \SRdildec,
(b) \SRdilprod, (c) \SRtridec and (d) \SRtriprod.
The corrections from a background-only fit are applied.
Only events with \(\dnn<0.5\) were used in the fit and in the plots.
The first bin contains all events below its lower boundary, while the last
bin contains all events falling above its upper boundary.
The \textit{Others} template contains various minor processes, which
individually only have small contributions.
The total statistical and systematic uncertainty is indicated by the hatched band.
The dashed line depicts the sum of all background processes prior to the
fit. The last bin includes the overflow.
}\label{fig:leading_NN_var_validation_postfit}
\end{figure}

Before the application of the NNs to the full collision data, modelling of the
input variables is checked in parts of the signal regions.
For this purpose, the events populating the respective signal regions up to a
threshold value of \(\dnn=0.5\) are selected.
The threshold of \(\dnn=0.5\) is chosen to maximise the available statistics while
avoiding a signal contamination of more than \SI{10}{\percent}.
The full fit is performed with both statistical and systematic uncertainties considered
in all control regions and the low-\dnn signal regions.
The corrections made by the fit are then applied to the input variables of the
various NNs, using only events with \(\dnn<0.5\), to check the variable modelling.
The resulting distributions of the most important NN input variables in each
signal region are shown in \Cref{fig:leading_NN_var_validation_postfit}.
Overall, no discrepancies between the post-fit distributions and data could be
observed in any of the input variables.

\subsection{Full fit to data}

\begin{figure}
\centering
\subfigure[]{ \includegraphics[width=0.4\linewidth]{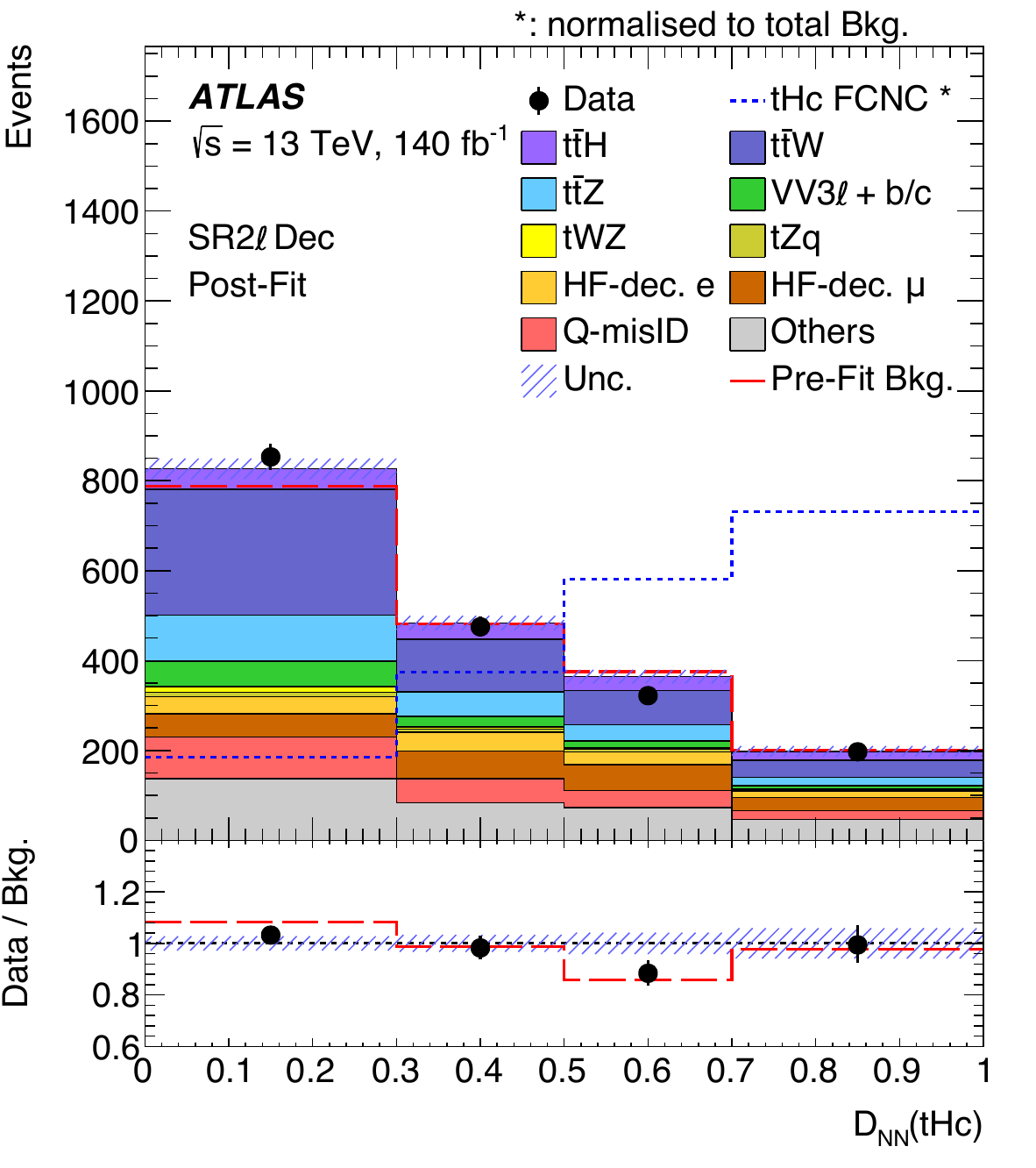} }
\subfigure[]{ \includegraphics[width=0.4\linewidth]{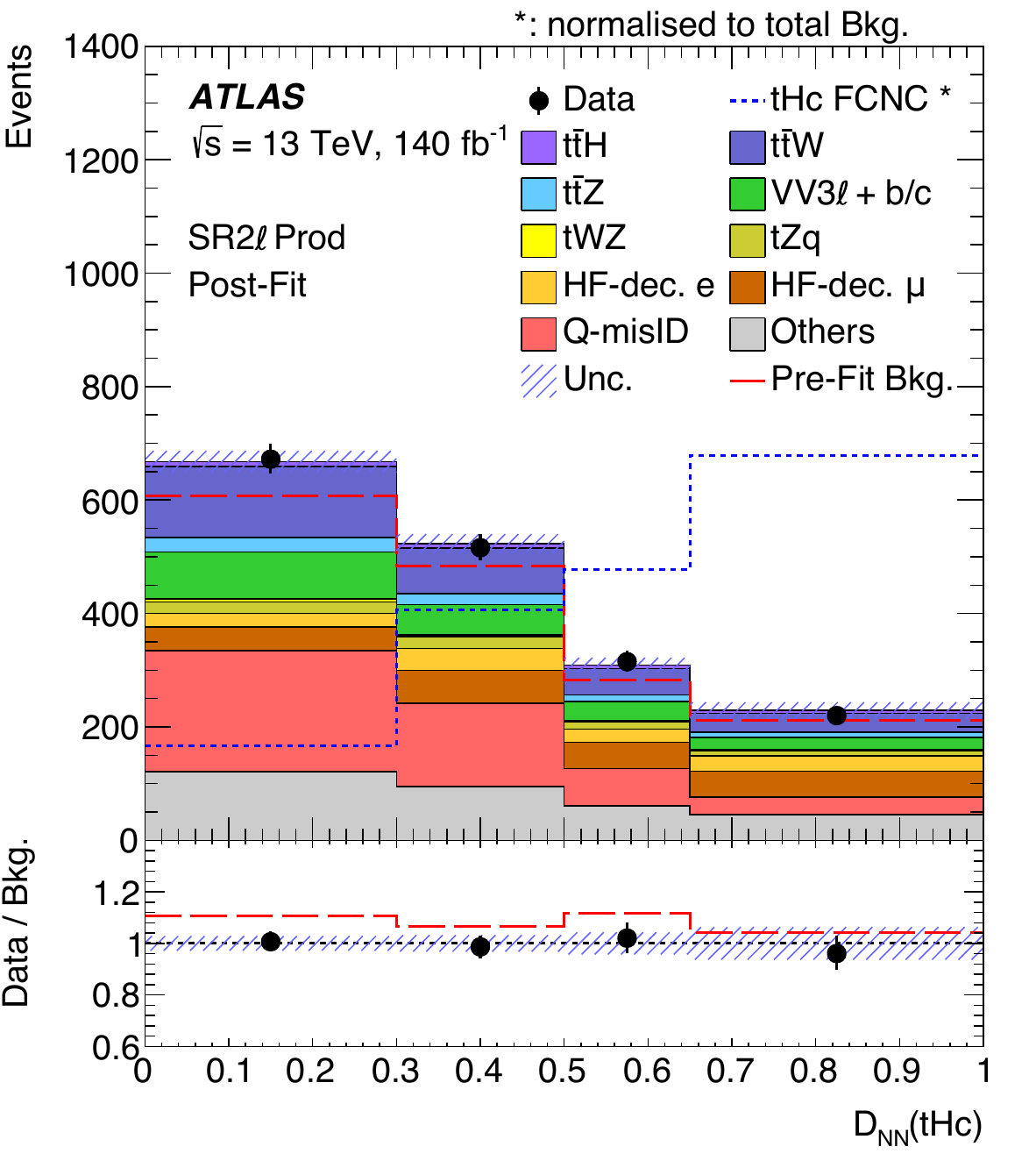} }
\subfigure[]{ \includegraphics[width=0.4\linewidth]{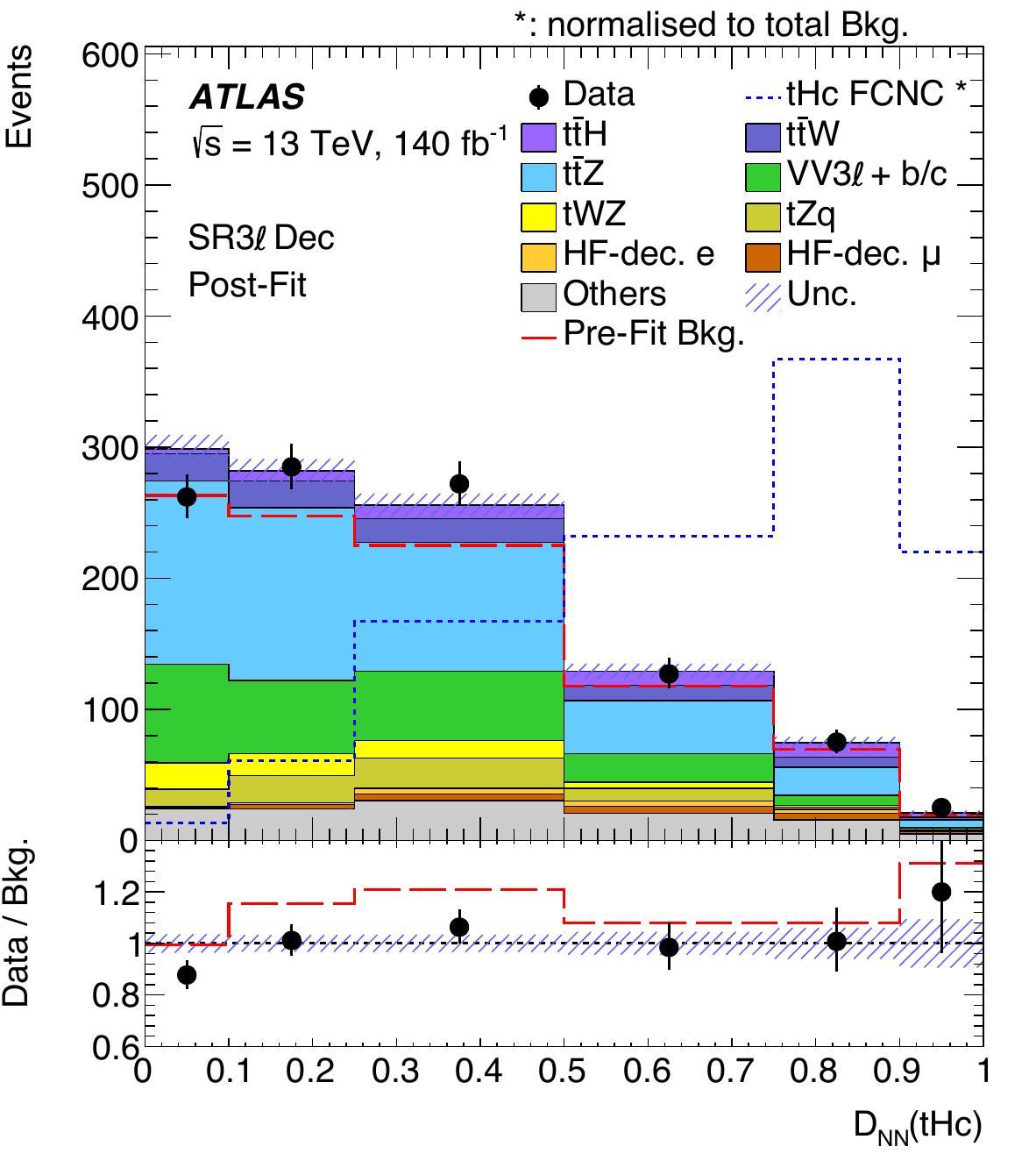} }
\subfigure[]{ \includegraphics[width=0.4\linewidth]{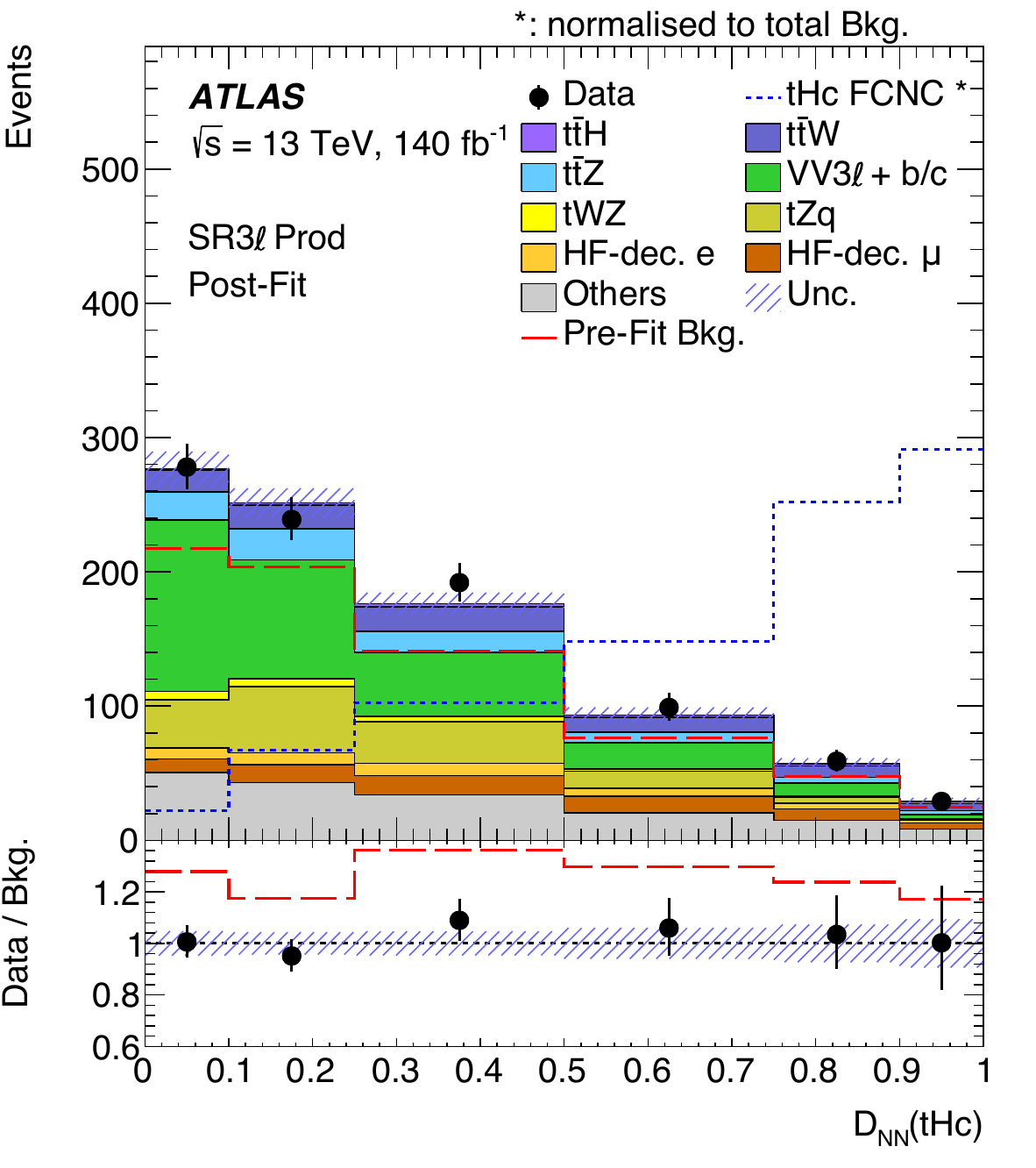} }
\caption{
The \dnn distributions in the (a) \SRdildec, (b) \SRdilprod, (c) \SRtridec and
(d) \SRtriprod, obtained from the signal-plus-background fit to data in the \tHc channel.
The \textit{Others} template contains various minor processes, which
individually only have small contributions.
The total statistical and systematic uncertainty is indicated by the hatched
band.
The dotted line represents the distribution of the signal, scaled to the
number of background events.
The dashed line depicts the sum of all background processes prior to the
fit.
}\label{fig:post_fit_NN}
\end{figure}

\begin{figure}
\centering
\includegraphics[width=0.9\linewidth]{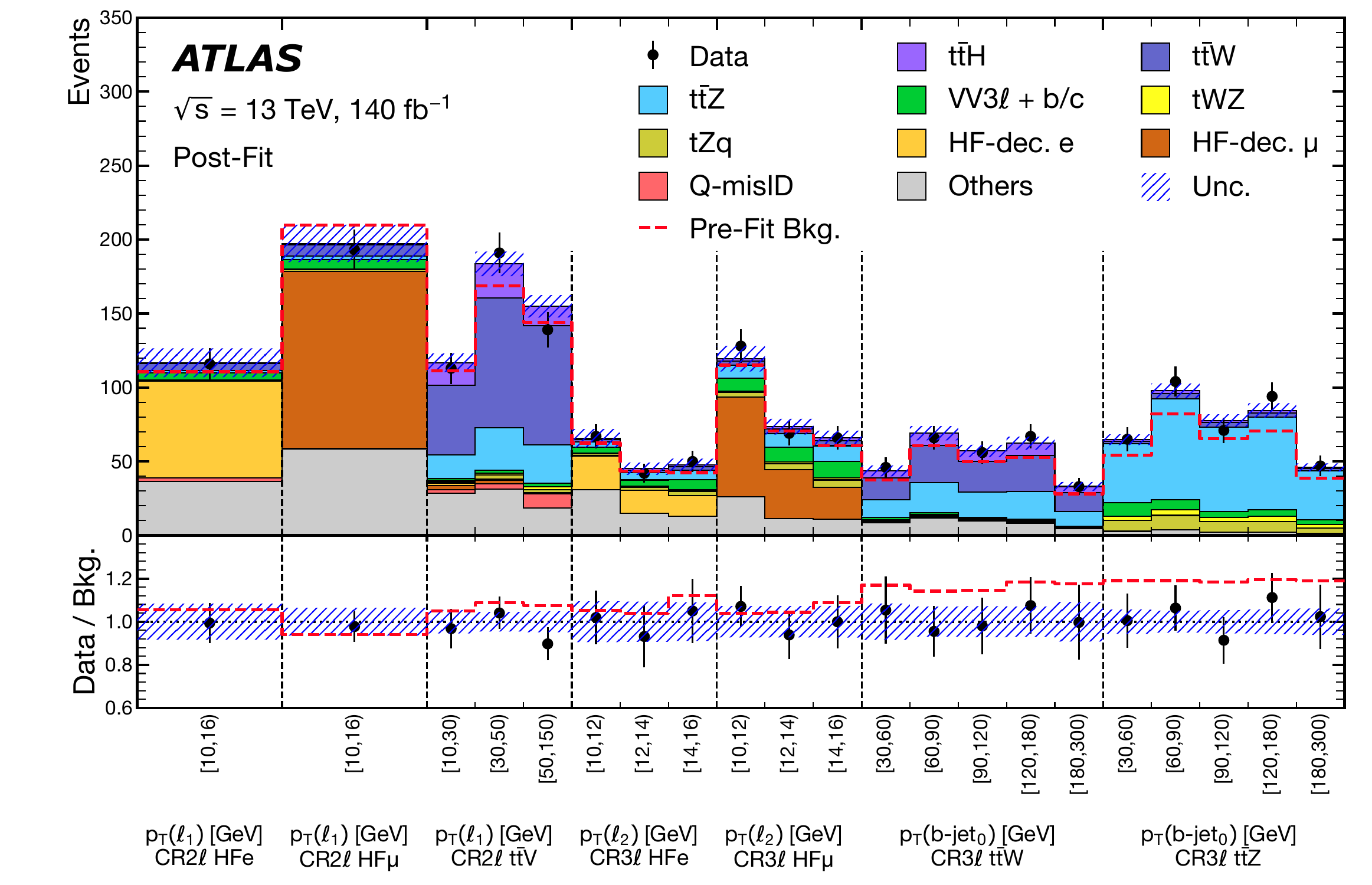}
\caption{
Summary plot of the fitted distributions in all control regions obtained from
the signal-plus-background fit to data in the \tHc channel.
The bin boundaries for each region are depicted by the parenthesised numbers.
The \textit{Others} template contains various minor processes, which
individually only have small contributions.
The total statistical and systematic uncertainty is indicated by the hatched
band.
The dashed line depicts the sum of all background processes prior to the
fit.
}\label{fig:post_fit_CR_summary}
\end{figure}

\Cref{fig:post_fit_NN} shows the \(\dnn(\tHc)\) distributions in all four signal regions
after the fit.
A summary plot of all \dil and \tri control regions is shown in
\Cref{fig:post_fit_CR_summary}.
Overall, a good agreement between MC and data is observed.
The event yields shown in \Cref{tab:post_fit_yields_2l,tab:post_fit_yields_3l} confirm this.
The best fit value of the normalisation of the \tHu (\tHc) signal is found to be
\(\mu_{\tHq}=-0.03\pm0.15(-0.08\pm0.19)\).
No pulls beyond $1\sigma$ or strong constraints are observed for any of the
nuisance parameters.
The observed post-fit values of all normalisation factors and nuisance parameters
are in agreement with the background-only fit consistent with the statistical uncertainties,
suggesting a good modelling of the most signal-sensitive \dnn bins.
The obtained normalisation factors for free-floating background processes are
listed in \Cref{tab:norm_factors}.
For the \HFe and \HFmu processes they are compatible with one.
A \SI{40}{\percent} increase can be observed for the \VVsplit{3}{b/c} template,
owing to the poorer modelling of $b$-jets by the shower generator.
The normalisations of the \ttW and \ttZ processes are increased by approximately
\SI{15}{\percent}, which is compatible at a $1\sigma$ level.
The increased normalisation of \ttW production is expected based on dedicated
measurements of the process,
while for \ttZ, the slightly higher normalisation comes from the inclusion of
events with low jet multiplicity.
Studies on the process confirm that for higher values of \njets the normalisation
is compatible with one.

As the best fit value of the signal strength is compatible with \(\mu=0\), upper
limits are determined with the \CLs method.
In addition to the nominal limits of the analysis, expected limits are determined
for a fit considering only statistical uncertainties, as well as for fits
in the \dil and \tri channels separately.
A comparison of the expected upper limits  on the branching ratio \(\BR(t\to Hq)\)
for each configuration is shown in \Cref{tab:exp_limit_comp}.
A degradation in the expected upper limit of approximately \(\SI{20}{\percent}\) can be
observed for both the signal processes when systematic uncertainties are included.
This shows that the sensitivity of this analysis is primarily limited by
available statistics.
When comparing the two final states with each other, a clear dominance of the
\dil final state can be observed.
Including the \tri final states yields an improvement of \SI{20}{\percent} in
the \tHu channel and \SI{13}{\percent} in the \tHc channel.

The observed upper limits on the branching ratio \(\BR(t\to Hq)\) together
with a reinterpretation as limits on the absolute value of the Wilson coefficient
\(|\Cuphi|\) for the EFT dimension-6 operators can be found in \Cref{tab:limits}.
The signal does not interfere with other SM processes,
meaning that the signal normalisation is proportional to \(|\Cuphi|^2\) and
limits can only be set on the coefficient's absolute value.
Additionally, owing to the averaging of the left-handed and the right-handed
signal component, these limits are set on the average of the two respective coefficients
\(|\CuphiRh|\) and \(|\CuphiLh|\).
The final observed (expected) upper limits on the branching ratio are
\(\BR(t\to Hu)<2.8\,(3.0)\times10^{-4}\) and \(\BR(t\to Hc)<3.3\,(3.8)\times10^{-4}\)
for a cutoff scale $\Lambda=\SI{1}{\TeV}$.
The limits on the Wilson coefficients amount to \(|\Cuphiavgu|<0.71\,(0.73)\) and
\(|\Cuphiavgc|<0.76\,(0.82)\).
Limits for each of the two signal couplings are determined with the assumption that the
other signal coupling do not contribute.

Compared with a previous ATLAS search for \tHq FCNC couplings in \dil/ \tri final
states using a partial Run 2 data sample of \SI{36}{\ifb}~\cite{TOPQ-2017-15},
a luminosity-adjusted improvement by a factor of two for the \tHc channel
and three for the \tHu channel can be observed.
This is determined by creating an Asimov-data sample, scaling
its luminosity to that of the previous analysis, and comparing the resulting
expected upper limits to the original analysis.
This improvement primarily results from using improved analysis techniques
such as new methods for event reconstruction and a more thorough multivariate
analysis, and more precise detector calibrations.
The additional improvement in the \tHu channel is achieved because of the
inclusion of the \(gq\to Ht\) production process in the analysis.
This process primarily plays a role in the \tHu channel because the up quark is
a valence quark of the proton.

\begin{table}
\centering
\caption{
The predicted and observed yields in all \dil regions of the analysis from the signal-plus-background fit to signal and control regions.
The pre-fit predictions for the two signal components are presented also,
scaled to a cross-section equivalent to a branching ratio \(\BR(t\to Hq)=\SI{0.1}{\percent}\).
The \textit{Others} template contains various minor processes, which
individually only have small contributions.
The uncertainties in non-signal MC yields reflect the sum in quadrature of all
post-fit systematic and statistical uncertainties.
The row labelled \textit{Post-fit BG} shows the sum of all above background (BG)
processes with post-fit uncertainties, while the row labelled \textit{Pre-fit BG} shows the total
pre-fit background yield with the corresponding pre-fit uncertainties.
Hyphens signify that the corresponding process does not contribute to the
given region.
}\label{tab:post_fit_yields_2l}
\begin{tabular}{
c
S[table-number-alignment = right, table-format=4.0]@{ }
c@{ }
S[table-number-alignment = left, table-format=2.0]
S[table-number-alignment = right, table-format=4.0]@{ }
c@{ }
S[table-number-alignment = left, table-format=2.0]
S[table-number-alignment = right, table-format=3.2]@{ }
c@{ }
S[table-number-alignment = left, table-format=2.2]
S[table-number-alignment = right, table-format=3.2]@{ }
c@{ }
S[table-number-alignment = left, table-format=2.2]
S[table-number-alignment = right, table-format=3.1]@{ }
c@{ }
S[table-number-alignment = left, table-format=2.1]
}
\toprule
Process                      & \multicolumn{3}{c}{\SRdildec} & \multicolumn{3}{c}{\SRdilprod} & \multicolumn{3}{c}{\CRdilHFe} & \multicolumn{3}{c}{\CRdilHFmu} & \multicolumn{3}{c}{\CRdilttV}                                                                                         \\
\midrule
\HFe                         & 122                           & $\pm$                          & 27                            & 113                            & $\pm$                         & 25 & 66   & $\pm$ & 13   & \multicolumn{3}{c}{--}                & 2.9  & $\pm$ & 0.9                \\
\HFmu                        & 201                           & $\pm$                          & 36                            & 192                            & $\pm$                         & 35 & 0.1  & $\pm$ & 0.02 & 120                   & $\pm$ & 22    & 5.6  & $\pm$ & 1.2 \\
\qmisid                      & 204                           & $\pm$                          & 16                            & 457                            & $\pm$                         & 35 & 2.4  & $\pm$ & 0.2  & \multicolumn{3}{c}{--}                & 15.5 & $\pm$ & 1.4                \\
\ttH                         & 132                           & $\pm$                          & 20                            & 27                             & $\pm$                         & 5  & 0.6  & $\pm$ & 0.1  & 1.0                   & $\pm$ & 0.2   & 51   & $\pm$ & 8   \\
\ttW                         & 512                           & $\pm$                          & 61                            & 285                            & $\pm$                         & 42 & 4.8  & $\pm$ & 0.9  & 7.5                   & $\pm$ & 1.4   & 216  & $\pm$ & 24  \\
\ttZ                         & 210                           & $\pm$                          & 21                            & 66                             & $\pm$                         & 9  & 1.5  & $\pm$ & 0.2  & 2.4                   & $\pm$ & 0.4   & 70   & $\pm$ & 6   \\
\VV $3\ell$+$b$/$c$          & 104                           & $\pm$                          & 20                            & 192                            & $\pm$                         & 32 & 4.7  & $\pm$ & 1.0  & 6.6                   & $\pm$ & 1.4   & 6.0  & $\pm$ & 1.2 \\
\tWZ                         & 23                            & $\pm$                          & 7                             & 12                             & $\pm$                         & 4  & 0.11 & $\pm$ & 0.04 & 0.17                  & $\pm$ & 0.06  & 3.6  & $\pm$ & 1.1 \\
\tZq                         & 26                            & $\pm$                          & 8                             & 63                             & $\pm$                         & 18 & 0.7  & $\pm$ & 0.2  & 1.1                   & $\pm$ & 0.3   & 5.8  & $\pm$ & 1.7 \\
Others                       & 340                           & $\pm$                          & 64                            & 322                            & $\pm$                         & 46 & 36   & $\pm$ & 8    & 59                    & $\pm$ & 20    & 79   & $\pm$ & 14  \\\midrule
Pre-fit BG                   & 1845                          & $\pm$                          & 91                            & 1585                           & $\pm$                         & 70 & 111  & $\pm$ & 11   & 210                   & $\pm$ & 32    & 424  & $\pm$ & 22  \\
Post-fit BG                  & 1874                          & $\pm$                          & 38                            & 1729                           & $\pm$                         & 36 & 117  & $\pm$ & 10   & 198                   & $\pm$ & 12    & 455  & $\pm$ & 17  \\\midrule
$\ttbar(t\to Hu)$            & 207                           & $\pm$                          & 22                            & 181                            & $\pm$                         & 10 & 3.4  & $\pm$ & 0.3  & 5.4                   & $\pm$ & 0.7   & 6.8  & $\pm$ & 0.6 \\
$ug\to Ht$                   & 31                            & $\pm$                          & 4                             & 68                             & $\pm$                         & 2  & 1.2  & $\pm$ & 0.1  & 2.1                   & $\pm$ & 0.2   & 1.1  & $\pm$ & 0.1 \\
$\ttbar(t\to Hc)$            & 196                           & $\pm$                          & 22                            & 180                            & $\pm$                         & 10 & 3.5  & $\pm$ & 0.4  & 5.9                   & $\pm$ & 0.7   & 13.4 & $\pm$ & 1.5 \\
$cg\to Ht$                   & 5                             & $\pm$                          & 1                             & 11                             & $\pm$                         & 1  & 0.2  & $\pm$ & 0.1  & 0.4                   & $\pm$ & 0.1   & 0.2  & $\pm$ & 0.1 \\\midrule
Data                         & \multicolumn{3}{c}{1847}      & \multicolumn{3}{c}{1723}       & \multicolumn{3}{c}{116}       & \multicolumn{3}{c}{193}        & \multicolumn{3}{c}{443}                                                                                               \\
\bottomrule
\end{tabular}
\end{table}

\begin{table}
\centering
\caption{
The predicted and observed yields in all \tri regions of the analysis from the signal-plus-background fit to signal and control regions.
The pre-fit predictions for the two signal components are presented as well,
scaled to a cross-section equivalent to a branching ratio \(\BR(t\to Hq)=\SI{0.1}{\percent}\).
The \textit{Others} template contains various minor processes, which
individually only have small contributions.
The uncertainties in non-signal MC yields reflect the sum in quadrature of all
post-fit systematic and statistical uncertainties.
The row labelled \textit{Post-fit BG} shows the sum of all above background (BG)
processes with post-fit uncertainties, while the row labelled \textit{Pre-fit BG} shows the total
pre-fit background yield with the corresponding pre-fit uncertainties.
Hyphens signify that the corresponding process does not contribute to the
given region.
}\label{tab:post_fit_yields_3l}
\begin{tabular}{
c
S[table-number-alignment = right, table-format=4.0]@{ }
c@{ }
S[table-number-alignment = left, table-format=2.0]
S[table-number-alignment = right, table-format=4.0]@{ }
c@{ }
S[table-number-alignment = left, table-format=2.0]
S[table-number-alignment = right, table-format=3.1]@{ }
c@{ }
S[table-number-alignment = left, table-format=2.1]
S[table-number-alignment = right, table-format=3.2]@{ }
c@{ }
S[table-number-alignment = left, table-format=2.2]
S[table-number-alignment = right, table-format=3.1]@{ }
c@{ }
S[table-number-alignment = left, table-format=2.1]
S[table-number-alignment = right, table-format=3.1]@{ }
c@{ }
S[table-number-alignment = left, table-format=2.1]
}
\toprule
Process             & \multicolumn{3}{c}{\SRtriprod} & \multicolumn{3}{c}{\SRtridec} & \multicolumn{3}{c}{\CRtrittW} & \multicolumn{3}{c}{\CRtrittZ} & \multicolumn{3}{c}{\CRtriHFe} & \multicolumn{3}{c}{\CRtriHFmu}                                                                                                      \\
\midrule
\HFe                & 38                             & $\pm$                         & 9                             & 14                            & $\pm$                         & 3                              & 1.3 & $\pm$ & 0.3 & 0.28 & $\pm$ & 0.09 & 53   & $\pm$ & 11  & \multicolumn{3}{c}{--}               \\
\HFmu               & 63                             & $\pm$                         & 11                            & 22                            & $\pm$                         & 4                              & 1.6 & $\pm$ & 0.3 & 0.37 & $\pm$ & 0.08 & 0.2  & $\pm$ & 0.1 & 122                   & $\pm$ & 19  \\
\qmisid             & \multicolumn{3}{c}{--}          & \multicolumn{3}{c}{--}         & \multicolumn{3}{c}{--}         & \multicolumn{3}{c}{--}         & \multicolumn{3}{c}{--}         & \multicolumn{3}{c}{--}                                                                                                               \\
\ttH                & 10                             & $\pm$                         & 2                             & 47                            & $\pm$                         & 7                              & 32  & $\pm$ & 5   & 6.7  & $\pm$ & 1.1  & 3.0  & $\pm$ & 0.5 & 5.2                   & $\pm$ & 0.9 \\
\ttW                & 77                             & $\pm$                         & 12                            & 80                            & $\pm$                         & 10                             & 98  & $\pm$ & 16  & 12.5 & $\pm$ & 1.6  & 5.8  & $\pm$ & 1.0 & 9.5                   & $\pm$ & 1.4 \\
\ttZ                & 75                             & $\pm$                         & 11                            & 438                           & $\pm$                         & 40                             & 78  & $\pm$ & 7   & 261  & $\pm$ & 20   & 14.7 & $\pm$ & 1.8 & 28                    & $\pm$ & 3   \\
\VV $3\ell$+$b$/$c$ & 296                            & $\pm$                         & 49                            & 215                           & $\pm$                         & 39                             & 4.8 & $\pm$ & 0.9 & 27   & $\pm$ & 5    & 15   & $\pm$ & 3   & 30                    & $\pm$ & 5   \\
\tWZ                & 19                             & $\pm$                         & 6                             & 57                            & $\pm$                         & 18                             & 2.9 & $\pm$ & 0.9 & 16   & $\pm$ & 5    & 1.9  & $\pm$ & 0.6 & 3.6                   & $\pm$ & 1.1 \\
\tZq                & 134                            & $\pm$                         & 38                            & 69                            & $\pm$                         & 20                             & 3.5 & $\pm$ & 1.0 & 35   & $\pm$ & 10   & 6.1  & $\pm$ & 1.8 & 12                    & $\pm$ & 3   \\
Others              & 171                            & $\pm$                         & 32                            & 119                           & $\pm$                         & 23                             & 43  & $\pm$ & 7   & 11.7 & $\pm$ & 1.5  & 59   & $\pm$ & 8   & 48                    & $\pm$ & 9   \\\midrule
Pre-fit BG          & 710                            & $\pm$                         & 48                            & 941                           & $\pm$                         & 42                             & 228 & $\pm$ & 17  & 312  & $\pm$ & 19   & 148  & $\pm$ & 11  & 248                   & $\pm$ & 16  \\
Post-fit BG         & 882                            & $\pm$                         & 28                            & 1061                          & $\pm$                         & 28                             & 265 & $\pm$ & 14  & 371  & $\pm$ & 17   & 159  & $\pm$ & 10  & 258                   & $\pm$ & 14  \\\midrule
$\ttbar(t\to Hu)$   & 26                             & $\pm$                         & 2                             & 39                            & $\pm$                         & 3                              & 1.2 & $\pm$ & 0.2 & 0.7  & $\pm$ & 0.1  & 4.6  & $\pm$ & 0.5 & 8.0                   & $\pm$ & 0.8 \\
$ug\to Ht$          & 14                             & $\pm$                         & 1                             & 7                             & $\pm$                         & 1                              & 0.4 & $\pm$ & 0.1 & 0.2  & $\pm$ & 0.1  & 1.2  & $\pm$ & 0.1 & 2.1                   & $\pm$ & 0.2 \\
$\ttbar(t\to Hc)$   & 27                             & $\pm$                         & 2                             & 37                            & $\pm$                         & 3                              & 4.3 & $\pm$ & 0.4 & 1.5  & $\pm$ & 0.1  & 4.4  & $\pm$ & 0.5 & 7.8                   & $\pm$ & 0.7 \\
$cg\to Ht$          & 2                              & $\pm$                         & 1                             & 1                             & $\pm$                         & 1                              & 0.1 & $\pm$ & 0.1 & 0.1  & $\pm$ & 0.1  & 0.2  & $\pm$ & 0.1 & 0.4                   & $\pm$ & 0.1 \\\midrule
Data                & \multicolumn{3}{c}{896}        & \multicolumn{3}{c}{1046}      & \multicolumn{3}{c}{268}       & \multicolumn{3}{c}{381}       & \multicolumn{3}{c}{159}       & \multicolumn{3}{c}{263}                                                                                                             \\
\bottomrule
\end{tabular}
\end{table}

\begin{table}
\centering
\caption{
Normalisation factors for various backgrounds determined from the signal-plus-background fit to
signal and control regions.
The results for the \tHu fit and the \tHc fit are shown.
}\label{tab:norm_factors}
\begin{tabular}{ccc}
\toprule
Process          & \tHu fit & \tHc fit \\
\midrule
\HFe             & $1.05\pm0.24$ & $1.02\pm0.23$ \\
\HFmu            & $0.94\pm0.18$ & $0.92\pm0.18$ \\
\VVsplit{3}{b/c} & $1.41\pm0.23$ & $1.37\pm0.24$ \\
\ttW             & $1.15\pm0.14$ & $1.19\pm0.14$ \\
\ttZ             & $1.16\pm0.11$ & $1.17\pm0.11$ \\
\bottomrule
\end{tabular}
\end{table}

\begin{table}
\centering
\caption{
The expected \SI{95}{\percent} upper limits on the
branching ratio \(\BR(t\to Hq)\) for the nominal and alternative fit
configurations.
One fit is performed in the full phase space considering only statistical
uncertainties.
Two other fits are performed using the full set of uncertainties in only
the \dil or \tri final states.
The \(\pm1\,\sigma\) interval of the expected limit is indicated by the upper
and lower indices.
For the \tHu channel, the assumption of \(\BR(t\to Hc)=0\) is made and
vice versa.
}\label{tab:exp_limit_comp}
\begin{tabular}{ccc}
\toprule
\multirow{2}{*}{Fit configuration} & \multicolumn{2}{c}{Expected \SI{95}{\percent} CL upper limits $/\,10^{-4}$ }                       \\
& \(\BR(t\to Hu)\)                                                         & \(\BR(t\to Hc)\)    \\
\midrule
Nominal fit                        & $3.0^{+1.2}_{-0.8}$                                                      & $3.8^{+1.5}_{-1.1}$ \\
\midrule
Statistical uncertainties only     & $2.6^{+1.1}_{-0.7}$                                                      & $3.3^{+1.2}_{-1.0}$ \\
\midrule
\dil final state only              & $3.6^{+1.5}_{-1.0}$                                                      & $4.3^{+1.9}_{-1.2}$ \\[0.2cm]
\tri final state only              & $6.5^{+2.7}_{-1.9}$                                                      & $8.9^{+3.7}_{-2.6}$ \\
\bottomrule
\end{tabular}
\end{table}

\begin{table}
\centering
\caption{
The observed (expected) \SI{95}{\percent} upper limits on the
branching ratio \(\BR(t\to Hq)\) and the absolute value of the Wilson
coefficient \(|\Cuphi|\) under the assumption of a cutoff-scale of $\Lambda=\SI{1}{\TeV}$.
For the \tHu channel, the assumption of \(\BR(t\to Hc)=0\) is made and
vice versa.
}\label{tab:limits}
\begin{tabular}{ccc}
\toprule
\multirow{2}{*}{Signal} & \multicolumn{2}{c}{Observed (expected) \SI{95}{\percent} CL upper limits }  \\
& \(\BR(t\to Hq)\)            & \(|\Cuphiavg|\)  \\
\midrule
\tHu   & \(2.8\,(3.0)\times10^{-4}\) & \(0.71\,(0.73)\) \\
\midrule
\tHc   & \(3.3\,(3.8)\times10^{-4}\) & \(0.76\,(0.82)\) \\
\bottomrule
\end{tabular}
\end{table}

% End of text imported from the .//tex/10_results.tex input file

%
\section{Combination of results with other searches}
%

% The next lines are included from the .//tex/11_combination.tex input file
The reported results are combined with the corresponding ATLAS
searches that use \(H \to \tau^+\tau^-\) decays with one or two hadronically decaying  $\tau$-leptons~\cite{TOPQ-2019-17},
and  \(H \to \bbbar\)~\cite{HDBS-2020-16} and \(H \to \gamma\gamma\)~\cite{TOPQ-2019-04_alt} decays.
For clarity, this analysis is referred to as the \(H\to VV^*\) channel in the
following, as the \(H\to WW^*/ZZ^*\) decay modes are by far the most dominant
ones contributing to \dil and \tri final states.
Orthogonality among all analyses is ensured by their respective event selections.
The \(H\to\tautau\) and the \(H\to VV^*\) analysis, specifically, are orthogonal
to each other, because the \(H\to\tautau\) analysis exclusively uses events with
exactly one or zero leptons.
The combination is performed through a simultaneous profile-likelihood fit of all
four analyses.
The correlations between the uncertainties in the different channels are assessed.
In each search, the dominant systematic uncertainties are different.
In addition, the \(H \to \tau^+\tau^-\), \(H \to \gamma\gamma\) and \(H\to VV^*\)
channels are dominated by the  statistical uncertainty in the data.
Therefore, the combination exhibits minimal sensitivity to possible
correlations of uncertainties across channels.
Specifically, uncertainties pertaining to luminosity, \pileup modelling, and jet
energy scale and resolution are correlated among the four channels.
The uncertainties related to \(b\)-tagging are correlated between the
\(H \to \tau^+\tau^-\), the \(H \to \bbbar\) and the \(H\to VV^*\) analyses,
but uncorrelated with the \(H\to\gamma\gamma\) analysis, which uses a simplified
\(b\)-tagging scheme.
The remaining uncertainties (mostly from experimental sources, and signal and
background modelling) are taken as uncorrelated.
Some of the sources of systematic uncertainties (especially related to
electron and muon identification and signal modelling) are common to the three
search channels, but are treated as uncorrelated due to slight differences in
their treatment by individual analyses.
The combined \(p\)-value for the \tHu fit  is \(0.134\), while that for
the \tHc fit it is \(0.117\), showing an overall good agreement
among the four analyses.

\begin{figure}
\begin{center}
\subfigure[]{\includegraphics[width=.49\textwidth]{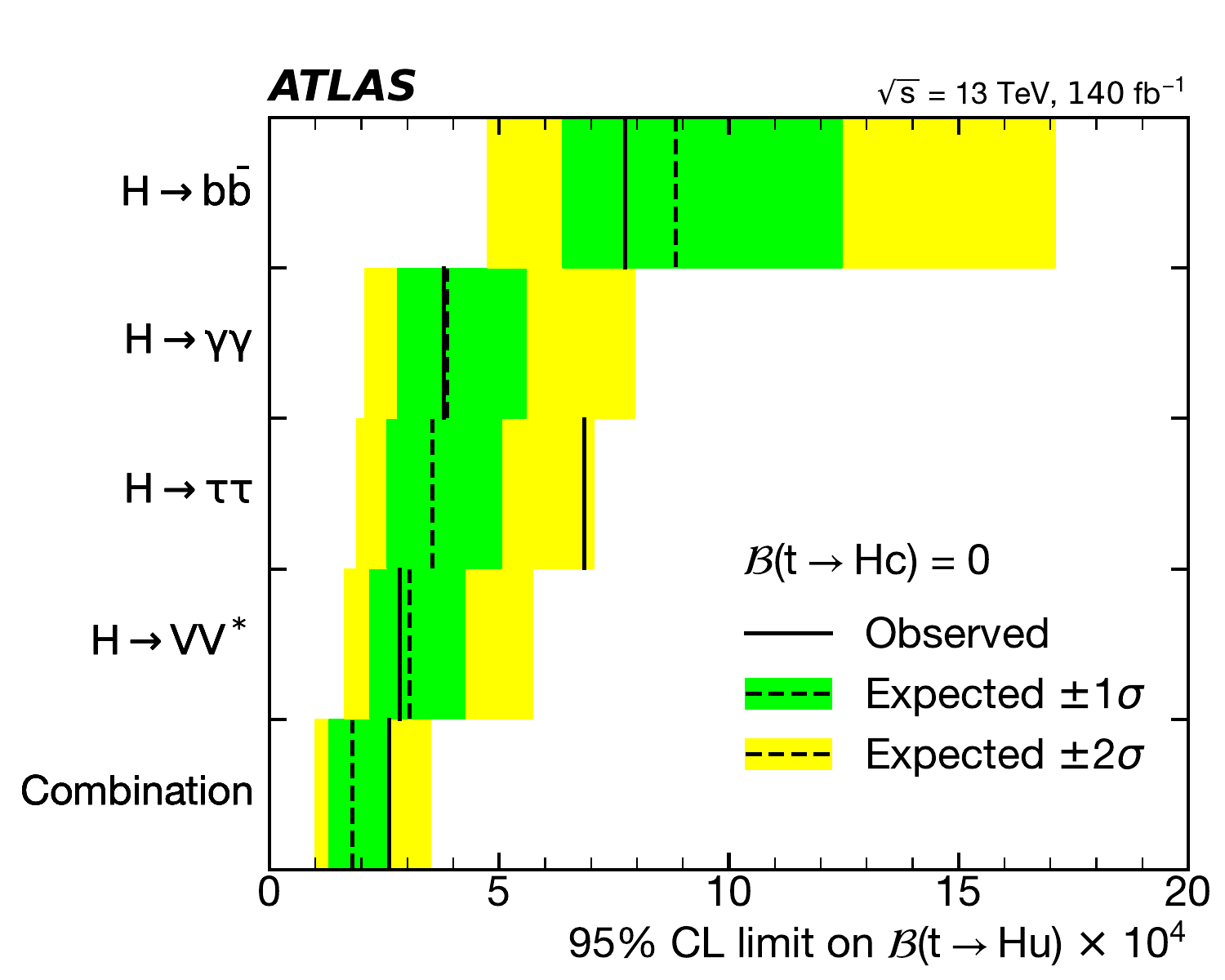} }
\subfigure[]{\includegraphics[width=.49\textwidth]{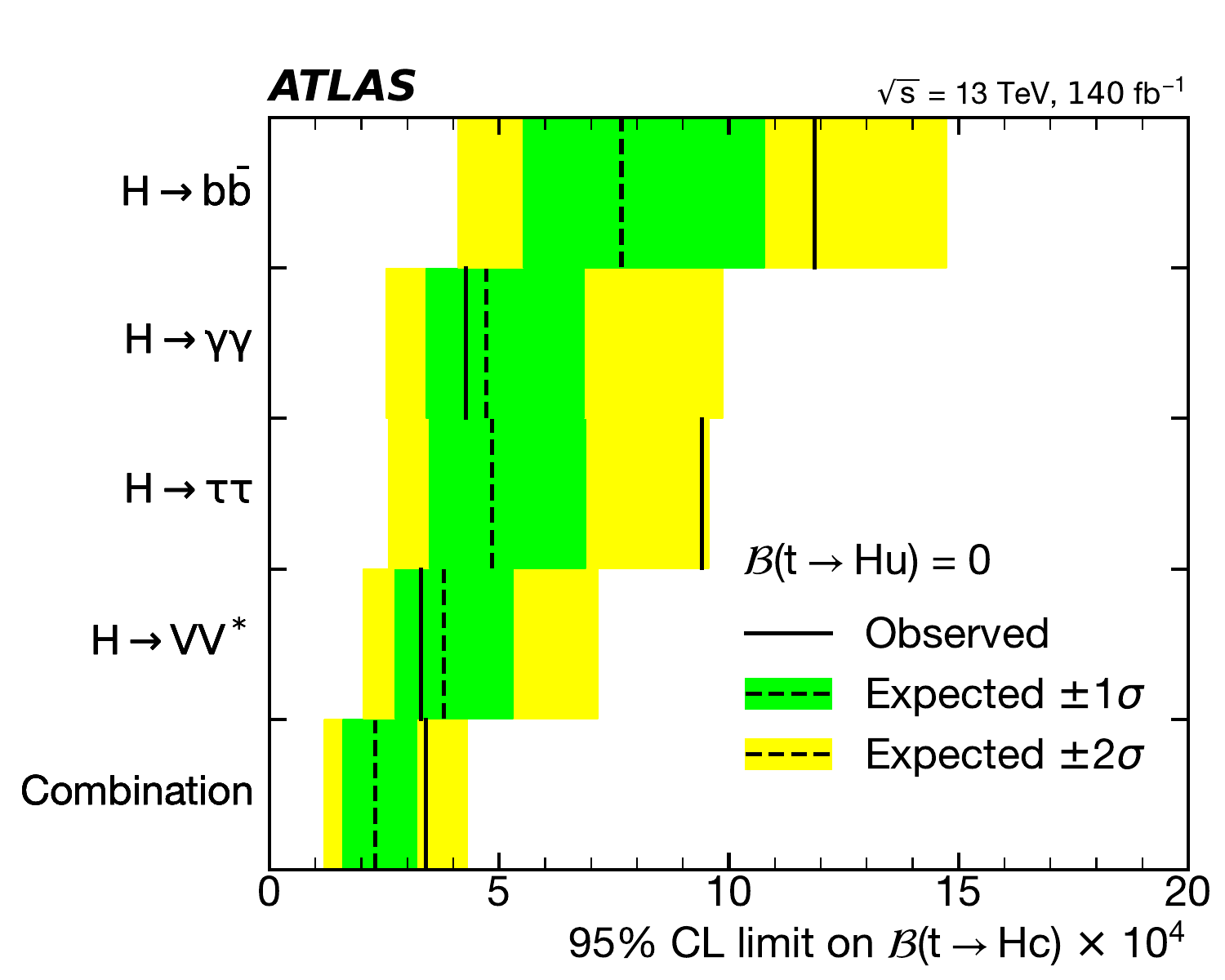} }
\end{center}
\caption{
The \SI{95}{\percent} CL upper limits on
(a) \(\BR(t \to Hu)\) assuming \(\BR(t \to Hc) = 0\) and
(b) \(\BR(t \to Hc)\) assuming \(\BR(t \to Hu) = 0\)
for the individual searches and their combination.
The observed limits (solid lines) are compared with the expected (median)
limits under the background-only hypothesis (dotted lines).
The surrounding shaded bands correspond to the \SI{68}{\percent} and \SI{95}{\percent}
CL intervals around the expected limits, denoted by \(\pm 1\sigma\) and
\(\pm 2 \sigma\), respectively.
The \(H\to VV^*\) also includes events from leptonic \(H\to\tau^+\tau^-\) decays.
}
\label{fig:combLim}
\end{figure}

The observed (expected) \SI{95}{\percent} CL combined upper limits on the branching
ratios are \(\BR(t\to Hu)<2.6\,(1.8)\times10^{-4}\)
and \(\BR(t\to Hc)<3.4\,(2.3)\times 10^{-4}\).
For the \tHc coupling the observed limit is larger than that of the individual
\(H\to VV^*\) analysis, because of the strong upward-fluctuation in the
\(H\to \tautau\) analysis.
A summary of the upper limits on the branching ratios from the individual
searches, and their combination, is given in \Cref{fig:combLim}.
In the EFT framework, the limits observed for the respective branching ratios
translate to a limit on the Wilson coefficients of the \tHu
dimension-6 operators of \(|\Cuphiavgu|< 0.68(0.56)\) at
\SI{95}{\percent} CL, assuming \(\Cuphiavgc = 0\), and for a mass scale $\Lambda =\SI{1}{\TeV}$.
The analogous limit on the Wilson coefficient of the \tHc dimension-6 operator
amounts to \(|\Cuphiavgc|< 0.78(0.64)\) assuming \(\Cuphiavgu = 0\).

% End of text imported from the .//tex/11_combination.tex input file

%

%
%
\FloatBarrier

\section{Conclusion}
%

% The next lines are included from the .//tex/12_conclusion.tex input file
A search is reported for FCNC couplings between the top quark, the
Higgs boson and a second up-type quark in final states containing either two
leptons of the same charge or three leptons.
This search uses \SI{140}{\ifb} of proton--proton collision data collected with the
ATLAS detector at the LHC between 2015 and 2018.
The FCNC processes are considered in \ttbar production where either the top quark
or the top antiquark decays via \(t\to Hq\), and in the production of a
single top quark  with a Higgs boson via the \(q\to tH\) FCNC process.
The results are compatible with the SM and no evidence of FCNC couplings is observed.
Upper limits at \SI{95}{\percent} CL are set on the branching ratio \(\BR(t\to Hq)\).
The observed (expected) upper limits on the branching ratio are
\(\BR(t\to Hu)<2.8\,(3.0)\times10^{-4}\) and \(\BR(t\to Hc)<3.3\,(3.8)\times10^{-4}\).
They are reinterpreted as limits on the average of the Wilson coefficients
for the left-handed and the right-handed dimension-6 operators modelling the
effective \tHq couplings with a new physics scale at \(\Lambda=\SI{1}{\TeV}\).
These limits amount to \(|\Cuphiavgu|<0.71\,(0.73)\) and
\(|\Cuphiavgc|<0.76\,(0.82)\).
These are the most stringent upper limits reported by any analysis searching for \tHq FCNC
couplings with the ATLAS detector.
They are also lower than any other search for these couplings in \dil and \tri final
states.

The results are combined with three other searches for \tHq FCNC couplings with
the ATLAS detector.
The combined observed (expected) limits set by all four analyses, considering
correlations between the individual searches, are
\(\BR(t\to Hu)<2.6\,(1.8)\times10^{-4}\) and \(\BR(t\to Hc)<3.4\,(2.3)\times10^{-4}\)
on the branching ratios, and \(|\Cuphiavgu|< 0.68(0.56)\) and
\(|\Cuphiavgc|< 0.78(0.64)\) on the dimension-6 Wilson coefficients.

% End of text imported from the .//tex/12_conclusion.tex input file

%

%
\section*{Acknowledgements}
%

% The next lines are included from the .//acknowledgements/Acknowledgements.tex input file
%
%

%
%

We thank CERN for the very successful operation of the LHC and its injectors, as well as the support staff at
CERN and at our institutions worldwide without whom ATLAS could not be operated efficiently.

The crucial computing support from all WLCG partners is acknowledged gratefully, in particular from CERN, the ATLAS Tier-1 facilities at TRIUMF/SFU (Canada), NDGF (Denmark, Norway, Sweden), CC-IN2P3 (France), KIT/GridKA (Germany), INFN-CNAF (Italy), NL-T1 (Netherlands), PIC (Spain), RAL (UK) and BNL (USA), the Tier-2 facilities worldwide and large non-WLCG resource providers. Major contributors of computing resources are listed in Ref.~\cite{ATL-SOFT-PUB-2023-001}.

We gratefully acknowledge the support of ANPCyT, Argentina; YerPhI, Armenia; ARC, Australia; BMWFW and FWF, Austria; ANAS, Azerbaijan; CNPq and FAPESP, Brazil; NSERC, NRC and CFI, Canada; CERN; ANID, Chile; CAS, MOST and NSFC, China; Minciencias, Colombia; MEYS CR, Czech Republic; DNRF and DNSRC, Denmark; IN2P3-CNRS and CEA-DRF/IRFU, France; SRNSFG, Georgia; BMBF, HGF and MPG, Germany; GSRI, Greece; RGC and Hong Kong SAR, China; ISF and Benoziyo Center, Israel; INFN, Italy; MEXT and JSPS, Japan; CNRST, Morocco; NWO, Netherlands; RCN, Norway; MEiN, Poland; FCT, Portugal; MNE/IFA, Romania; MESTD, Serbia; MSSR, Slovakia; ARRS and MIZ\v{S}, Slovenia; DSI/NRF, South Africa; MICINN, Spain; SRC and Wallenberg Foundation, Sweden; SERI, SNSF and Cantons of Bern and Geneva, Switzerland; MOST, Taipei; TENMAK, T\"urkiye; STFC, United Kingdom; DOE and NSF, United States of America.

Individual groups and members have received support from BCKDF, CANARIE, CRC and DRAC, Canada; CERN-CZ, PRIMUS 21/SCI/017 and UNCE SCI/013, Czech Republic; COST, ERC, ERDF, Horizon 2020, ICSC-NextGenerationEU and Marie Sk{\l}odowska-Curie Actions, European Union; Investissements d'Avenir Labex, Investissements d'Avenir Idex and ANR, France; DFG and AvH Foundation, Germany; Herakleitos, Thales and Aristeia programmes co-financed by EU-ESF and the Greek NSRF, Greece; BSF-NSF and MINERVA, Israel; Norwegian Financial Mechanism 2014-2021, Norway; NCN and NAWA, Poland; La Caixa Banking Foundation, CERCA Programme Generalitat de Catalunya and PROMETEO and GenT Programmes Generalitat Valenciana, Spain; G\"{o}ran Gustafssons Stiftelse, Sweden; The Royal Society and Leverhulme Trust, United Kingdom.

In addition, individual members wish to acknowledge support from CERN: European Organization for Nuclear Research (CERN PJAS); Chile: Agencia Nacional de Investigaci\'on y Desarrollo (FONDECYT 1190886, FONDECYT 1210400, FONDECYT 1230812, FONDECYT 1230987); China: Chinese Ministry of Science and Technology (MOST-2023YFA1605700), National Natural Science Foundation of China (NSFC - 12175119, NSFC 12275265, NSFC-12075060); Czech Republic: PRIMUS Research Programme (PRIMUS/21/SCI/017); EU: H2020 European Research Council (ERC - 101002463); European Union: European Research Council (ERC - 948254, ERC 101089007), Horizon 2020 Framework Programme (MUCCA - CHIST-ERA-19-XAI-00), European Union, Future Artificial Intelligence Research (FAIR-NextGenerationEU PE00000013), Italian Center for High Performance Computing, Big Data and Quantum Computing (ICSC, NextGenerationEU); France: Agence Nationale de la Recherche (ANR-20-CE31-0013, ANR-21-CE31-0013, ANR-21-CE31-0022), Investissements d'Avenir Labex (ANR-11-LABX-0012); Germany: Baden-Württemberg Stiftung (BW Stiftung-Postdoc Eliteprogramme), Deutsche Forschungsgemeinschaft (DFG - 469666862, DFG - CR 312/5-2); Italy: Istituto Nazionale di Fisica Nucleare (ICSC, NextGenerationEU); Japan: Japan Society for the Promotion of Science (JSPS KAKENHI JP21H05085, JSPS KAKENHI JP22H01227, JSPS KAKENHI JP22H04944, JSPS KAKENHI JP22KK0227); Netherlands: Netherlands Organisation for Scientific Research (NWO Veni 2020 - VI.Veni.202.179); Norway: Research Council of Norway (RCN-314472); Poland: Polish National Agency for Academic Exchange (PPN/PPO/2020/1/00002/U/00001), Polish National Science Centre (NCN 2021/42/E/ST2/00350, NCN OPUS nr 2022/47/B/ST2/03059, NCN UMO-2019/34/E/ST2/00393, UMO-2020/37/B/ST2/01043, UMO-2021/40/C/ST2/00187, UMO-2022/47/O/ST2/00148); Slovenia: Slovenian Research Agency (ARIS grant J1-3010); Spain: BBVA Foundation (LEO22-1-603), Generalitat Valenciana (Artemisa, FEDER, IDIFEDER/2018/048), Ministry of Science and Innovation (MCIN \& NextGenEU PCI2022-135018-2, MICIN \& FEDER PID2021-125273NB, RYC2019-028510-I, RYC2020-030254-I, RYC2021-031273-I, RYC2022-038164-I), PROMETEO and GenT Programmes Generalitat Valenciana (CIDEGENT/2019/023, CIDEGENT/2019/027); Sweden: Swedish Research Council (VR 2018-00482, VR 2022-03845, VR 2022-04683, VR grant 2021-03651), Knut and Alice Wallenberg Foundation (KAW 2017.0100, KAW 2018.0157, KAW 2018.0458, KAW 2019.0447, KAW 2022.0358); Switzerland: Swiss National Science Foundation (SNSF - PCEFP2\_194658); United Kingdom: Leverhulme Trust (Leverhulme Trust RPG-2020-004), Royal Society (NIF-R1-231091); United States of America: U.S. Department of Energy (ECA DE-AC02-76SF00515), Neubauer Family Foundation.

%
%

% End of text imported from the .//acknowledgements/Acknowledgements.tex input file

%
\clearpage
\appendix
\part*{Appendix}
\addcontentsline{toc}{part}{Appendix}

% The next lines are included from the .//tex/Appendix_NN_var_tables.tex input file
\Cref{tab:NNvars_SR3lprod,tab:NNvars_SR3ldec,tab:NNvars_SR2lprod,tab:NNvars_SR2ldec} list the input variables to the NNs used in the four signal regions, approximately ordered by the increase in significance provided by each variable. The exact order differs among the two signal processes.
\begin{table}[H]
\centering
\caption{
List of input variables to the NN in the \SRtriprod, approximately
ordered by the increase in significance provided by each variable.
The exact order differs among the two signal processes.
}\label{tab:NNvars_SR3lprod}
\begin{tabular}{ll}
\toprule
Variable                                              & Description                                                                                                          \\
\midrule
\( m(\ell_{\text{OS}}, \ell_{\text{SS,1}})\)          & \makecell*[{{p{0.75\linewidth}}}]{ Invariant mass of the opposite-charge and the subleading-\pT same-charge lepton }                      \\
\( m(\ell_{\text{OS}}, \ell_{\text{SS,0}})\)          & \makecell*[{{p{0.75\linewidth}}}]{ Invariant mass of the opposite-charge and the leading-\pT same-charge lepton }                         \\
\( m(\ell_t,b_t)\)                                    & \makecell*[{{p{0.75\linewidth}}}]{ Invariant mass of the \(b\)-tagged jet and the lepton assigned to the top-quark decay }                \\
\( \njets\)                                           & \makecell*[{{p{0.75\linewidth}}}]{ Number of jets}                                                                                    \\
\( H_T(\text{jets})\)                                 & \makecell*[{{p{0.75\linewidth}}}]{ Scalar sum of the \pT of all jets }                                                                           \\
\( m(\tSM, H)\)                                       & \makecell*[{{p{0.75\linewidth}}}]{ Invariant mass of the RJR top quark decaying via \(t\to Wb\) and the Higgs boson }                 \\
\( \Delta R(\ell_{\text{SS,0}}, \ell_{\text{SS,1}})\) & \makecell*[{{p{0.75\linewidth}}}]{ Angular separation between the leading and subleading-\pT same-charge lepton}                  \\
\( m(\ell_{H,0}, \ell_{H,1})\)                        & \makecell*[{{p{0.75\linewidth}}}]{ Invariant mass of the two leptons assigned to the Higgs-boson decay }                                  \\
\( m(b\text{-jet}, \ell_{\text{SS,0}})\)              & \makecell*[{{p{0.75\linewidth}}}]{ Invariant mass of the \(b\)-tagged jet and the leading-\pT same-charge lepton }                        \\
\( \Delta R(\ell_t, b_t)\)                            & \makecell*[{{p{0.75\linewidth}}}]{ Angular separation between the \(b\)-tagged jet and the lepton assigned to the top-quark decay }       \\
\( \pT(\tSM)\)                                        & \makecell*[{{p{0.75\linewidth}}}]{ Transverse momentum of the RJR top quark decaying via \(t\to Wb\) }                                    \\
\( \pT(b\text{-jet})\)                                & \makecell*[{{p{0.75\linewidth}}}]{ Transverse momentum of the \(b\)-tagged jet }                                                          \\
\( \eta(\ell_{\text{SS,1}})\)                         & \makecell*[{{p{0.75\linewidth}}}]{ Pseudorapidity of the subleading-\pT same-charge lepton }                                              \\
\( \pT(\ell_{\text{SS,1}})\)                          & \makecell*[{{p{0.75\linewidth}}}]{ Transverse momentum of the subleading-\pT same-charge lepton }                                         \\
\( m(H, \ell_{\text{SS,1}})\)                         & \makecell*[{{p{0.75\linewidth}}}]{ Invariant mass of the RJR Higgs boson and the subleading-\pT same-charge lepton }                      \\
\( \Delta R(\tSM,\ell_{\text{OS}})\)                  & \makecell*[{{p{0.75\linewidth}}}]{ Angular separation between the RJR top quark decaying via \(t\to Wb\) and the opposite-charge lepton } \\
\( \Delta R(H,\ell_{\text{OS}})\)                     & \makecell*[{{p{0.75\linewidth}}}]{ Angular separation between the RJR Higgs boson and the opposite-charge lepton }                        \\
\( \Delta R(\ell_{\text{OS}, \ell_{\text{SS,1}}})\)   & \makecell*[{{p{0.75\linewidth}}}]{ Angular separation between the opposite-charge and the subleading-\pT same-charge lepton }             \\
\bottomrule
\end{tabular}
\end{table}

\begin{table}
\centering
\caption{
List of input variables to the NN in the \SRtridec, approximately
ordered by the increase in significance provided by each variable.
The exact order differs among the two signal processes.
Variables labelled NICE were reconstructed with a fulfilled NICE Reco condition.
}\label{tab:NNvars_SR3ldec}
\begin{tabular}{ll}
\toprule
Variable                                           & Description                                                                                                             \\
\midrule
\(m(\ell_{\text{OS}}, \ell_{\text{SS,1}})\)        & \makecell*[{{p{0.75\linewidth}}}]{ Invariant mass of the opposite-charge and the subleading-\pT same-charge lepton }                         \\
\(m(\ell_{\text{OS}}, \ell_{\text{SS,0}})\)        & \makecell*[{{p{0.75\linewidth}}}]{ Invariant mass of the opposite-charge and the leading-\pT same-charge lepton }                            \\
NICE \(m(\ell_t, b_t)\)                            & \makecell*[{{p{0.75\linewidth}}}]{ Invariant mass of the \(b\)-tagged jet and the lepton assigned to the top-quark decay with a fulfilled NICE Reco condition} \\
\(H_T(\text{jets})\)                               & \makecell*[{{p{0.75\linewidth}}}]{ Scalar sum of the \pT of all jets }                                                                              \\
\(m(b\text{-jet}, \ell_{\text{SS,0}})\)            & \makecell*[{{p{0.75\linewidth}}}]{ Invariant mass of the \(b\)-tagged jet and the leading-\pT same-charge lepton }                           \\
\(m(\tSM, H)\)                                     & \makecell*[{{p{0.75\linewidth}}}]{ Invariant mass of the RJR top quark decaying via \(t\to Wb\) and the RJR Higgs boson }                    \\
\(m(\ell_{H,0}, \ell_{H,1})\)                      & \makecell*[{{p{0.75\linewidth}}}]{ Invariant mass of the two leptons assigned to the Higgs-boson decay }                                     \\
\(m(H,\ell_{\text{SS,1}})\)                        & \makecell*[{{p{0.75\linewidth}}}]{ Invariant mass of the RJR Higgs boson and the subleading-\pT same-charge lepton }                         \\
\(\Delta R(b\text{-jet}, \tSM)\)                   & \makecell*[{{p{0.75\linewidth}}}]{ Angular separation between the \(b\)-tagged jet and the RJR top quark decaying via \(t\to Wb\) }          \\
\(m(\ell_0, \tSM)\)                                & \makecell*[{{p{0.75\linewidth}}}]{ Invariant mass of the leading-\pT lepton and the RJR top quark decaying via \(t\to Wb\) }                 \\
\(\pT(\tSM)\)                                      & \makecell*[{{p{0.75\linewidth}}}]{ Transverse momentum of the RJR top quark decaying via \(t\to Wb\) }                                       \\
\(m(\tSM, \ell_{\text{SS,1}})\)                    & \makecell*[{{p{0.75\linewidth}}}]{ Invariant mass of the RJR top quark decaying via \(t\to Wb\) and the subleading-\pT same-charge lepton }  \\
\(\Delta R(\ell_{\text{OS}}, \ell_{\text{SS,0}})\) & \makecell*[{{p{0.75\linewidth}}}]{ Angular separation between the opposite-charge and the leading-\pT same-charge lepton }                   \\
\(\pT(\ell_{\text{OS}})\)                          & \makecell*[{{p{0.75\linewidth}}}]{ Transverse momentum of the opposite-charge lepton }                                                       \\
\(m(b\text{-jet}, \ell_{\text{OS}})\)              & \makecell*[{{p{0.75\linewidth}}}]{ Invariant mass of the \(b\)-tagged jet and the opposite-charge lepton }                                   \\
\(m(b\text{-jet}, H)\)                             & \makecell*[{{p{0.75\linewidth}}}]{ Invariant mass of the \(b\)-tagged jet and the RJR Higgs boson }                                          \\
\(\pT(\ell_2)\)                                    & \makecell*[{{p{0.75\linewidth}}}]{ Transverse momentum of the third-leading-\pT lepton }                                                     \\
\(\eta(\ell_0)\)                                   & \makecell*[{{p{0.75\linewidth}}}]{ Pseudorapidity of the leading-\pT lepton }                                                                \\
\(m(W_t)\)                                         & \makecell*[{{p{0.75\linewidth}}}]{ Mass of the RJR \(W\) boson from the top-quark decay }                                                    \\
\(m(\ell_t, b_t)\)                                 & \makecell*[{{p{0.75\linewidth}}}]{ Invariant mass of the \(b\)-tagged jet and the lepton assigned to the top-quark decay }                   \\
\bottomrule
\end{tabular}
\end{table}

\begin{table}
\centering
\caption{
List of input variables to the NN in the \SRdilprod, approximately
ordered by the increase in significance provided by each variable.
The exact order differs among the various signal processes.
}\label{tab:NNvars_SR2lprod}
\begin{tabular}{ll}
\toprule
Variable                     & Description                                                                                                                               \\
\midrule
\(m(\ell_1, H) \)            & \makecell*[{{p{0.8\linewidth}}}]{ Invariant mass of the subleading-\pT lepton and the RJR Higgs boson }                                   \\
\(\njets \)                  & \makecell*[{{p{0.8\linewidth}}}]{ Number of jets }                                                                                    \\
\(m(b\text{-jet}, \tSM) \)   & \makecell*[{{p{0.8\linewidth}}}]{ Invariant mass of the \(b\)-tagged jet and the RJR top quark decaying via \(t\to Wb\) }                 \\
\(m(H,b\text{-jet}) \)       & \makecell*[{{p{0.8\linewidth}}}]{ Invariant mass of the RJR Higgs boson and the \(b\)-tagged jet }                                        \\
\(\pT(\whad) \)              & \makecell*[{{p{0.8\linewidth}}}]{ Transverse momentum of the hadronically decaying RJR \(W\) boson }                                      \\
\(\Delta R(\ell_1, H) \)     & \makecell*[{{p{0.8\linewidth}}}]{ Angular separation between the subleading-\pT lepton and the RJR Higgs boson }                          \\
\(m(\whad) \)                & \makecell*[{{p{0.8\linewidth}}}]{ Mass of the hadronically decaying RJR \(W\) boson }                                                     \\
\(\pT(\ell_1) \)             & \makecell*[{{p{0.8\linewidth}}}]{ Transverse momentum of the subleading-\pT lepton }                                                      \\
\(\eta(\ell_1) \)            & \makecell*[{{p{0.8\linewidth}}}]{ Pseudorapidity of the subleading-\pT lepton }                                                           \\
\(\Delta R(H,W_t) \)         & \makecell*[{{p{0.8\linewidth}}}]{ Angular separation between the RJR Higgs boson and the RJR \(W\) boson from the top-quark decay }       \\
\(\Delta R(\ell_0,\ell_1) \) & \makecell*[{{p{0.8\linewidth}}}]{ Angular separation between leading and subleading-\pT lepton }                                  \\
\(m(\ell_1,b\text{-jet}) \)  & \makecell*[{{p{0.8\linewidth}}}]{ Invariant mass of the subleading-\pT lepton and the \(b\)-tagged jet }                                  \\
\(\eta(b\text{-jet}) \)      & \makecell*[{{p{0.8\linewidth}}}]{ Pseudorapidity of the \(b\)-tagged jet }                                                                \\
\(\Delta R(\ell_0,\tSM) \)   & \makecell*[{{p{0.8\linewidth}}}]{ Angular separation between the leading-\pT lepton and the RJR top quark decaying via \(t\to Wb\) }      \\
\(\MET   \)                  & \makecell*[{{p{0.8\linewidth}}}]{ Missing transverse momentum}                                                                            \\
f\/l.\((\ell_0)\)            & \makecell*[{{p{0.8\linewidth}}}]{ Flavour of the leading-\pT lepton}                                                                      \\
\(\eta(\ell_0) \)            & \makecell*[{{p{0.8\linewidth}}}]{ Pseudorapidity of the leading-\pT lepton }                                                              \\
\(\pT(\ell_0) \)             & \makecell*[{{p{0.8\linewidth}}}]{ Transverse momentum of the leading-\pT lepton }                                                         \\
\(\Delta R(\ell_1, \tSM) \)  & \makecell*[{{p{0.8\linewidth}}}]{ Angular separation between the subleading-\pT lepton and the RJR top quark decaying via \(t\to Wb\) }   \\
\(m(H,W_t) \)                & \makecell*[{{p{0.8\linewidth}}}]{ Invariant mass of the RJR Higgs boson and the RJR \(W\) boson from the top-quark decay }                \\
\(\Delta R(\ell_1,W_t) \)    & \makecell*[{{p{0.8\linewidth}}}]{ Angular separation between the subleading-\pT lepton and the RJR \(W\) boson from the top-quark decay } \\
\(m(\ell_0,H) \)             & \makecell*[{{p{0.8\linewidth}}}]{ Invariant mass of the leading-\pT lepton and the RJR Higgs boson }                                      \\
\(\pT(b\text{-jet}) \)       & \makecell*[{{p{0.8\linewidth}}}]{ Transverse momentum of the \(b\)-tagged jet }                                                           \\
\bottomrule
\end{tabular}
\end{table}

\begin{table}
\centering
\caption{
List of input variables to the NN in the \SRdildec, approximately
ordered by the increase in significance provided by each variable.
The exact order differs among the various signal processes.
}\label{tab:NNvars_SR2ldec}
\begin{tabular}{ll}
\toprule
Variable                            & Description                                                                                                                   \\
\midrule
\(H_T(\text{jets})\)                & \makecell*[{{p{0.8\linewidth}}}]{ Scalar sum of the \pT of all jets }                                                                                    \\
\(m(\ell_0,b\text{-jet})\)          & \makecell*[{{p{0.8\linewidth}}}]{ Invariant mass of the leading-\pT lepton and the \(b\)-tagged jet }                                             \\
\(\Delta R(\ell_1, H)\)             & \makecell*[{{p{0.8\linewidth}}}]{ Angular separation between the subleading-\pT lepton and the RJR Higgs boson }                                  \\
\(\pT(\ell_1)\)                     & \makecell*[{{p{0.8\linewidth}}}]{ Transverse momentum of the subleading-\pT lepton }                                                              \\
\(\mclosejets\)                     & \makecell*[{{p{0.8\linewidth}}}]{ Invariant mass of the two non-\(b\)-tagged jets with the smallest \(\Delta R\)}                                 \\
\(m(\tSM,l\text{-jet}_0)\)          & \makecell*[{{p{0.8\linewidth}}}]{ Invariant mass of the RJR top quark decaying via \(t\to Wb\) and the leading-\pT non-\(b\)-tagged jet }         \\
\(\eta(\ell_1)\)                    & \makecell*[{{p{0.8\linewidth}}}]{ Pseudorapidity of the subleading-\pT lepton }                                                                   \\
\(\Delta R(\ell_0,l\text{-jet}_1)\) & \makecell*[{{p{0.8\linewidth}}}]{ Angular separation between the leading-\pT lepton and the subleading-\pT non-\(b\)-tagged jet }                 \\
\(m(\ell_1,l\text{-jet}_0)\)        & \makecell*[{{p{0.8\linewidth}}}]{ Invariant mass of the subleading-\pT lepton and the leading-\pT non-\(b\)-tagged jet }                          \\
\(m(\ell_0,l\text{-jet}_0)\)        & \makecell*[{{p{0.8\linewidth}}}]{ Invariant mass of the leading-\pT lepton and the leading-\pT non-\(b\)-tagged jet }                             \\
\(\Delta R(\ell_0,l\text{-jet}_2)\) & \makecell*[{{p{0.8\linewidth}}}]{ Angular separation between the leading-\pT lepton and the third-leading-\pT non-\(b\)-tagged jet }              \\
\(\Delta R(\ell_1,l\text{-jet}_2)\) & \makecell*[{{p{0.8\linewidth}}}]{ Angular separation between the subleading-\pT lepton and the third-leading-\pT non-\(b\)-tagged jet }           \\
\(m(\tFCNC,l\text{-jet}_0)\)        & \makecell*[{{p{0.8\linewidth}}}]{ Invariant mass of the RJR top quark decaying via \(t\to Hq\)  and the leading-\pT non-\(b\)-tagged jet }        \\
\(m(\ell_1,l\text{-jet}_1)\)        & \makecell*[{{p{0.8\linewidth}}}]{ Invariant mass of the subleading-\pT lepton and the subleading-\pT non-\(b\)-tagged jet }                       \\
\(m(\ell_1,\tFCNC)\)                & \makecell*[{{p{0.8\linewidth}}}]{ Invariant mass of the subleading-\pT lepton and the RJR top quark decaying via \(t\to Hq\) }                    \\
\(m(W_t,\whad)\)                    & \makecell*[{{p{0.8\linewidth}}}]{ Invariant mass of the RJR \(W\) boson from the top-quark decay and the hadronically decaying RJR \(W\) boson }  \\
\(\Delta R(\ell_0,l\text{-jet}_0)\) & \makecell*[{{p{0.8\linewidth}}}]{ Angular separation between the leading-\pT lepton and the leading-\pT non-\(b\)-tagged jet }                    \\
\(m(\ell_1,b\text{-jet})\)          & \makecell*[{{p{0.8\linewidth}}}]{ Invariant mass of the subleading-\pT lepton and the \(b\)-tagged jet }                                          \\
\(\njets\)                          & \makecell*[{{p{0.8\linewidth}}}]{ Number of jets }                                                                                            \\
\(m(H,b\text{-jet})\)               & \makecell*[{{p{0.8\linewidth}}}]{ Invariant mass of the RJR Higgs boson and the \(b\)-tagged jet }                                                \\
\(H_T(\ell_0,\ell_1)\)              & \makecell*[{{p{0.8\linewidth}}}]{ Scalar sum of the \pT of all leptons }                                                                                 \\
\(\pT(\ell_0)\)                     & \makecell*[{{p{0.8\linewidth}}}]{ Transverse momentum of the leading-\pT lepton }                                                                 \\
\(m(W_t,\tFCNC)\)                   & \makecell*[{{p{0.8\linewidth}}}]{ Invariant mass of the RJR \(W\) boson from the top-quark decay and the RJR top quark decaying via \(t\to Hq\) } \\

\bottomrule
\end{tabular}
\end{table}

% End of text imported from the .//tex/Appendix_NN_var_tables.tex input file

%

\FloatBarrier
\printbibliography

@Article{LUCID-2,
author = {G. Avoni and M. Bruschi and G. Cabras and D. Caforio and N. Dehghanian and A. Floderus and B. Giacobbe and F. Giannuzzi and F. Giorgi and P. Grafström and V. Hedberg and F. Lasagni Manghi and S. Meneghini and J. Pinfold and E. Richards and C. Sbarra and N. Semprini Cesari and A. Sbrizzi and R. Soluk and G. Ucchielli and S. Valentinetti and O. Viazlo and M. Villa and C. Vittori and R. Vuillermet and A. Zoccoli},
title = "{The new LUCID-2 detector for luminosity measurement and monitoring in ATLAS}",
journal = "JINST",
volume = "13",
year = "2018",
pages = "P07017",
doi ="10.1088/1748-0221/13/07/p07017"
}

@article{Gutschow:2018tuk,
    author = {G\"utschow, Christian and Lindert, Jonas M. and Sch\"onherr, Marek},
    title = "{Multi-jet merged top-pair production including electroweak corrections}",
    eprint = "1803.00950",
    archivePrefix = "arXiv",
    primaryClass = "hep-ph",
    reportNumber = "CERN-TH-18-039, CERN-TH-2018-039, IPPP/18/13, MCnet-18-05, IPPP-18-13, MCNET-18-05",
    doi = "10.1140/epjc/s10052-018-5804-2",
    journal = "Eur. Phys. J. C",
    volume = "78",
    number = "4",
    pages = "317",
    year = "2018"
}

@article{Frederix:2021agh,
    author = "Frederix, Rikkert and Tsinikos, Ioannis",
    title = "{On improving NLO merging for $ \mathrm{t}\overline{\mathrm{t}}\mathrm{W} $ production}",
    eprint = "2108.07826",
    archivePrefix = "arXiv",
    primaryClass = "hep-ph",
    reportNumber = "LU-TP 21-34",
    doi = "10.1007/JHEP11(2021)029",
    journal = "JHEP",
    volume = "11",
    pages = "029",
    year = "2021"
}

@Article{Feindt:2006pm,
     author    = "Feindt, M. and Kerzel, U.",
     title     = "{The NeuroBayes neural network package}",
     journal   = "Nucl.\ Instrum.\ Meth. A",
     volume    = "559",
     year      = "2006",
     pages     = "190-194",
     doi       = "10.1016/j.nima.2005.11.166",
     SLACcitation  = "%%CITATION = NUIMA,A559,190;%%"
}

@booklet{feindt-2004,
     author    = "Michael Feindt",
     title     = "{A Neural Bayesian Estimator for Conditional Probability Densities}",
     year      = "2004",
     eprint    = "physics/0402093",
     primaryClass = "physics.data-an",
     archivePrefix = "arXiv",
     SLACcitation  = "%%CITATION = 1007.3178;%%"
}

@article{RJR,
 	author = "Jackson, Paul and Rogan, Christopher",
 	title = "{Recursive Jigsaw Reconstruction: HEP event analysis in the presence of kinematic and combinatoric ambiguities}",
 	eprint = "1705.10733",
 	archivePrefix = "arXiv",
 	primaryClass = "hep-ph",
 	doi = "10.1103/PhysRevD.96.112007",
 	journal = "Phys. Rev. D",
 	volume = "96",
 	number = "11",
 	pages = "112007",
 	year = "2017"
 }

@article{ALLOUL20142250,
title = {FeynRules  2.0 — A complete toolbox for tree-level phenomeno\-logy},
journal = {Comput. Phys. Commun.},
volume = {185},
number = {8},
pages = {2250-2300},
year = {2014},
	eprint = "1310.1921",
	archivePrefix = "arXiv",
	primaryClass = "hep-ph",
doi = {https://doi.org/10.1016/j.cpc.2014.04.012},
%url = {https://www.sciencedirect.com/science/article/pii/S0010465514001350},
author = {Adam Alloul and Neil D. Christensen and Céline Degrande and Claude Duhr and Benjamin Fuks},
keywords = {Model building, Feynman rules, Monte Carlo programs}
}

@article{Hoche:2021mkv,
	author = {H\"oche, Stefan and Mrenna, Stephen and Payne, Shay and Preuss, Christian Tobias and Skands, Peter},
	title = "{A Study of QCD Radiation in VBF Higgs Production with Vincia and Pythia}",
	eprint = "2106.10987",
	archivePrefix = "arXiv",
	primaryClass = "hep-ph",
	reportNumber = "FERMILAB-PUB-21-289-SCD-T, MCNET-21-11",
	doi = "10.21468/SciPostPhys.12.1.010",
	journal = "SciPost Phys.",
	volume = "12",
	number = "1",
	pages = "010",
	year = "2022"
}

@article{GIM_mechanism,
  title = {Weak Interactions with Lepton-Hadron Symmetry},
  author = {Glashow, S. L. and Iliopoulos, J. and Maiani, L.},
  journal = {Phys. Rev. D},
  volume = {2},
  number = {7},
  pages = {1285--1292},
  numpages = {0},
  year = {1970},
  month = {Oct},
  publisher = {American Physical Society},
  doi = {10.1103/PhysRevD.2.1285},
  %url = {https://link.aps.org/doi/10.1103/PhysRevD.2.1285}
}

@article{Eilam:1990zc,
      author         = "Eilam, G. and Hewett, J. L. and Soni, A.",
      title          = "{Rare decays of the top quark in the standard and two-Higgs-doublet models}",
      journal        = "Phys. Rev. D",
      volume         = "44",
      year           = "1991",
      pages          = "1473",
      doi            = "10.1103/PhysRevD.44.1473",
      note           = "Erratum: \href{http://dx.doi.org/10.1103/PhysRevD.59.039901}{Phys. Rev. D \textbf{59} (1998) 039901}",
      reportNumber   = "MAD-PH-596, TECHNION-PH-90-32, BNL-45778",
      SLACcitation   = "%%CITATION = PHRVA,D44,1473;%%"
}

@article{Mele:1998ag,
      author         = "Mele, B. and Petrarca, S. and Soddu, A.",
      title          = "{A new evaluation of the $t \to cH$ decay width in the
                        standard model}",
      journal        = "Phys. Lett. B",
      volume         = "435",
      year           = "1998",
      pages          = "401",
      doi            = "10.1016/S0370-2693(98)00822-3",
      eprint         = "hep-ph/9805498",
      archivePrefix  = "arXiv",
      primaryClass   = "hep-ph",
      reportNumber   = "ROME1-1207-98",
      SLACcitation   = "%%CITATION = HEP-PH/9805498;%%"
}

@article{AguilarSaavedra:2004wm,
      author         = "Aguilar-Saavedra, J. A.",
      title          = "{Top flavor-changing neutral interactions: theoretical
                        expectations and experimental detection}",
      journal        = "Acta  Phys. Polon. B",
      volume         = "35",
      pages          = "2695",
      year           = "2004",
      eprint         = "hep-ph/0409342",
      archivePrefix  = "arXiv",
      primaryClass   = "hep-ph",
      SLACcitation   = "%%CITATION = HEP-PH/0409342;%%",
      doi            = "10.48550/arXiv.hep-ph/0409342"

}

@article{Zhang:2013xya,
      author         = "Zhang, Cen and Maltoni, Fabio",
      title          = "{Top-quark decay into Higgs boson and a light quark at
                        next-to-leading order in QCD}",
      journal        = "Phys. Rev. D",
      volume         = "88",
      pages          = "054005",
      doi            = "10.1103/PhysRevD.88.054005",
      year           = "2013",
      eprint         = "1305.7386",
      archivePrefix  = "arXiv",
      primaryClass   = "hep-ph",
      reportNumber   = "CP3-13-26",
      SLACcitation   = "%%CITATION = ARXIV:1305.7386;%%",
}

@article{AguilarSaavedra:2002kr,
      author         = "Aguilar-Saavedra, J. A.",
      title          = "{Effects of mixing with quark singlets}",
      journal        = "Phys. Rev. D",
      volume         = "67",
      year           = "2003",
      pages          = "035003",
      doi            = "10.1103/PhysRevD.67.035003",
      note           = "Erratum: \href{http://dx.doi.org/10.1103/PhysRevD.69.099901}{Phys. Rev. D \textbf{69} (2004) 099901}",
      eprint         = "hep-ph/0210112",
      archivePrefix  = "arXiv",
      primaryClass   = "hep-ph",
      reportNumber   = "FISIST-16-2002-CFIF",
      SLACcitation   = "%%CITATION = HEP-PH/0210112;%%"
}

@article{Branco:2hdm2012,
      author         = "Branco, G. C. and Ferreira, P. M. and Lavoura, L. and Rebelo, M. N. and Sher, Marc and Silva, Joao P.",
      title          = "{Theory and phenomenology of two-Higgs-doublet models}",
      journal        = "Phys. Rep.",
      volume         = "516",
      year           = "2012",
      pages          = "1",
      doi            = "10.1016/j.physrep.2012.02.002",
      eprint         = "1106.0034",
      archivePrefix  = "arXiv",
      primaryClass   = "hep-ph",
      SLACcitation   = "%%CITATION = ARXIV:1106.0034;%%"
}

@article{Bejar:2000ub,
      author         = "B\'{e}jar, Santi and Guasch, Jaume and Sol{\`a}, Joan",
      title          = "{Loop induced flavor changing neutral decays of the top
                        quark in a general two-Higgs-doublet model}",
      journal        = "Nucl. Phys. B",
      volume         = "600",
      year           = "2001",
      pages          = "21",
      doi            = "10.1016/S0550-3213(01)00044-X",
      eprint         = "hep-ph/0011091",
      archivePrefix  = "arXiv",
      primaryClass   = "hep-ph",
      reportNumber   = "UAB-FT-491, KA-TP-22-2000",
      SLACcitation   = "%%CITATION = HEP-PH/0011091;%%"
}

@article{Guasch:1999jp,
      author         = "Guasch, Jaume and Sol{\`a}, Joan",
      title          = "{FCNC top quark decays in the MSSM: a door to SUSY physics in high
                        luminosity colliders?}",
      journal        = "Nucl. Phys. B",
      volume         = "562",
      year           = "1999",
      pages          = "3",
      doi            = "10.1016/S0550-3213(99)00579-9",
      eprint         = "hep-ph/9906268",
      archivePrefix  = "arXiv",
      primaryClass   = "hep-ph",
      reportNumber   = "UAB-FT-429, KA-TP-6-99",
      SLACcitation   = "%%CITATION = HEP-PH/9906268;%%"
}

@article{Cao:2007dk,
      author         = "Cao, J. J. and Eilam, G. and Frank, M. and Hikasa, K. and
                        Liu, G. L. and Turan, I. and Yang, J. M.",
      title          = "{Supersymmetry-induced flavor-changing neutral-current top-quark processes at the CERN Large Hadron Collider}",
      journal        = "Phys. Rev. D",
      volume         = "75",
      year           = "2007",
      pages          = "075021",
      doi            = "10.1103/PhysRevD.75.075021",
      eprint         = "hep-ph/0702264",
      archivePrefix  = "arXiv",
      primaryClass   = "hep-ph",
      reportNumber   = "CUMQ-HEP-144",
      SLACcitation   = "%%CITATION = HEP-PH/0702264;%%"
}

@article{Cao:2014udj,
      author         = "Cao, Junjie and Han, Chengcheng and Wu, Lei and Yang, Jin
                        Min and Zhang, Mengchao",
      title          = "{SUSY induced top quark FCNC decay $t \rightarrow { ch}$
                        after Run I of LHC}",
      journal        = "Eur. Phys. J. C",
      volume         = "74",
      year           = "2014",
      pages          = "3058",
      doi            = "10.1140/epjc/s10052-014-3058-1",
      eprint         = "1404.1241",
      archivePrefix  = "arXiv",
      primaryClass   = "hep-ph",
      SLACcitation   = "%%CITATION = ARXIV:1404.1241;%%"
}

@article{Eilam:2001dh,
      author         = "Eilam, G. and Gemintern, A. and Han, Tao and Yang, J. M.
                        and Zhang, X.",
      title          = "{Top-quark rare decay $t \to ch$ in R-parity-violating
                        SUSY}",
      journal        = "Phys. Lett. B",
      volume         = "510",
      year           = "2001",
      pages          = "227",
      doi            = "10.1016/S0370-2693(01)00598-6",
      eprint         = "hep-ph/0102037",
      archivePrefix  = "arXiv",
      primaryClass   = "hep-ph",
      reportNumber   = "MADPH-01-1212",
      SLACcitation   = "%%CITATION = HEP-PH/0102037;%%"
}

@article{Azatov:2009na,
      author         = "Azatov, Aleksandr and Toharia, Manuel and Zhu, Lijun",
      title          = "{Higgs mediated flavor changing neutral currents in warped extra dimensions}",
      journal        = "Phys. Rev. D",
      volume         = "80",
      year           = "2009",
      pages          = "035016",
      doi            = "10.1103/PhysRevD.80.035016",
      eprint         = "0906.1990",
      archivePrefix  = "arXiv",
      primaryClass   = "hep-ph",
      reportNumber   = "UMD-PP-09-039",
      SLACcitation   = "%%CITATION = ARXIV:0906.1990;%%"
}

@article{Cheng:1987rs,
      author         = "Cheng, T. P. and Sher, Marc",
      title          = "{Mass-matrix ansatz and flavor nonconservation in models
                        with multiple Higgs doublets}",
      journal        = "Phys. Rev. D",
      volume         = "35",
      year           = "1987",
      pages          = "3484",
      doi            = "10.1103/PhysRevD.35.3484",
      reportNumber   = "WU-TH-87-1",
      SLACcitation   = "%%CITATION = PHRVA,D35,3484;%%"
}

@article{Baum:2008qm,
      author         = "Baum, Itzhak and Eilam, Gad and Bar-Shalom, Shaouly",
      title          = "{Scalar flavor changing neutral currents and rare top
                        quark decays in a two Higgs doublet model "for the top
                        quark"}",
      journal        = "Phys. Rev. D",
      volume         = "77",
      year           = "2008",
      pages          = "113008",
      doi            = "10.1103/PhysRevD.77.113008",
      eprint         = "0802.2622",
      archivePrefix  = "arXiv",
      primaryClass   = "hep-ph",
      reportNumber   = "UCI-TR-2007-51",
      SLACcitation   = "%%CITATION = ARXIV:0802.2622;%%"
}

@article{Chen:2013qta,
      author         = "Chen, Kai-Feng and Hou, Wei-Shu and Kao, Chung and Kohda,
                        Masaya",
      title          = "{When the Higgs meets the top: search for $t \to c h^0$ at
                        the LHC}",
      journal        = "Phys. Lett. B",
      volume         = "725",
      year           = "2013",
      pages          = "378",
      doi            = "10.1016/j.physletb.2013.07.060",
      eprint         = "1304.8037",
      archivePrefix  = "arXiv",
      primaryClass   = "hep-ph",
      SLACcitation   = "%%CITATION = ARXIV:1304.8037;%%"
}

@article{Chiang:2015cba,
      author         = "Chiang, Cheng-Wei and Fukuda, Hajime and Takeuchi,
                        Michihisa and Yanagida, Tsutomu T.",
      title          = "{Flavor-changing neutral-current decays in top-specific variant axion model}",
      journal        = "JHEP",
      volume         = "11",
      year           = "2015",
      pages          = "057",
      doi            = "10.1007/JHEP11(2015)057",
      eprint         = "1507.04354",
      archivePrefix  = "arXiv",
      primaryClass   = "hep-ph",
      reportNumber   = "IPMU-15-0111",
      SLACcitation   = "%%CITATION = ARXIV:1507.04354;%%"
}

@article{Crivellin:2015hha,
      author         = "Crivellin, Andreas and Heeck, Julian and Stoffer, Peter",
      title          = "{Perturbed Lepton-Specific Two-Higgs-Doublet Model Facing Experimental Hints for Physics beyond the Standard Model}",
      journal        = "Phys. Rev. Lett.",
      volume         = "116",
      year           = "2016",
      pages          = "081801",
      doi            = "10.1103/PhysRevLett.116.081801",
      eprint         = "1507.07567",
      archivePrefix  = "arXiv",
      primaryClass   = "hep-ph",
      reportNumber   = "CERN-PH-TH-2015-168, ULB-15-13",
      SLACcitation   = "%%CITATION = ARXIV:1507.07567;%%"
}

@article{Botella:2015hoa,
      author         = "Botella, F. J. and Branco, G. C. and Nebot, M. and
                        Rebelo, M. N.",
      title          = "{Flavour-changing Higgs couplings in a class of two Higgs doublet models}",
      journal        = "Eur. Phys. J. C",
      volume         = "76",
      year           = "2016",
      pages          = "161",
      doi            = "10.1140/epjc/s10052-016-3993-0",
      eprint         = "1508.05101",
      archivePrefix  = "arXiv",
      primaryClass   = "hep-ph",
      reportNumber   = "IFIC-15-62",
      SLACcitation   = "%%CITATION = ARXIV:1508.05101;%%"
}

@article{Gori:2017tvg,
      author         = "Gori, Stefania and Grojean, Christophe and Juste, Aurelio
                        and Paul, Ayan",
      title          = "{Heavy Higgs searches: flavour matters}",
      journal        = "JHEP",
      volume         = "01",
      year           = "2018",
      pages          = "108",
      doi            = "10.1007/JHEP01(2018)108",
      eprint         = "1710.03752",
      archivePrefix  = "arXiv",
      primaryClass   = "hep-ph",
      reportNumber   = "DESY-17-103",
      SLACcitation   = "%%CITATION = ARXIV:1710.03752;%%"
}

@article{Chiang:2017fjr,
      author         = "Chiang, Cheng-Wei and Fukuda, Hajime and Takeuchi,
                        Michihisa and Yanagida, Tsutomu T.",
      title          = "{Current status of top-specific variant axion model}",
      journal        = "Phys. Rev. D",
      volume         = "97",
      year           = "2018",
      pages          = "035015",
      doi            = "10.1103/PhysRevD.97.035015",
      eprint         = "1711.02993",
      archivePrefix  = "arXiv",
      primaryClass   = "hep-ph",
      reportNumber   = "IPMU-17-0147",
      SLACcitation   = "%%CITATION = ARXIV:1711.02993;%%"
}

@article{Greljo:2014dka,
      author         = "Greljo, Admir and Kamenik, Jernej F. and Kopp, Joachim",
      title          = "{Disentangling flavor violation in the top-Higgs sector
                        at the LHC}",
      journal        = "JHEP",
      volume         = "07",
      year           = "2014",
      pages          = "046",
      doi            = "10.1007/JHEP07(2014)046",
      eprint         = "1404.1278",
      archivePrefix  = "arXiv",
      primaryClass   = "hep-ph",
      SLACcitation   = "%%CITATION = ARXIV:1404.1278;%%"
}

@Article{fcnc_production_theory,
    author         = "{C. Degrande, F. Maltoni, J. Wang and C. Zhang}",
    title          = "{Automatic computations at next-to-leading order in QCD for top-quark flavor-changing neutral processes}",
    journal        = "Phys. Rev. D",
    volume         = "91",
    year           = "2015",
    pages          = "034024",
    doi            = "10.1103/PhysRevD.91.034024",
    eprint         = "1412.5594",
    year           = "2015",
    archivePrefix  = "arXiv",
    primaryClass   = "hep-ex",
}

@article{Zboson_mass,
  title = {Review of Particle Physics},
  author = {{Particle Data Group} and Tanabashi, M. and others},
  %author = {Tanabashi, M. and Hagiwara, K. and Hikasa, K. and Nakamura, K. and Sumino, Y. and Takahashi, F. and Tanaka, J. and Agashe, K. and Aielli, G. and Amsler, C. and Antonelli, M. and Asner, D. M. and Baer, H. and Banerjee, Sw. and Barnett, R. M. and Basaglia, T. and Bauer, C. W. and Beatty, J. J. and Belousov, V. I. and Beringer, J. and Bethke, S. and Bettini, A. and Bichsel, H. and Biebel, O. and Black, K. M. and Blucher, E. and Buchmuller, O. and Burkert, V. and Bychkov, M. A. and Cahn, R. N. and Carena, M. and Ceccucci, A. and Cerri, A. and Chakraborty, D. and Chen, M.-C. and Chivukula, R. S. and Cowan, G. and Dahl, O. and D'Ambrosio, G. and Damour, T. and de Florian, D. and de Gouv\^ea, A. and DeGrand, T. and de Jong, P. and Dissertori, G. and Dobrescu, B. A. and D'Onofrio, M. and Doser, M. and Drees, M. and Dreiner, H. K. and Dwyer, D. A. and Eerola, P. and Eidelman, S. and Ellis, J. and Erler, J. and Ezhela, V. V. and Fetscher, W. and Fields, B. D. and Firestone, R. and Foster, B. and Freitas, A. and Gallagher, H. and Garren, L. and Gerber, H.-J. and Gerbier, G. and Gershon, T. and Gershtein, Y. and Gherghetta, T. and Godizov, A. A. and Goodman, M. and Grab, C. and Gritsan, A. V. and Grojean, C. and Groom, D. E. and Gr\"unewald, M. and Gurtu, A. and Gutsche, T. and Haber, H. E. and Hanhart, C. and Hashimoto, S. and Hayato, Y. and Hayes, K. G. and Hebecker, A. and Heinemeyer, S. and Heltsley, B. and Hern\'andez-Rey, J. J. and Hisano, J. and H\"ocker, A. and Holder, J. and Holtkamp, A. and Hyodo, T. and Irwin, K. D. and Johnson, K. F. and Kado, M. and Karliner, M. and Katz, U. F. and Klein, S. R. and Klempt, E. and Kowalewski, R. V. and Krauss, F. and Kreps, M. and Krusche, B. and Kuyanov, Yu. V. and Kwon, Y. and Lahav, O. and Laiho, J. and Lesgourgues, J. and Liddle, A. and Ligeti, Z. and Lin, C.-J. and Lippmann, C. and Liss, T. M. and Littenberg, L. and Lugovsky, K. S. and Lugovsky, S. B. and Lusiani, A. and Makida, Y. and Maltoni, F. and Mannel, T. and Manohar, A. V. and Marciano, W. J. and Martin, A. D. and Masoni, A. and Matthews, J. and Mei\ss{}ner, U.-G. and Milstead, D. and Mitchell, R. E. and M\"onig, K. and Molaro, P. and Moortgat, F. and Moskovic, M. and Murayama, H. and Narain, M. and Nason, P. and Navas, S. and Neubert, M. and Nevski, P. and Nir, Y. and Olive, K. A. and Pagan Griso, S. and Parsons, J. and Patrignani, C. and Peacock, J. A. and Pennington, M. and Petcov, S. T. and Petrov, V. A. and Pianori, E. and Piepke, A. and Pomarol, A. and Quadt, A. and Rademacker, J. and Raffelt, G. and Ratcliff, B. N. and Richardson, P. and Ringwald, A. and Roesler, S. and Rolli, S. and Romaniouk, A. and Rosenberg, L. J. and Rosner, J. L. and Rybka, G. and Ryutin, R. A. and Sachrajda, C. T. and Sakai, Y. and Salam, G. P. and Sarkar, S. and Sauli, F. and Schneider, O. and Scholberg, K. and Schwartz, A. J. and Scott, D. and Sharma, V. and Sharpe, S. R. and Shutt, T. and Silari, M. and Sj\"ostrand, T. and Skands, P. and Skwarnicki, T. and Smith, J. G. and Smoot, G. F. and Spanier, S. and Spieler, H. and Spiering, C. and Stahl, A. and Stone, S. L. and Sumiyoshi, T. and Syphers, M. J. and Terashi, K. and Terning, J. and Thoma, U. and Thorne, R. S. and Tiator, L. and Titov, M. and Tkachenko, N. P. and T\"ornqvist, N. A. and Tovey, D. R. and Valencia, G. and Van de Water, R. and Varelas, N. and Venanzoni, G. and Verde, L. and Vincter, M. G. and Vogel, P. and Vogt, A. and Wakely, S. P. and Walkowiak, W. and Walter, C. W. and Wands, D. and Ward, D. R. and Wascko, M. O. and Weiglein, G. and Weinberg, D. H. and Weinberg, E. J. and White, M. and Wiencke, L. R. and Willocq, S. and Wohl, C. G. and Womersley, J. and Woody, C. L. and Workman, R. L. and Yao, W.-M. and Zeller, G. P. and Zenin, O. V. and Zhu, R.-Y. and Zhu, S.-L. and Zimmermann, F. and Zyla, P. A. and Anderson, J. and Fuller, L. and Lugovsky, V. S. and Schaffner, P.},
  collaboration = {Particle Data Group},
  journal = {Phys. Rev. D},
  volume = {98},
  number = {3},
  pages = {030001},
  numpages = {1898},
  year = {2018},
  month = {Aug},
  publisher = {American Physical Society},
  doi = {10.1103/PhysRevD.98.030001},
  %url = {https://link.aps.org/doi/10.1103/PhysRevD.98.030001}
}

@article{FCNC_EFT,
  title = {Global approach to top-quark flavor-changing interactions},
  author = {Durieux, Gauthier and Maltoni, Fabio and Zhang, Cen},
  journal = {Phys. Rev. D},
  volume = {91},
  pages = {074017},
  numpages = {16},
  year = {2015},
  month = {Apr},
  publisher = {American Physical Society},
  doi = {10.1103/PhysRevD.91.074017},
  eprint={1412.7166},
  archivePrefix={arXiv},
  primaryClass={hep-ph},
}

@Article{HDBS-2020-16,
    author         = "{ATLAS Collaboration}",
    title          = "{Search for a new scalar resonance in flavour-changing neutral-current top-quark decays \(t \to qX\ (q=u,c\)), with \(X\to b\bar{b}\), in proton--proton collisions at \(\sqrt{s} = 13\,\text{TeV}\) with the ATLAS detector}",
    journal        = "JHEP",
    volume         = "07",
    year           = "2023",
    pages          = "199",
    doi            = "10.1007/JHEP07(2023)199",
    reportNumber   = "CERN-EP-2022-176",
    eprint         = "2301.03902",
    archivePrefix  = "arXiv",
    primaryClass   = "hep-ex",
}

@Article{TOPQ-2019-04_alt,
      author="{ATLAS Collaboration}",
      title={Search for flavour-changing neutral $tqH$ interactions with $H\rightarrow \gamma\gamma$ in $pp$ collisions at $\sqrt{s}$ = 13 TeV using the ATLAS detector}, 
      journal        = "JHEP",
      volume         = "12",
      year           = "2023",
      pages          = "195",
      doi            = "10.1007/jhep12(2023)195",
      reportNumber   = "CERN-EP-2023-204",
      eprint         = "2309.12817",
      archivePrefix  = "arXiv",
      primaryClass   = "hep-ex"
}

@incollection{Feed_forward_NN,
  address = {Cambridge, MA},
  author = {Rumelhart, David E. and Hinton, Geoffrey E. and Williams, Ronald J.},
  booktitle = {Parallel Distributed Processing: Explorations in the Microstructure of Cognition, {V}olume 1: {F}oundations},
  description = {idsia},
  editor = {Rumelhart, David E. and Mcclelland, James L.},
  pages = {318--362},
  publisher = {MIT Press},
  title = {Learning Internal Representations by Error Propagation},
  year = 1986
}

@article{Kidonakis:2015nna,
    author = "Kidonakis, Nikolaos",
    title = "{Theoretical results for electroweak boson and single-top production}",
    eprint = "1506.04072",
    archivePrefix = "arXiv",
    primaryClass = "hep-ph",
    doi = "10.22323/1.247.0170",
    journal = "PoS",
    volume = "DIS2015",
    pages = "170",
    year = 2016
}

@misc{ATLAS-TOPQ-2019-30,
      title={Measurement of the total and differential cross-sections of $t\bar{t}W$ production in $pp$ collisions at $\sqrt{s}=13$ TeV with the ATLAS detector},
      author="{ATLAS Collaboration}",
      year={2024},
      eprint={2401.05299},
      archivePrefix={arXiv},
      primaryClass={hep-ex}
}

@misc{ATLAS-TOPQ-2020-20,
      title={Inclusive and differential cross-section measurements of $t\bar{t}Z$ production in $pp$ collisions at $\sqrt{s}=13$ TeV with the ATLAS detector, including EFT and spin-correlation interpretations}, 
      author="{ATLAS Collaboration}",
      year={2023},
      eprint={2312.04450},
      archivePrefix={arXiv},
      primaryClass={hep-ex}
}

@article{PhysRevD.100.015003,
  title = {Rare top decays as probes of flavorful Higgs bosons},
  author = {Altmannshofer, Wolfgang and Maddock, Brian and Tuckler, Douglas},
  journal = {Phys. Rev. D},
  volume = {100},
  number = {1},
  pages = {015003},
  numpages = {16},
  year = {2019},
  month = {Jul},
  publisher = {American Physical Society},
  doi = {10.1103/PhysRevD.100.015003},
eprint={1904.10956},
archivePrefix={arXiv},
primaryClass={hep-ph}
}

@Article{10.21468/SciPostPhysCodeb.8,
	author = "Bierlich, Christian and others",
 	title = "{A comprehensive guide to the physics and usage of PYTHIA 8.3}",
 	eprint = "2203.11601",
 	archivePrefix = "arXiv",
 	primaryClass = "hep-ph",
 	reportNumber = "LU-TP 22-16, MCNET-22-04, FERMILAB-PUB-22-227-SCD",
 	doi = "10.21468/SciPostPhysCodeb.8",
 	journal = "SciPost Phys. Codeb.",
 	volume = "2022",
 	pages = "8",
 	year = "2022"

}

@article{ATLAS:2024cmj,
	author = "{ATLAS Collaboration}",
    title = "{The performance of missing transverse momentum reconstruction and its significance with the ATLAS detector using \(140\,\text{fb}^{-1}\)
of \(\sqrt{s} = 13\,\text{TeV}\) \(pp\) collisions}",
	eprint = "2402.05858",
	archivePrefix = "arXiv",
	primaryClass = "hep-ex",
	reportNumber = "CERN-EP-2024-023",
	month = "2",
	year = "2024"
}

@Article{CMS-TOP-20-007_fixed,
    author         = "{CMS Collaboration}",
    title          = "{Search for flavor-changing neutral current interactions of the top quark and Higgs boson in final states with two photons in proton--proton collisions at \(\sqrt{s} = 13\,\text{TeV}\)}",
    year           = "2022",
    reportNumber   = "CERN-EP-2021-201",
    eprint         = "2111.02219",
    archivePrefix  = "arXiv",
    primaryClass   = "hep-ex",
    journal   = "JHEP",
    publisher = "Springer Science and Business Media LLC",
    number    =  {2},
    month     =  {2},
    volume    = "02",
    year      =  {2022},
    pages={169},
    doi       = {10.1007/JHEP02(2022)169}

}

@Article{Cacciari:2008gp,
     author    = "Cacciari, Matteo and Salam, Gavin P. and Soyez, Gregory",
     title     = "{The anti-\(k_{t}\) jet clustering algorithm}",
     journal   = "JHEP",
     volume    = "04",
     year      = "2008",
     pages     = "063",
     eprint    = "0802.1189",
     archivePrefix = "arXiv",
     primaryClass  =  "hep-ph",
     doi       = "10.1088/1126-6708/2008/04/063",
     SLACcitation  = "%%CITATION = 0802.1189;%%"
}

@Article{Fastjet,
      author         = "Cacciari, Matteo and Salam, Gavin P. and Soyez, Gregory",
      title          = "{FastJet user manual}",
      journal        = "Eur. Phys. J. C",
      volume         = "72",
      year           = "2012",
      pages          = "1896",
      doi            = "10.1140/epjc/s10052-012-1896-2",
      eprint         = "1111.6097",
      archivePrefix  = "arXiv",
      primaryClass   = "hep-ph",
      reportNumber   = "CERN-PH-TH-2011-297",
      SLACcitation   = "%%CITATION = ARXIV:1111.6097;%%"
}

@Article{Butterworth:2015oua,
      author         = "Butterworth, Jon and others",
      title          = "{PDF4LHC recommendations for LHC Run II}",
      journal        = "J. Phys. G",
      volume         = "43",
      year           = "2016",
      pages          = "023001",
      doi            = "10.1088/0954-3899/43/2/023001",
      eprint         = "1510.03865",
      archivePrefix  = "arXiv",
      primaryClass   = "hep-ph",
      reportNumber   = "OUTP-15-17P, SMU-HEP-15-12, TIF-UNIMI-2015-14,
                        LCTS-2015-27, CERN-PH-TH-2015-249",
      SLACcitation   = "%%CITATION = ARXIV:1510.03865;%%"
}

@article{Ball:2012cx,
      author         = "{NNPDF Collaboration} and Ball, Richard D. and others",
      title          = "{Parton distributions with LHC data}",
      journal        = "Nucl. Phys. B",
      volume         = "867",
      year           = "2013",
      pages          = "244",
      doi            = "10.1016/j.nuclphysb.2012.10.003",
      eprint         = "1207.1303",
      archivePrefix  = "arXiv",
      primaryClass   = "hep-ph",
      reportNumber   = "EDINBURGH-2012-08, IFUM-FT-997, FR-PHENO-2012-014,
                        RWTH-TTK-12-25, CERN-PH-TH-2012-037, SFB-CPP-12-47\,
                        --CERN-PH-TH-2012-037",
      SLACcitation   = "%%CITATION = ARXIV:1207.1303;%%"
}

@article{Ball:2014uwa,
      author         = "{NNPDF Collaboration} and Ball, Richard D. and others",
      title          = "{Parton distributions for the LHC run II}",
      journal        = "JHEP",
      volume         = "04",
      year           = "2015",
      pages          = "040",
      doi            = "10.1007/JHEP04(2015)040",
      eprint         = "1410.8849",
      archivePrefix  = "arXiv",
      primaryClass   = "hep-ph",
      reportNumber   = "EDINBURGH-2014-15, IFUM-1034-FT, CERN-PH-TH-2013-253,
                        OUTP-14-11P, CAVENDISH-HEP-14-11",
      SLACcitation   = "%%CITATION = ARXIV:1410.8849;%%"
}

@article{Sjostrand:2014zea,
      author         = "Sj{\"o}strand, Torbj{\"o}rn and Ask, Stefan and Christiansen,
                        Jesper R. and Corke, Richard and Desai, Nishita and Ilten,
                        Philip and Mrenna, Stephen and Prestel, Stefan and
                        Rasmussen, Christine O. and Skands, Peter Z.",
      title          = "{An introduction to PYTHIA 8.2}",
      journal        = "Comput. Phys. Commun.",
      volume         = "191",
      year           = "2015",
      pages          = "159",
      doi            = "10.1016/j.cpc.2015.01.024",
      eprint         = "1410.3012",
      archivePrefix  = "arXiv",
      primaryClass   = "hep-ph",
      reportNumber   = "LU-TP-14-36, MCNET-14-22, CERN-PH-TH-2014-190,
                        FERMILAB-PUB-14-316-CD, DESY-14-178, SLAC-PUB-16122,
                        --FERMILAB-PUB-14-316-CD",
      SLACcitation   = "%%CITATION = ARXIV:1410.3012;%%"
}

@Article{Lange:2001uf,
      author         = "Lange, D. J.",
      title          = "{The EvtGen particle decay simulation package}",
      booktitle      = "{Proceedings, 7th International Conference on B physics
                        at hadron machines (BEAUTY 2000)}",
      journal        = "Nucl. Instrum. Meth. A",
      volume         = "462",
      year           = "2001",
      pages          = "152",
      doi            = "10.1016/S0168-9002(01)00089-4",
      SLACcitation   = "%%CITATION = NUIMA,A462,152;%%"
}

@Article{Alwall:2014hca,
      author         = "Alwall, J. and Frederix, R. and Frixione, S. and Hirschi,
                        V. and Maltoni, F. and Mattelaer, O. and Shao, H. -S. and
                        Stelzer, T. and Torrielli, P. and Zaro, M.",
      title          = "{The automated computation of tree-level and
                        next-to-leading order differential cross sections, and
                        their matching to parton shower simulations}",
      journal        = "JHEP",
      volume         = "07",
      year           = "2014",
      pages          = "079",
      doi            = "10.1007/JHEP07(2014)079",
      eprint         = "1405.0301",
      archivePrefix  = "arXiv",
      primaryClass   = "hep-ph",
      reportNumber   = "CERN-PH-TH-2014-064, CP3-14-18, LPN14-066, MCNET-14-09,
                        ZU-TH-14-14",
      SLACcitation   = "%%CITATION = ARXIV:1405.0301;%%"
}

@article{Artoisenet:2012st,
      author         = "Artoisenet, Pierre and Frederix, Rikkert and Mattelaer,
                        Olivier and Rietkerk, Robbert",
      title          = "{Automatic spin-entangled decays of heavy resonances in
                        Monte Carlo simulations}",
      journal        = "JHEP",
      volume         = "03",
      pages          = "015",
      doi            = "10.1007/JHEP03(2013)015",
      year           = "2013",
      eprint         = "1212.3460",
      archivePrefix  = "arXiv",
      primaryClass   = "hep-ph",
      reportNumber   = "NIKHEF-2012-021, CERN-PH-TH-2012-329",
      SLACcitation   = "%%CITATION = ARXIV:1212.3460;%%",
}

@article{Frixione:2007zp,
      author         = "Frixione, Stefano and Laenen, Eric and Motylinski,
                        Patrick and Webber, Bryan R.",
      title          = "{Angular correlations of lepton pairs from vector boson
                        and top quark decays in Monte Carlo simulations}",
      journal        = "JHEP",
      volume         = "04",
      year           = "2007",
      pages          = "081",
      doi            = "10.1088/1126-6708/2007/04/081",
      eprint         = "hep-ph/0702198",
      archivePrefix  = "arXiv",
      OPTprimaryClass   = "HEP-PH",
      reportNumber   = "CAVENDISH-HEP-07-01, GEF-TH-09-2007, ITP-UU-07-10,
                        NIKHEF-2007-004",
      SLACcitation   = "%%CITATION = HEP-PH/0702198;%%"
}

@Article{Bahr:2008pv,
      author         = "B{\"a}hr, M. and others",
      title          = "{Herwig++ physics and manual}",
      journal        = "Eur. Phys. J. C",
      volume         = "58",
      year           = "2008",
      pages          = "639",
      doi            = "10.1140/epjc/s10052-008-0798-9",
      eprint         = "0803.0883",
      archivePrefix  = "arXiv",
      primaryClass   = "hep-ph",
      reportNumber   = "CERN-PH-TH-2008-038, CAVENDISH-HEP-08-03, KA-TP-05-2008,
                        DCPT-08-22, IPPP-08-11, CP3-08-05",
      SLACcitation   = "%%CITATION = ARXIV:0803.0883;%%"
}

@Article{Bellm:2015jjp,
      author         = "Bellm, Johannes and others",
      title          = "{Herwig 7.0/Herwig++ 3.0 release note}",
      journal        = "Eur. Phys. J. C",
      volume         = "76",
      year           = "2016",
      number         = "4",
      pages          = "196",
      doi            = "10.1140/epjc/s10052-016-4018-8",
      eprint         = "1512.01178",
      archivePrefix  = "arXiv",
      primaryClass   = "hep-ph",
      reportNumber   = "CERN-PH-TH-2015-289, MAN-HEP-2015-15, IFJPAN-IV-2015-13,
                        HERWIG-2015-01, KA-TP-18-2015, DCPT-15-142, MCNET-15-28,
                        IPPP-15-71, --HERWIG-2015-01",
      SLACcitation   = "%%CITATION = ARXIV:1512.01178;%%"
}

@Article{Nason:2004rx,
      author         = "Nason, Paolo",
      title          = "{A new method for combining NLO QCD with shower Monte Carlo algorithms}",
      journal        = "JHEP",
      volume         = "11",
      pages          = "040",
      doi            = "10.1088/1126-6708/2004/11/040",
      year           = "2004",
      eprint         = "hep-ph/0409146",
      archivePrefix  = "arXiv",
}

@Article{Frixione:2007vw,
      author         = "Frixione, Stefano and Nason, Paolo and Oleari, Carlo",
      title          = "{Matching NLO QCD computations with parton shower
                        simulations: the POWHEG method}",
      journal        = "JHEP",
      volume         = "11",
      pages          = "070",
      doi            = "10.1088/1126-6708/2007/11/070",
      year           = "2007",
      eprint         = "0709.2092",
      archivePrefix  = "arXiv",
      primaryClass   = "hep-ph",
}

@Article{Alioli:2010xd,
      author         = "Alioli, Simone and Nason, Paolo and Oleari, Carlo and Re,
                        Emanuele",
      title          = "{A general framework for implementing NLO calculations in
                        shower Monte Carlo programs: the POWHEG BOX}",
      journal        = "JHEP",
      volume         = "06",
      pages          = "043",
      doi            = "10.1007/JHEP06(2010)043",
      year           = "2010",
      eprint         = "1002.2581",
      archivePrefix  = "arXiv",
      primaryClass   = "hep-ph",
}

@article{Alioli:2009je,
      author         = "Alioli, Simone and Nason, Paolo and Oleari, Carlo and Re,
                        Emanuele",
      title          = "{NLO single-top production matched with shower in POWHEG:
                        \(s\)- and \(t\)-channel contributions}",
      journal        = "JHEP",
      volume         = "09",
      year           = "2009",
      pages          = "111",
      doi            = "10.1088/1126-6708/2009/09/111",
      eprint         = "0907.4076",
      archivePrefix  = "arXiv",
      primaryClass   = "hep-ph",
      SLACcitation   = "%%CITATION = ARXIV:0907.4076;%%",
      related        = "Alioli:2009je-err",
      relatedstring  = "Erratum:",
}

@article{Frederix:2012dh,
      author         = "Frederix, Rikkert and Re, Emanuele and Torrielli, Paolo",
      title          = "{Single-top \(t\)-channel hadroproduction in the four-flavour
                        scheme with POWHEG and aMC@NLO}",
      journal        = "JHEP",
      volume         = "09",
      year           = "2012",
      pages          = "130",
      doi            = "10.1007/JHEP09(2012)130",
      eprint         = "1207.5391",
      archivePrefix  = "arXiv",
      primaryClass   = "hep-ph",
      reportNumber   = "ZU-TH-14-12, IPPP-12-54, DCPT-12-108,
                        CERN-PH-TH-2012-206, MCNET-12-10, LPN12-084",
      SLACcitation   = "%%CITATION = ARXIV:1207.5391;%%"
}

@article{Aliev:2010zk,
      author         = "Aliev, M. and Lacker, H. and Langenfeld, U. and Moch, S.
                        and Uwer, P. and Wiedermann, M.",
      title          = "{HATHOR -- HAdronic Top and Heavy quarks crOss section
                        calculatoR}",
      journal        = "Comput. Phys. Commun.",
      volume         = "182",
      year           = "2011",
      pages          = "1034-1046",
      doi            = "10.1016/j.cpc.2010.12.040",
      eprint         = "1007.1327",
      archivePrefix  = "arXiv",
      primaryClass   = "hep-ph",
      reportNumber   = "DESY-10-091, HU-EP-10-33, SFB-CPP-10-60",
      SLACcitation   = "%%CITATION = ARXIV:1007.1327;%%"
}

@article{Re:2010bp,
      author         = "Re, Emanuele",
      title          = "{Single-top \(Wt\)-channel production matched with parton
                        showers using the POWHEG method}",
      journal        = "Eur. Phys. J. C",
      volume         = "71",
      year           = "2011",
      pages          = "1547",
      doi            = "10.1140/epjc/s10052-011-1547-z",
      eprint         = "1009.2450",
      archivePrefix  = "arXiv",
      primaryClass   = "hep-ph",
      reportNumber   = "IPPP-10-74, DCPT-10-148",
      SLACcitation   = "%%CITATION = ARXIV:1009.2450;%%"
}

@article{Kant:2014oha,
      author         = "Kant, P. and Kind, O. M. and Kintscher, T. and Lohse, T.
                        and Martini, T. and Mölbitz, S. and Rieck, P. and Uwer, P.",
      title          = "{HatHor for single top-quark production: Updated
                        predictions and uncertainty estimates for single top-quark
                        production in hadronic collisions}",
      journal        = "Comput. Phys. Commun.",
      volume         = "191",
      year           = "2015",
      pages          = "74-89",
      doi            = "10.1016/j.cpc.2015.02.001",
      eprint         = "1406.4403",
      archivePrefix  = "arXiv",
      primaryClass   = "hep-ph",
      reportNumber   = "HU-EP-14-22",
      SLACcitation   = "%%CITATION = ARXIV:1406.4403;%%"
}

@article{Frixione:2007nw,
      author         = "Frixione, Stefano and Ridolfi, Giovanni and Nason, Paolo",
      title          = "{A positive-weight next-to-leading-order Monte Carlo for
                        heavy flavour hadroproduction}",
      journal        = "JHEP",
      volume         = "09",
      pages          = "126",
      doi            = "10.1088/1126-6708/2007/09/126",
      year           = "2007",
      eprint         = "0707.3088",
      archivePrefix  = "arXiv",
      primaryClass   = "hep-ph",
      SLACcitation   = "%%CITATION = ARXIV:0707.3088;%%",
}

@article{Frederix:2017wme,
      author         = "Frederix, Rikkert and Pagani, Davide and Zaro, Marco",
      title          = "{Large NLO corrections in \(t\bar{t}W^{\pm}\) and
                        \(t\bar{t}t\bar{t}\) hadroproduction from supposedly
                        subleading EW contributions}",
      journal        = "JHEP",
      volume         = "02",
      year           = "2018",
      pages          = "031",
      doi            = "10.1007/JHEP02(2018)031",
      eprint         = "1711.02116",
      archivePrefix  = "arXiv",
      primaryClass   = "hep-ph",
      SLACcitation   = "%%CITATION = ARXIV:1711.02116;%%"
}

@article{deFlorian:2016spz,
      author         = "de Florian, D. and others",
      title          = "{Handbook of LHC Higgs Cross Sections: 4. Deciphering the
                        Nature of the Higgs Sector}",
      collaboration  = "LHC Higgs Cross Section Working Group",
      doi            = "10.23731/CYRM-2017-002",
      year           = "2017",
      eprint         = "1610.07922",
      archivePrefix  = "arXiv",
      primaryClass   = "hep-ph",
      reportNumber   = "FERMILAB-FN-1025-T, CERN-2017-002-M",
      SLACcitation   = "%%CITATION = ARXIV:1610.07922;%%"
}

@Article{Gleisberg:2008ta,
      author         = "Gleisberg, T. and H{\"o}che, S. and Krauss, F. and Sch{\"o}nherr, M. and Schumann, S. and Siegert, F. and Winter J.",
      title          = "{Event generation with SHERPA 1.1}",
      journal        = "JHEP",
      volume         = "02",
      pages          = "007",
      doi            = "10.1088/1126-6708/2009/02/007",
      year           = "2009",
      eprint         = "0811.4622",
      archivePrefix  = "arXiv",
      primaryClass   = "hep-ph",
}

@Article{Gleisberg:2008fv,
      author         = "Gleisberg, Tanju and H{\"o}che, Stefan",
      title          = "{Comix, a new matrix element generator}",
      journal        = "JHEP",
      volume         = "12",
      pages          = "039",
      doi            = "10.1088/1126-6708/2008/12/039",
      year           = "2008",
      eprint         = "0808.3674",
      archivePrefix  = "arXiv",
      primaryClass   = "hep-ph",
}

@Article{Schumann:2007mg,
      author         = "Schumann, Steffen and Krauss, Frank",
      title          = "{A parton shower algorithm based on Catani--Seymour dipole factorisation}",
      journal        = "JHEP",
      volume         = "03",
      pages          = "038",
      doi            = "10.1088/1126-6708/2008/03/038",
      year           = "2008",
      eprint         = "0709.1027",
      archivePrefix  = "arXiv",
      primaryClass   = "hep-ph",
}

@article{Hoeche:2012yf,
      author         = "H{\"o}che, Stefan and Krauss, Frank and Sch{\"o}nherr, Marek and
                        Siegert, Frank",
      title          = "{QCD matrix elements + parton showers. The NLO case}",
      journal        = "JHEP",
      volume         = "04",
      year           = "2013",
      pages          = "027",
      doi            = "10.1007/JHEP04(2013)027",
      eprint         = "1207.5030",
      archivePrefix  = "arXiv",
      primaryClass   = "hep-ph",
      reportNumber   = "SLAC-PUB-15191, IPPP-12-52, DCPT-12-104, LPN12-081,
                        FR-PHENO-2012-017, MCNET-12-09, --FR-PHENO-2012-017",
      SLACcitation   = "%%CITATION = ARXIV:1207.5030;%%"
}

@article{Bothmann:2019yzt,
      author = "Bothmann, Enrico and others",
      title = "{Event generation with Sherpa 2.2}",
      journal = "SciPost Phys.",
      volume = "7",
      year = "2019",
      number = "3",
      pages = "034",
      doi = "10.21468/SciPostPhys.7.3.034",
      reportNumber = "FERMILAB-PUB-19-218-T, SLAC-PUB-17433, IPPP/19/42, MCNET-19-11",
      eprint = "1905.09127",
      archivePrefix = "arXiv",
      primaryClass = "hep-ph",
}

@article{Winter:2003tt,
      author         = "Winter, Jan-Christopher and Krauss, Frank and Soff,
                        Gerhard",
      title          = "{A modified cluster-hadronisation model}",
      journal        = "Eur. Phys. J. C",
      volume         = "36",
      year           = "2004",
      pages          = "381-395",
      doi            = "10.1140/epjc/s2004-01960-8",
      eprint         = "hep-ph/0311085",
      archivePrefix  = "arXiv",
      XprimaryClass   = "hep-ph",
      reportNumber   = "CERN-TH-2003-272",
      SLACcitation   = "%%CITATION = HEP-PH/0311085;%%"
}

@article{Cascioli:2011va,
      author         = "Cascioli, Fabio and Maierh{\"o}fer, Philipp and Pozzorini, Stefano",
      title          = "{Scattering Amplitudes with Open Loops}",
      journal        = "Phys. Rev. Lett.",
      volume         = "108",
      year           = "2012",
      pages          = "111601",
      doi            = "10.1103/PhysRevLett.108.111601",
      eprint         = "1111.5206",
      archivePrefix  = "arXiv",
      primaryClass   = "hep-ph",
      reportNumber   = "ZU-TH-23-11, LPN11-66",
      SLACcitation   = "%%CITATION = ARXIV:1111.5206;%%"
}

@article{Buccioni:2019sur,
    author = "Buccioni, Federico and Lang, Jean-Nicolas and Lindert, Jonas M. and Maierh{\"o}fer, Philipp and Pozzorini, Stefano and Zhang, Hantian and Zoller, Max F.",
    title = "{OpenLoops 2}",
    eprint = "1907.13071",
    archivePrefix = "arXiv",
    primaryClass = "hep-ph",
    reportNumber = "IPPP/19/62, FR-PHENO-2019-12, PSI-PR-19-15, ZU-TH 37/19",
    doi = "10.1140/epjc/s10052-019-7306-2",
    journal = "Eur. Phys. J. C",
    volume = "79",
    number = "10",
    pages = "866",
    year = "2019"
}

@article{Denner:2016kdg,
      author         = "Denner, Ansgar and Dittmaier, Stefan and Hofer, Lars",
      title          = "{\textsc{Collier}: A fortran-based complex one-loop library in
                        extended regularizations}",
      journal        = "Comput. Phys. Commun.",
      volume         = "212",
      year           = "2017",
      pages          = "220-238",
      doi            = "10.1016/j.cpc.2016.10.013",
      eprint         = "1604.06792",
      archivePrefix  = "arXiv",
      primaryClass   = "hep-ph",
      reportNumber   = "FR-PHENO-2016-003, ICCUB-16-016",
      SLACcitation   = "%%CITATION = ARXIV:1604.06792;%%"
}

@article{Kallweit:2015dum,
    author = "Kallweit, Stefan and Lindert, Jonas M. and Maierh{\"o}fer, Philipp and Pozzorini, Stefano and Sch{\"o}nherr, Marek",
    title = "{NLO QCD+EW predictions for V + jets including off-shell vector-boson decays and multijet merging}",
    eprint = "1511.08692",
    archivePrefix = "arXiv",
    primaryClass = "hep-ph",
    reportNumber = "DCPT-15-140, FR-PHENO-2015-014, IPPP-15-70, MCNET-15-23, ZU-TH-41-15, MITP-15-108",
    doi = "10.1007/JHEP04(2016)021",
    journal = "JHEP",
    volume = "04",
    pages = "021",
    year = "2016"
}

@article{Czakon:2011xx,
  author         = "Czakon, Michal and Mitov, Alexander",
  title          = "{Top++: A program for the calculation of the top-pair cross-section at hadron colliders}",
  journal        = "Comput. Phys. Commun.",
  volume         = "185",
  year           = "2014",
  pages          = "2930",
  doi            = "10.1016/j.cpc.2014.06.021",
  eprint         = "1112.5675",
  archivePrefix  = "arXiv",
  primaryClass   = "hep-ph",
  reportNumber   = "CERN-PH-TH-2011-303, TTK-11-58",
}

@article{Agostinelli:2002hh,
      author         = "Agostinelli, S. and others",
      title          = "{\textsc{Geant4} -- a simulation toolkit}",
      journal        = "Nucl. Instrum. Meth. A",
      volume         = "506",
      year           = "2003",
      pages          = "250",
      doi            = "10.1016/S0168-9002(03)01368-8",
      reportNumber   = "SLAC-PUB-9350, FERMILAB-PUB-03-339",
      SLACcitation   = "%%CITATION = NUIMA,A506,250;%%"
}

@Article{Read:2002hq,
    Author = {Read, Alexander L.},
    Journal = {J. Phys. G},
    Pages = {2693},
    doi = "10.1088/0954-3899/28/10/313",
    Title = "{Presentation of search results: the \(CL_S\) technique}",
    Volume = {28},
    Year = {2002}
}

@Article{Cowan:2010js,
   Author = {Cowan, Glen and Cranmer, Kyle and Gross, Eilam and Vitells, Ofer},
   Title = {{Asymptotic formulae for likelihood-based tests of new physics}},
   Journal = {Eur. Phys. J. C},
   Volume = {71},
   Year = {2011},
   Pages = {1554},
   doi = "10.1140/epjc/s10052-011-1554-0",
   Eprint = "1007.1727",
   Archiveprefix = {arXiv},
   Primaryclass = {physics.data-an},
   related = "Cowan:2010js-err",
   relatedstring  = "Erratum:",
}

@Booklet{RooFit,
      author         = "Verkerke, Wouter and Kirkby, David",
      title          = "{The RooFit toolkit for data modeling}",
      year           = "2003",
      eprint         = "physics/0306116",
      archivePrefix  = "arXiv",
      primaryClass   = "physics.data-an",
}

@article{Barlow:1993dm,
    author = "Barlow, Roger and Beeston, Christine",
    title = "{Fitting using finite Monte Carlo samples}",
    reportNumber = "MAN-HEP-93-1",
    doi = "10.1016/0010-4655(93)90005-W",
    journal = "Comput. Phys. Commun.",
    volume = "77",
    pages = "219",
    year = "1993"
}

@Article{PERF-2007-01,
    author         = "{ATLAS Collaboration}",
    title          = "{The ATLAS Experiment at the CERN Large Hadron Collider}",
    journal        = "JINST",
    volume         = "3",
    year           = "2008",
    pages          = "S08003",
    doi            = "10.1088/1748-0221/3/08/S08003",
    primaryClass   = "hep-ex",
}

@Article{SOFT-2010-01,
    author         = "{ATLAS Collaboration}",
    title          = "{The ATLAS Simulation Infrastructure}",
    journal        = "Eur. Phys. J. C",
    volume         = "70",
    year           = "2010",
    pages          = "823",
    doi            = "10.1140/epjc/s10052-010-1429-9",
    eprint         = "1005.4568",
    archivePrefix  = "arXiv",
    primaryClass   = "physics.ins-det",
}

@Article{HIGG-2012-27,
    author         = "{ATLAS Collaboration}",
    title          = "{Observation of a new particle in the search for the Standard Model Higgs boson with the ATLAS detector at the LHC}",
    journal        = "Phys. Lett. B",
    volume         = "716",
    year           = "2012",
    pages          = "1",
    doi            = "10.1016/j.physletb.2012.08.020",
    reportNumber   = "CERN-PH-EP-2012-218",
    eprint         = "1207.7214",
    archivePrefix  = "arXiv",
    primaryClass   = "hep-ex",
}

@Article{PERF-2014-03,
    author         = "{ATLAS Collaboration}",
    title          = "{Performance of pile-up mitigation techniques for jets in \(pp\) collisions at \(\sqrt{s} = 8\,\text{TeV}\) using the ATLAS detector}",
    journal        = "Eur. Phys. J. C",
    volume         = "76",
    year           = "2016",
    pages          = "581",
    doi            = "10.1140/epjc/s10052-016-4395-z",
    reportNumber   = "CERN-PH-EP-2015-206",
    eprint         = "1510.03823",
    archivePrefix  = "arXiv",
    primaryClass   = "hep-ex",
}

@Article{PERF-2014-07,
    author         = "{ATLAS Collaboration}",
    title          = "{Topological cell clustering in the ATLAS calorimeters and its performance in LHC Run~1}",
    journal        = "Eur. Phys. J. C",
    volume         = "77",
    year           = "2017",
    pages          = "490",
    doi            = "10.1140/epjc/s10052-017-5004-5",
    reportNumber   = "CERN-PH-EP-2015-304",
    eprint         = "1603.02934",
    archivePrefix  = "arXiv",
    primaryClass   = "hep-ex",
}

@Article{PERF-2015-09,
    author         = "{ATLAS Collaboration}",
    title          = "{Jet reconstruction and performance using particle flow with the ATLAS Detector}",
    journal        = "Eur. Phys. J. C",
    volume         = "77",
    year           = "2017",
    pages          = "466",
    doi            = "10.1140/epjc/s10052-017-5031-2",
    reportNumber   = "CERN-EP-2017-024",
    eprint         = "1703.10485",
    archivePrefix  = "arXiv",
    primaryClass   = "hep-ex",
}

@Article{HIGG-2016-26,
    author         = "{ATLAS Collaboration}",
    title          = "{Search for top quark decays \(t\rightarrow qH\), with \(H\rightarrow\gamma\gamma\), in \(\sqrt{s} = 13\,\text{TeV}\) \(pp\) collisions using the ATLAS detector}",
    journal        = "JHEP",
    volume         = "10",
    year           = "2017",
    pages          = "129",
    doi            = "10.1007/JHEP10(2017)129",
    reportNumber   = "CERN-EP-2017-118",
    eprint         = "1707.01404",
    archivePrefix  = "arXiv",
    primaryClass   = "hep-ex",
}

@Article{TOPQ-2016-14,
    author         = "{ATLAS Collaboration}",
    title          = "{Measurement of the production cross-section of a single top quark in association with a \(Z\) boson in proton--proton collisions at \(13\,\text{TeV}\) with the ATLAS detector}",
    journal        = "Phys. Lett. B",
    volume         = "780",
    year           = "2018",
    pages          = "557",
    doi            = "10.1016/j.physletb.2018.03.023",
    reportNumber   = "CERN-EP-2017-188",
    eprint         = "1710.03659",
    archivePrefix  = "arXiv",
    primaryClass   = "hep-ex",
}

@Article{TRIG-2016-01,
    author         = "{ATLAS Collaboration}",
    title          = "{Performance of the ATLAS trigger system in 2015}",
    journal        = "Eur. Phys. J. C",
    volume         = "77",
    year           = "2017",
    pages          = "317",
    doi            = "10.1140/epjc/s10052-017-4852-3",
    reportNumber   = "CERN-EP-2016-241",
    eprint         = "1611.09661",
    archivePrefix  = "arXiv",
    primaryClass   = "hep-ex",
}

@Article{HIGG-2017-02,
    author         = "{ATLAS Collaboration}",
    title          = "{Evidence for the associated production of the Higgs boson and a top quark pair with the ATLAS detector}",
    journal        = "Phys. Rev. D",
    volume         = "97",
    year           = "2018",
    pages          = "072003",
    doi            = "10.1103/PhysRevD.97.072003",
    reportNumber   = "CERN-EP-2017-281",
    eprint         = "1712.08891",
    archivePrefix  = "arXiv",
    primaryClass   = "hep-ex",
}

@Article{TOPQ-2017-07,
    author         = "{ATLAS Collaboration}",
    title          = "{Search for top-quark decays \(t \to Hq\) with \(36\,\text{fb}^{-1}\) of \(pp\) collision data at \(\sqrt{s} = 13\,\text{TeV}\) with the ATLAS detector}",
    journal        = "JHEP",
    volume         = "05",
    year           = "2019",
    pages          = "123",
    doi            = "10.1007/JHEP05(2019)123",
    reportNumber   = "CERN-EP-2018-295",
    eprint         = "1812.11568",
    archivePrefix  = "arXiv",
    primaryClass   = "hep-ex",
}

@Article{TOPQ-2017-15,
    author         = "{ATLAS Collaboration}",
    title          = "{Search for flavor-changing neutral currents in top quark decays \(t \rightarrow Hc\) and \(t \rightarrow Hu\) in multilepton final states in proton--proton collisions at \(\sqrt{s} = 13\,\text{TeV}\) with the ATLAS detector}",
    journal        = "Phys. Rev. D",
    volume         = "98",
    year           = "2018",
    pages          = "032002",
    doi            = "10.1103/PhysRevD.98.032002",
    reportNumber   = "CERN-EP-2018-067",
    eprint         = "1805.03483",
    archivePrefix  = "arXiv",
    primaryClass   = "hep-ex",
}

@Article{DAPR-2018-01,
    author         = "{ATLAS Collaboration}",
    title          = "{ATLAS data quality operations and performance for 2015--2018 data-taking}",
    journal        = "JINST",
    volume         = "15",
    year           = "2020",
    pages          = "P04003",
    doi            = "10.1088/1748-0221/15/04/P04003",
    reportNumber   = "CERN-EP-2019-207",
    eprint         = "1911.04632",
    archivePrefix  = "arXiv",
    primaryClass   = "physics.ins-det",
}

@Article{EGAM-2018-01,
    author         = "{ATLAS Collaboration}",
    title          = "{Electron and photon performance measurements with the ATLAS detector using the 2015--2017 LHC proton--proton collision data}",
    journal        = "JINST",
    volume         = "14",
    year           = "2019",
    pages          = "P12006",
    doi            = "10.1088/1748-0221/14/12/P12006",
    reportNumber   = "CERN-EP-2019-145",
    eprint         = "1908.00005",
    archivePrefix  = "arXiv",
    primaryClass   = "hep-ex",
}

@Article{FTAG-2018-01,
    author         = "{ATLAS Collaboration}",
    title          = "{ATLAS \(b\)-jet identification performance and efficiency measurement with \(t\bar{t}\) events in \(pp\) collisions at \(\sqrt{s} = 13\,\text{TeV}\)}",
    journal        = "Eur. Phys. J. C",
    volume         = "79",
    year           = "2019",
    pages          = "970",
    doi            = "10.1140/epjc/s10052-019-7450-8",
    reportNumber   = "CERN-EP-2019-132",
    eprint         = "1907.05120",
    archivePrefix  = "arXiv",
    primaryClass   = "hep-ex",
}

@Article{JETM-2018-05,
    author         = "{ATLAS Collaboration}",
    title          = "{Jet energy scale and resolution measured in proton--proton collisions at \(\sqrt{s} = 13\,\text{TeV}\) with the ATLAS detector}",
    journal        = "Eur. Phys. J. C",
    volume         = "81",
    year           = "2021",
    pages          = "689",
    doi            = "10.1140/epjc/s10052-021-09402-3",
    reportNumber   = "CERN-EP-2020-083",
    eprint         = "2007.02645",
    archivePrefix  = "arXiv",
    primaryClass   = "hep-ex",
}

@Article{MUON-2018-03,
    author         = "{ATLAS Collaboration}",
    title          = "{Muon reconstruction and identification efficiency in ATLAS using the full Run~2 \(pp\) collision data set at \(\sqrt{s} = 13\,\text{TeV}\)}",
    journal        = "Eur. Phys. J. C",
    volume         = "81",
    year           = "2021",
    pages          = "578",
    doi            = "10.1140/epjc/s10052-021-09233-2",
    reportNumber   = "CERN-EP-2020-199",
    eprint         = "2012.00578",
    archivePrefix  = "arXiv",
    primaryClass   = "hep-ex",
}

@Article{PIX-2018-001,
    author         = "Abbott, B. and others",
    title          = "{Production and integration of the ATLAS Insertable B-Layer}",
    journal        = "JINST",
    volume         = "13",
    year           = "2018",
    pages          = "T05008",
    doi            = "10.1088/1748-0221/13/05/T05008",
    eprint         = "1803.00844",
    archivePrefix  = "arXiv",
    primaryClass   = "physics.ins-det",
}

@Article{TRIG-2018-01,
    author         = "{ATLAS Collaboration}",
    title          = "{Performance of the ATLAS muon triggers in Run~2}",
    journal        = "JINST",
    volume         = "15",
    year           = "2020",
    pages          = "P09015",
    doi            = "10.1088/1748-0221/15/09/p09015",
    reportNumber   = "CERN-EP-2020-031",
    eprint         = "2004.13447",
    archivePrefix  = "arXiv",
    primaryClass   = "physics.ins-det",
}

@Article{TRIG-2018-05,
    author         = "{ATLAS Collaboration}",
    title          = "{Performance of electron and photon triggers in ATLAS during LHC Run~2}",
    journal        = "Eur. Phys. J. C",
    volume         = "80",
    year           = "2020",
    pages          = "47",
    doi            = "10.1140/epjc/s10052-019-7500-2",
    reportNumber   = "CERN-EP-2019-169",
    eprint         = "1909.00761",
    archivePrefix  = "arXiv",
    primaryClass   = "hep-ex",
}

@Article{FTAG-2019-02,
    author         = "{ATLAS Collaboration}",
    title          = "{Calibration of the light-flavour jet mistagging efficiency of the \(b\)-tagging algorithms with \(Z\)+jets events using \(139\,\text{fb}^{-1}\) of ATLAS proton--proton collision data at \(\sqrt{s} = 13\,\text{TeV}\)}",
    journal        = "Eur. Phys. J. C",
    volume         = "83",
    year           = "2023",
    pages          = "728",
    doi            = "10.1140/epjc/s10052-023-11736-z",
    reportNumber   = "CERN-EP-2022-211",
    eprint         = "2301.06319",
    archivePrefix  = "arXiv",
    primaryClass   = "hep-ex",
}

@Article{FTAG-2019-07,
    author         = "{ATLAS Collaboration}",
    title          = "{ATLAS flavour-tagging algorithms for the LHC Run~2 \(pp\) collision dataset}",
    journal        = "Eur. Phys. J. C",
    volume         = "83",
    year           = "2023",
    pages          = "681",
    doi            = "10.1140/epjc/s10052-023-11699-1",
    reportNumber   = "CERN-EP-2022-226",
    eprint         = "2211.16345",
    archivePrefix  = "arXiv",
    primaryClass   = "physics.data-an",
}

@Article{TOPQ-2019-17,
    author         = "{ATLAS Collaboration}",
    title          = "{Search for flavour-changing neutral current interactions of the top quark and the Higgs boson in events with a pair of \(\tau\)-leptons in \(pp\) collisions at \(\sqrt{s} = 13\,\text{TeV}\) with the ATLAS detector}",
    journal        = "JHEP",
    volume         = "06",
    year           = "2023",
    pages          = "155",
    doi            = "10.1007/JHEP06(2023)155",
    reportNumber   = "CERN-EP-2022-123",
    eprint         = "2208.11415",
    archivePrefix  = "arXiv",
    primaryClass   = "hep-ex",
}

@Article{FTAG-2020-08,
    author         = "{ATLAS Collaboration}",
    title          = "{Measurement of the \(c\)-jet mistagging efficiency in \(t\bar{t}\) events using \(pp\) collision data at \(\sqrt{s} = 13\,\text{TeV}\) collected with the ATLAS detector}",
    journal        = "Eur. Phys. J. C",
    volume         = "82",
    year           = "2022",
    pages          = "95",
    doi            = "10.1140/epjc/s10052-021-09843-w",
    reportNumber   = "CERN-EP-2021-166",
    eprint         = "2109.10627",
    archivePrefix  = "arXiv",
    primaryClass   = "hep-ex",
}

@Article{DAPR-2021-01,
    author         = "{ATLAS Collaboration}",
    title          = "{Luminosity determination in \(pp\) collisions at \(\sqrt{s} = 13\,\text{TeV}\) using the ATLAS detector at the LHC}",
    journal        = "Eur. Phys. J. C",
    volume         = "83",
    year           = "2023",
    pages          = "982",
    doi            = "10.1140/epjc/s10052-023-11747-w",
    reportNumber   = "CERN-EP-2022-281",
    eprint         = "2212.09379",
    archivePrefix  = "arXiv",
    primaryClass   = "hep-ex",
}

@Article{EGAM-2021-01,
    author         = "{ATLAS Collaboration}",
    title          = "{Electron and photon efficiencies in LHC Run~2 with the ATLAS experiment}",
    year           = "2023",
    reportNumber   = "CERN-EP-2023-182",
    eprint         = "2308.13362",
    archivePrefix  = "arXiv",
    primaryClass   = "hep-ex",
}

@Article{HIGG-2021-23,
    author         = "{ATLAS Collaboration}",
    title          = "{A detailed map of Higgs boson interactions by the ATLAS experiment ten years after the discovery}",
    journal        = "Nature",
    volume         = "607",
    year           = "2022",
    pages          = "52--59",
    doi            = "10.1038/s41586-022-04893-w",
    reportNumber   = "CERN-EP-2022-057",
    eprint         = "2207.00092",
    archivePrefix  = "arXiv",
    primaryClass   = "hep-ex",
    related        = "HIGG-2021-23-err",
    relatedstring  = "Erratum:",
}

@Article{TOPQ-2021-08,
    author         = "{ATLAS Collaboration}",
    title          = "{Observation of four-top-quark production in the multilepton final state with the ATLAS detector}",
    journal        = "Eur. Phys. J. C",
    volume         = "83",
    year           = "2023",
    pages          = "496",
    doi            = "10.1140/epjc/s10052-023-11573-0",
    reportNumber   = "CERN-EP-2023-055",
    eprint         = "2303.15061",
    archivePrefix  = "arXiv",
    primaryClass   = "hep-ex",
}

@Booklet{ATL-SOFT-PUB-2021-001,
    author         = "{ATLAS Collaboration}",
    title          = "{The ATLAS Collaboration Software and Firmware}",
    howpublished   = "{ATL-SOFT-PUB-2021-001}",
    url            = "https://cds.cern.ch/record/2767187",
    year           = "2021",
}

@Booklet{ATL-SOFT-PUB-2023-001,
    author         = "{ATLAS Collaboration}",
    title          = "{ATLAS Computing Acknowledgements}",
    howpublished   = "{ATL-SOFT-PUB-2023-001}",
    url            = "https://cds.cern.ch/record/2869272",
    year           = "2023",
}

@Report{ATLAS-TDR-19,
    author         = "{ATLAS Collaboration}",
    title          = "{ATLAS Insertable B-Layer: Technical Design Report}",
    type           = "ATLAS-TDR-19; CERN-LHCC-2010-013",
    year           = "2010",
    url            = "https://cds.cern.ch/record/1291633",
    related        = "ATLAS-TDR-19-addm",
    relatedstring  = "Addendum:",
}

@Article{CMS-HIG-12-028,
    author         = "{CMS Collaboration}",
    title          = "{Observation of a new boson at a mass of 125 GeV with the CMS experiment at the LHC}",
    journal        = "Phys. Lett. B",
    volume         = "716",
    year           = "2012",
    pages          = "30",
    doi            = "10.1016/j.physletb.2012.08.021",
    reportNumber   = "CERN-PH-EP-2012-220",
    eprint         = "1207.7235",
    archivePrefix  = "arXiv",
    primaryClass   = "hep-ex",
}

@Article{CMS-TOP-17-003,
    author         = "{CMS Collaboration}",
    title          = "{Search for the flavor-changing neutral current interactions of the top quark and the Higgs boson which decays into a pair of \(b\) quarks at \(\sqrt{s} = 13\,\text{TeV}\)}",
    journal        = "JHEP",
    volume         = "06",
    year           = "2018",
    pages          = "102",
    doi            = "10.1007/JHEP06(2018)102",
    reportNumber   = "CERN-EP-2017-309",
    eprint         = "1712.02399",
    archivePrefix  = "arXiv",
    primaryClass   = "hep-ex",
}

@Article{CMS-TOP-19-002,
    author         = "{CMS Collaboration}",
    title          = "{Search for flavor-changing neutral current interactions of the top quark and the Higgs boson decaying to a bottom quark-antiquark pair at \(\sqrt{s}\) =\(13\,\text{TeV}\)}",
    journal        = "JHEP",
    volume         = "02",
    year           = "2022",
    pages          = "169",
    doi            = "10.1007/JHEP02(2022)169",
    reportNumber   = "CERN-EP-2021-241",
    eprint         = "2112.09734",
    archivePrefix  = "arXiv",
    primaryClass   = "hep-ex",
}

@Article{CMS-HIG-22-001,
    author         = "{CMS Collaboration}",
    title          = "{A portrait of the Higgs boson by the CMS experiment ten years after the discovery}",
    journal        = "Nature",
    volume         = "607",
    year           = "2022",
    pages          = "60--68",
    doi            = "10.1038/s41586-022-04892-x",
    reportNumber   = "CERN-EP-2022-039",
    eprint         = "2207.00043",
    archivePrefix  = "arXiv",
    primaryClass   = "hep-ex",
    related        = "CMS-HIG-22-001-err",
    relatedstring  = "Erratum:",
}

@Booklet{ATL-PHYS-PUB-2010-013,
    author         = "{ATLAS Collaboration}",
    title          = "{The simulation principle and performance of the ATLAS fast calorimeter simulation FastCaloSim}",
    howpublished   = "{ATL-PHYS-PUB-2010-013}",
    url            = "https://cds.cern.ch/record/1300517",
    year           = "2010",
}

@Booklet{ATL-PHYS-PUB-2014-021,
    author         = "{ATLAS Collaboration}",
    title          = "{ATLAS Pythia~8 tunes to \(7~\text{TeV}\) data}",
    howpublished   = "{ATL-PHYS-PUB-2014-021}",
    url            = "https://cds.cern.ch/record/1966419",
    year           = "2014",
}

@Booklet{ATL-PHYS-PUB-2015-026,
    author         = "{ATLAS Collaboration}",
    title          = "{Vertex Reconstruction Performance of the ATLAS Detector at \(\sqrt{s} = 13~\text{TeV}\)}",
    howpublished   = "{ATL-PHYS-PUB-2015-026}",
    url            = "https://cds.cern.ch/record/2037717",
    year           = "2015",
}

@Booklet{ATL-PHYS-PUB-2016-017,
    author         = "{ATLAS Collaboration}",
    title          = "{The Pythia~8 A3 tune description of ATLAS minimum bias and inelastic measurements incorporating the Donnachie--Landshoff diffractive model}",
    howpublished   = "{ATL-PHYS-PUB-2016-017}",
    url            = "https://cds.cern.ch/record/2206965",
    year           = "2016",
}

@Booklet{ATL-PHYS-PUB-2016-020,
    author         = "{ATLAS Collaboration}",
    title          = "{Studies on top-quark Monte Carlo modelling for Top2016}",
    howpublished   = "{ATL-PHYS-PUB-2016-020}",
    url            = "https://cds.cern.ch/record/2216168",
    year           = "2016",
}
\clearpage
% ATLAS Collaboration author list
% Reference date of TOPQ-2020-23 is 2023-12-12
% Author list last updated on date 27-JUN-24
% Data extracted on 27-Jun-2024 for paper reference TOPQ-2020-23
% at 9:07am
 
\begin{flushleft}
\hypersetup{urlcolor=black}
{\Large The ATLAS Collaboration}

\bigskip

\AtlasOrcid[0000-0002-6665-4934]{G.~Aad}$^\textrm{\scriptsize 103}$,
\AtlasOrcid[0000-0001-7616-1554]{E.~Aakvaag}$^\textrm{\scriptsize 16}$,
\AtlasOrcid[0000-0002-5888-2734]{B.~Abbott}$^\textrm{\scriptsize 121}$,
\AtlasOrcid[0000-0002-1002-1652]{K.~Abeling}$^\textrm{\scriptsize 55}$,
\AtlasOrcid[0000-0001-5763-2760]{N.J.~Abicht}$^\textrm{\scriptsize 49}$,
\AtlasOrcid[0000-0002-8496-9294]{S.H.~Abidi}$^\textrm{\scriptsize 29}$,
\AtlasOrcid[0009-0003-6578-220X]{M.~Aboelela}$^\textrm{\scriptsize 44}$,
\AtlasOrcid[0000-0002-9987-2292]{A.~Aboulhorma}$^\textrm{\scriptsize 35e}$,
\AtlasOrcid[0000-0001-5329-6640]{H.~Abramowicz}$^\textrm{\scriptsize 152}$,
\AtlasOrcid[0000-0002-1599-2896]{H.~Abreu}$^\textrm{\scriptsize 151}$
% ...
\bigskip
\\

% ...
$^{16}$Department for Physics and Technology, University of Bergen, Bergen; Norway.\\
% ...
$^{29}$Physics Department, Brookhaven National Laboratory, Upton NY; United States of America.\\
% ...
$^{35}$$^{(a)}$Facult\'e des Sciences Ain Chock, R\'eseau Universitaire de Physique des Hautes Energies - Universit\'e Hassan II, Casablanca;$^{(b)}$Facult\'{e} des Sciences, Universit\'{e} Ibn-Tofail, K\'{e}nitra;$^{(c)}$Facult\'e des Sciences Semlalia, Universit\'e Cadi Ayyad, LPHEA-Marrakech;$^{(d)}$LPMR, Facult\'e des Sciences, Universit\'e Mohamed Premier, Oujda;$^{(e)}$Facult\'e des sciences, Universit\'e Mohammed V, Rabat;$^{(f)}$Institute of Applied Physics, Mohammed VI Polytechnic University, Ben Guerir; Morocco.\\
% ...
$^{44}$Physics Department, Southern Methodist University, Dallas TX; United States of America.\\
% ...
$^{49}$Fakult\"{a}t Physik , Technische Universit{\"a}t Dortmund, Dortmund; Germany.\\
% ...
$^{55}$II. Physikalisches Institut, Georg-August-Universit\"{a}t G\"ottingen, G\"ottingen; Germany.\\
% ...
$^{103}$CPPM, Aix-Marseille Universit\'e, CNRS/IN2P3, Marseille; France.\\
% ...
$^{121}$Homer L. Dodge Department of Physics and Astronomy, University of Oklahoma, Norman OK; United States of America.\\
% ...
$^{151}$Department of Physics, Technion, Israel Institute of Technology, Haifa; Israel.\\
$^{152}$Raymond and Beverly Sackler School of Physics and Astronomy, Tel Aviv University, Tel Aviv; Israel.\\
% ...
$^{a}$ Also Affiliated with an institute covered by a cooperation agreement with CERN.\\
$^{b}$ Also at An-Najah National University, Nablus; Palestine.\\
$^{c}$ Also at Borough of Manhattan Community College, City University of New York, New York NY; United States of America.\\
$^{d}$ Also at Center for High Energy Physics, Peking University; China.\\
$^{e}$ Also at Center for Interdisciplinary Research and Innovation (CIRI-AUTH), Thessaloniki; Greece.\\
$^{f}$ Also at Centro Studi e Ricerche Enrico Fermi; Italy.\\
% ...

\end{flushleft}

% Created with Glance <Atlas.Glance@cern.ch>

%

%
%

%

%
%
%
%

\end{document}